\documentclass[12pt,preprint]{aastex}
\usepackage{graphicx}
\usepackage{epstopdf}
\usepackage{natbib}
\usepackage{amsmath}
\usepackage{amssymb}
\shorttitle{Dust Haze}
\shortauthors{Hiranaka et al.}

\newcommand{\noprint}[1]{}
\newcommand{\figsetstart}{{\bf Fig. Set} }
\newcommand{\figsetend}{}
\newcommand{\figsetgrpstart}{}
\newcommand{\figsetgrpend}{}
\newcommand{\figsetnum}[1]{{\bf #1.}}
\newcommand{\figsettitle}[1]{ {\bf #1} }
\newcommand{\figsetgrpnum}[1]{\noprint{#1}}
\newcommand{\figsetgrptitle}[1]{\noprint{#1}}
\newcommand{\figsetplot}[1]{\noprint{#1}}
\newcommand{\figsetgrpnote}[1]{\noprint{#1}}
\begin{document}
\title{Exploring the Role of Sub-micron Sized Dust Grains in the Atmospheres of Red L0 -- L6 Dwarfs}
\author{\altaffilmark{1, 2, 3}{Kay Hiranaka}}
\altaffiltext{1}{Hunter College, Department of Physics and Astronomy, City University of New York, 695 Park Ave, New York, NY 10065, U.S.A.} 
\altaffiltext{2}{CUNY Graduate Center,  City University of New York, 365 Fifth Avenue, New York, NY 10016, U.S.A.}
\email{khiranak@hunter.cuny.edu} 
\author{\altaffilmark{1, 2, 3, 4}{Kelle L. Cruz}}
\altaffiltext{3}{American Museum of Natural History, Department of Astrophysics, Central Park West at 79th Street, New York, NY, 10024, U.S.A.}
\altaffiltext{4}{Visiting Astronomer at the Infrared Telescope Facility, which is operated by the University of Hawaii under Cooperative Agreement no. NNX-08AE38A with the National Aeronautics and Space Administration, Science Mission Directorate, Planetary Astronomy Program.}
%\email{kellecruz@gmail.com}
%\email{mark.s.marley@nasa.gov}                                    
\author{\altaffilmark{3, 5}{Stephanie T. Douglas}}
\altaffiltext{5}{Columbia University, Department of Astronomy, 550 West 120th Street, Mail Code 5246, New York, NY, 10027, U.S.A.}
\author{\altaffilmark{6}{Mark S. Marley}}
\altaffiltext{6}{NASA Ames Research Center, MS-245-3, Moffett Field, CA 94035, U.S.A.}
\and
\author{\altaffilmark{1, 3, 7}{Vivienne F. Baldassare}}
\altaffiltext{7}{University of Michigan, Department of Astronomy, 1085 S. University, Ann Arbor, MI, 48109, U.S.A.}

\begin{abstract}

%There is a population of 'red' L dwarfs that have redder $J - K_s$ colors than normal objects. Red L dwarfs include young, low-gravity objects, which are systematically red. The observed reddening in L dwarfs is not well explained by current atmosphere models.
We examine the hypothesis that the red near-infrared colors of some L dwarfs could be explained by a ``dust haze" of small particles in their upper atmospheres. This dust haze would exist in conjunction with the clouds found in dwarfs with more typical colors. We developed a model which uses Mie theory and the Hansen particle size distributions to reproduce the extinction due to the proposed dust haze. We apply our method to 23 young L dwarfs and 23 red field L dwarfs. We constrain the properties of the dust haze including particle size distribution and column density using Markov-Chain Monte Carlo methods. We find that sub-micron range silicate grains reproduce the observed reddening.
% We also find that Hansen particle size distributions reproduce the shape of the observed reddening better than power law \textbf{or Gaussian} particle size distributions. 
Current brown dwarf atmosphere models include large grain (1--100~$\mu m$) dust clouds but not submicron dust grains. Our results provide a strong proof of concept and motivate a combination of large and small dust grains in brown dwarf atmosphere models.
%that are not currently in the models can explain the red colors of some L dwarfs.

%Red L dwarf spectra have excess flux in the near IR which is suspected to be caused by additional dust in the upper atmosphere. These cold small particles in the dust haze do not emit in the near IR but redden the emergent spectra.  In this paper, we test the hypothesis that the red L dwarfs have an extra dust haze of small particles above the regular clouds. We constrained the properties of this dust haze including particle size distribution, column density, and column mass. We found that this observed reddening can be reproduced by Mie theory and Hansen distribution with effective radius $\sim 0.2-0.3 \micron$, which is much smaller than the particle sizes that have been adopted in the atmosphere models. The derived column mass of the dust haze is consistent with the typical brown dwarf atmosphere.
\end{abstract}


\section{Introduction and Problem}

Brown dwarfs are substellar objects with intermediate masses bridging the stellar and planetary regimes. Brown dwarfs do not have high enough mass to sustain hydrogen fusion in their cores, so they keep cooling with time. As brown dwarfs age, they also contract and their surface gravities increase. The cool temperatures of brown dwarfs allow condensates to form in their atmospheres, which shape their emergent spectra \citep{Burrows01}.
 
%As shown in Figure~1 in \citet{faherty13}, 
There is a wide spread in $J - K_s$ colors of L dwarfs of a given spectral type. So-called `red' L dwarfs have redder-than-average near-infrared (NIR) colors, and have notably redder spectral slopes through the NIR than typical L dwarfs \citep{faherty13}.
% as shown in Figure~$\ref{fig: spec}$. 
Low-gravity objects tend to be systematically redder but some field-aged L dwarfs also have redder colors. 
Red, low-gravity L dwarfs can be considered `exoplanet analogs' and are of particular interest \citep{Kirkpatrick08, cruz09, Faherty09, faherty13}.
% observationally showed that low-gravity L dwarfs and giant exoplanets have similar dusty atmospheres.
% In Figure~\ref{fig: model}, we illustrate our conception of the dusty atmosphere of a red L dwarf and the atmosphere of a normal field-age L dwarf.
%This observed reddening is not yet well understood and it cannot be reproduced by current atmosphere models. Some of the red L dwarfs are suspected to be young and low-gravity but gravity alone cannot explain the reddening. 

While the range of NIR colors of L dwarfs has not been fully explained, it is commonly attributed to variation in metallicity, gravity, and/or cloud properties \citep{saumon08, Marley12}. Brown dwarfs are thought to have refractory particles typically ranging 1--100 $\mu m$ in size in their atmospheres and these particles 
%cause extinction and 
shape their emergent spectra \citep{AM01, Cushing08}. 
%Clouds of refractory particles typically 1--100 $\mu m$ in the atmospheres of brown dwarfs cause extinction and shape their emergent spectra \citep{Marley99, Ackerman01, Cushing08}. 
%Small, ISM-size dust grains will scatter bluer light more efficiently than redder light in the NIR, causing the observed reddening.  
%(more about how clouds affect spectra)
%There are several atmosphere models that attempt to reproduce the observed spectra of L dwarfs (ref).
Current brown dwarf atmosphere models either use these large dust grains that are organized into discrete cloud layers, or small submicron grains that are mixed throughout the atmosphere \citep{Allard01}. These atmosphere models reproduce the general trend of L dwarf spectra but fail to give a reasonable explanation for the observed reddening in the NIR.
Model spectral energy distribution (SED) fitting sometimes gives unrealistically low gravities and small $f_{\rm sed}$ (cloud sedimentation efficiency parameter) values \citep{AM01, Cushing08} and radii too small compared to evolutionary models \citep{Liu13} for red L dwarfs.
A better dust treatment is needed that can account for the observed reddening in L dwarf spectra.
%expand for thesis

In efforts to understand the red objects empirically, it has been found that dust described by the interstellar reddening law can de-redden red L dwarf spectra to look like standard objects \citep{Looper10, Marocco14}. Interstellar reddening is the extinction of starlight caused by the interstellar dust. The interstellar grains, which have radii less than 1 $\mu m$, suppress blue light more effectively than red light. As a result, distant stars look redder than they actually are. The interstellar reddening law (extinction law) is an empirical relationship between the extinction at any wavelength $A(\lambda)$ and the visual extinction $A(V)$ \citep{CCM89}. 
%These grains shape the spectra by scattering and absorbing the emergent light. 
%Despite the use of the interstellar reddening law to de-redden brown dwarf spectra not physically motivated, 
We do not expect significant interstellar reddening in brown dwarfs since they are so close to the Sun on a galactic scale, so the use of the interstellar reddening law to de-redden brown dwarf spectra is not physically motivated.
However, ISM-like grains ($< 1~\micron$) in the atmospheres of red L dwarfs might produce results similar to the interstellar reddening law and explain the observed reddening.

\citet{Looper10} de-redden the optical spectra of TWA 30 (young M5 star) using the interstellar reddening law described by \citet{CCM89}. Although the reddening seen towards TWA 30 may be due to dust in the disk within the line of sight, \citet{Looper10} demonstrated that the interstellar reddening law can be successfully used in this case and it can be useful for dealing with reddening due to small grains other than the interstellar medium (ISM). \citet{Marocco14} de-redden the spectra of several L dwarfs including ULAS J222711$-$004547 (L7pec) using two interstellar extinction curves \citep{CCM89, Fitz99} and found that sub-micron size dust grains can explain the reddening effect. 
%In both cases, the spectra have been successfully de-reddened by the interstellar reddening. 
Furthermore, \citet{Cushing06} ascribe a flattening in the spectra of mid-type L dwarfs seen at 9--11 $\micron$ to a population of small silicate grains above the main cloud deck.

% law even though small grains are not in the atmosphere models of brown dwarfs. 
These three results 
%stand in contrast to our expectations based on the atmosphere models of brown dwarfs and 
imply that brown dwarfs with red spectral energy distributions may have small grains like the ISM which are smaller than 1 $\micron$ in addition to larger grains ranging from 1--100 $\micron$ currently included in the models in their atmospheres that scatter and absorb the emergent light. 

%Spectra of young brown dwarfs show gravity sensitive features such as peaky H band in addition to the reddening due to extra dust. 

In this work, we develop a prescription for a dust haze in L dwarf atmospheres and test whether it can account for the characteristics
of red L dwarf spectra.  By constraining the nature of this dust haze we aim to better understand the physical cause of the reddening in brown dwarfs. %Our analysis is similar to \citet{Marocco14} but our method is unique and different. 
Independently, two previous studies explored a similar dust haze analysis, \citet{Marocco14} for brown dwarfs and \citet{Bonn15} for directly imaged exoplanets. 
%The latter study closely followed the approach of \citet{Marocco14} so we do not consider it further here.
\citet{Marocco14}, \citet{Bonn15}, and this work are all motivated by the success of the interstellar reddening law in de-reddening unusually red L dwarf spectra and aim to explain the observed reddening by introducing a layer of small dust grains in the upper atmospheres.
%Although our analyses are similar, our method is unique and different from that of \citet{Marocco14}. 
Despite the similar ideas and concepts, the method we describe here is distinct from those previous studies as described in $\S\ref{sec: Model}$
%, we use different grain size distributions and grain species to model the dust haze. Also, we carry out Bayesian analysis to constrain the properties of the dust haze as shown in $\S\ref{sec: Method}$, while \citet{Marocco14} perform $\chi^2$ minimization. In 
and $\S\ref{subsec: Comparison}$.
%, we further discuss the similarities and differences between the two studies.
%\section{Hypothesis and Method}
%\label{sec: Hypothesis}
%We hypothesize that the red L dwarfs have an additional dust haze of small particles comparable to ISM in size in their atmospheres. 
%We constrained the particle size distribution by matching the shapes of the forsterite extinction curves and the observed reddening curves. 

The sample of L dwarfs studied in this analysis is presented in $\S\ref{sec: Data}$. In $\S\ref{sec: reddening}$, we present our method of using spectral observations to estimate the reddening. We explain our dust haze model in detail in $\S\ref{sec: Model}$, and model fitting method in $\S\ref{sec: Method}$. Finally, we present our results in $\S\ref{sec: Results}$ and conclusion in $\S\ref{sec: Conclusion}$.

\section{Sample and Spectral Observations}
\label{sec: Data}
%\section{Sample}
%\label{sec: sample}

In order to study the observed reddening, we compiled a sample of low-resolution NIR spectra of 23 L dwarfs with low gravity features in the optical \citep{cruz09} and 23 red field L dwarfs \citep{Kirkpatrick10}. 

Our red field objects have spectral features indicative of field gravity with $J-K$ colors redder than the spectral standards \citep{Kirkpatrick10}. This definition of `red' is specific to the purpose of this analysis. For example, \citet{faherty13} defined `red' as having a redder $J-K$ color than the mean $J-K$ of normal objects in the spectral type as opposed to comparing to the spectral standard. Therefore, red objects in our sample may be different from objects that are defined red by \citet{faherty13} or other papers. The objects in our sample are listed in Table 1.
%(Describe objects in more detail?)

% Kelle's text
%\subsection{New Near-Infrared Spectra}
Our sample includes 18 new spectra of red L dwarfs obtained with the SpeX spectrograph on the Infrared Telescope Facility (IRTF) \citep{Rayner03}. Observations were obtained over 28 nights during 2003--2011. The targets and observation dates are listed in Table~\ref{table: data}. All objects were observed in clear and dry conditions.
The targets were observed dithered pairs (ABBA) to enable pair-wise subtraction.
We used the 0.5$\arcsec$ slit and prism-dispersed mode to obtain $\lambda/\Delta\lambda \thickapprox $~120 spectra covering 0.7-2.5~$\mu$m. The data were reduced using the SpeXtool package \citep{Cushing04},
nearby A0 V stars were observed for flux calibration and telluric correction \citep{Vacca03}, and internal flat field and Ar arc lamp exposures were obtained for pixel response and wavelength calibration. 
%We reduced data using the SpeXtool package 

\section{Estimating the Observed Reddening}
\label{sec: reddening}
In order to estimate the observed reddening, 
%we treat the overall shapes of the emergent spectra of both red L dwarfs and field L dwarf spectral standards to be identical, $\it{except}$ for the reddening. 
%Therefore, our analysis is independent of physical properties of L dwarfs such as metallicity, radius, and the properties of non-haze clouds. 
%W
we compared red L dwarfs (including young and field) to the field spectral standards. 
%Figure~$\ref{fig: model} illustrates a conceptual representation of 
% to isolate and characterize the observed reddening. 
%Table~$\ref{table: data}$ summarizes the objects we studied in this analysis. 
The observed reddening was obtained by dividing the spectrum of the field standard by the spectrum of the red L dwarf. 
The top panel of Figure~$\ref{fig: spec}$ shows spectra of a red L dwarf and a spectral standard, illustrating the redder spectral slope of the red object. The bottom pane is the ratio of the two spectra and visualizes the estimated observed reddening of the red object.
%Our analysis was conducted independent of \citet{Marocco14}, whose work is similar to ours but different in some ways as described in Subsection~$\ref{subsec: Comparison}$.
%In an attempt to measure the effect of the dust haze, we compared low-resolution NIR spectra of red objects to NIR standards in order to isolate the reddening. 
%As illustrated in Figure~$\ref{fig: model}$, we assume that red L dwarfs and field L dwarfs are identical except for the dust haze so that red L dwarf spectra are a composite of normal L dwarf spectra and the reddening effect of dust hazes. Thus, we compared the spectrum of each red L dwarf to the spectrum of the NIR standard of the corresponding spectral type to isolate the reddening. 
%Then we took the log of the flux ratio, which corresponds to the reddening. %by an additional opacity due to the proposed dust haze. 
%The continuum of the observed reddening shows a power law shape.
The small-scale features seen in the observed reddening are due to gravity-sensitive spectral features such as FeH, VO, and the triangular-shaped H band \citep{Kirkpatrick06, Allers13}. 
%These gravity-sensitive spectral features appear different in low-gravity and field-gravity L dwarfs. 
%Gravity-sensitive spectral features are characteristic of low gravity objects only and not seen in field L dwarf spectra. 
We assume these features are not caused by reddening, so we treat the overall shape of the observed reddening as a smooth curve. %which can be reproduced by extinction due to the proposed dust haze.

%We isolated the broad wavelength effects of the reddening (i.e. the continuum) from the small-scale features. 
% fitted a power law curve to the flux ratio (green dashed lines in Figure~$\ref{fig: fratio}$) using the least squares method. 

%We examine these power law fits as "inferred reddening curves" in the following sections.

\section{Modeling the Observed Reddening}
\label{sec: Model}
In this paper, we develop a prescription for a hypothesized dust haze of small particles in the atmospheres of the so-called `red' L dwarfs to explain the observed reddening in their spectral energy distributions (SEDs). In order to model the observed reddening, we used Mie theory to calculate the `raw' extinction coefficients due to forsterite grains. Then we averaged the raw extinction coefficients over various particle size distributions to calculate the `effective' extinction coefficients and to generate a model grid to compare with the observed reddening. 
%Mie scattering, effective extinction coefficients, particle size distribution, observed reddening, 

\subsection{Dust Haze Prescription}
In Figure~$\ref{fig: model}$, we show an illustration of the proposed dust haze in the upper atmospheres of red L dwarfs. Our model prescribes that the dust haze must be high in the atmospheres so that it is too cool (since the temperature decreases as the altitude increases) to radiate significantly in the NIR. The prescribed dust haze lies above the main cloud deck, so the dust grains in the haze affect the emergent spectra. 
Our prescription does not have thickness or height, therefore we are not constraining the position or dimensions of the dust haze any further than lying above the main cloud deck.
%except that it should be high enough (and therefore cool enough) not to radiate in the NIR, only scatter.

%We hypothesize that the upper atmospheres of red L dwarfs have a ``dust haze" of small particles (comparable to the sub-micron size of the ISM) to explain the reddening observed in the SEDs of L dwarfs. Our proposed dust haze could explain the observed reddening in young L dwarfs and so-called `red' field L dwarfs. In Figure~$\ref{fig: model}$, we show an illustration of the proposed dust haze in the upper atmospheres of red L dwarfs.
The dust haze grains were modeled by forsterite grains. As shown by \citet{Lodders06}, forsterite ($\rm Mg_{2}SiO_4$) is thought to exist in L dwarf atmospheres among other dust species such as corundum ($\rm Al_{2}O_3$), enstatite ($\rm MgSiO_3$), and iron. Corundum condenses at higher temperatures in the atmospheres of late M dwarfs. Liquid iron and silicates condense in early L dwarfs between 1600 -- 1840 K. Since iron grains form deeper in the atmosphere, the dust is most likely silicate. The silicate grains in the L dwarf atmospheres are thought to be a mixture of forsterite and enstatite. Since the extinction curves of forsterite and enstatite have similar shapes, forsterite was used in our analysis.

Extinction is the sum of absorption and scattering, and is the fraction of incoming light that gets affected by interactions with particles. 
Reddening is a type of extinction, which occurs when extinction is more effective at shorter (bluer) wavelengths than at longer (redder) wavelengths, and has the effect of making the spectral slope redder.
We used Mie theory to model the reddening effect of the proposed dust haze on the emergent spectra of red L dwarfs. Mie theory applies when the scattering particle is spherical and its size is similar to the wavelength of the scattered light, which is appropriate for sub-micron size grains in the NIR. For larger particles, Mie scattering is independent of wavelength while for smaller particles, it is wavelength dependent. Mie scattering reduces to strongly wavelength dependent Rayleigh scattering when particle sizes are much smaller than wavelength. 
We employed a Mie code, described in  \citet{Toon&Ackerman81}, and the refractive indices of forsterite (G. Sloan, pers. comm.) to compute the `raw' extinction coefficients, $Q_{\rm ext}(r, \lambda)$, for particles of radii between 0.01 and 10~$\micron$.
%We {\bf computed} Mie extinction curves to model the observed reddening curves.

% in the Mie scattering regime, so we explored forsterite as our "test particle" for the proposed dust haze in the upper atmospheres of red L dwarfs. 
%The observed reddening curves (Figure~$\ref{fig: spec}$) can be replicated by forsterite extinction curves assuming that the forsterite grains are spherical.
%The dust haze grains were approximated by forsterite grains which are available in brown dwarf atmospheres (\citet{Lodders06}).  among other particles such as iron and other silicates. The observed reddening curves (Figure~$\ref{fig: spec}$) can be replicated by forsterite extinction curves assuming that the forsterite grains are spherical. The properties of the forsterite dust haze, in particular the particle size distribution, change the shape of the forsterite extinction curves. 

%As described in $\S\ref{sec: Hypothesis}$, we utilized extinction coefficients for forsterite particles according to Mie theory in order to model the observed reddening curves (see $\S\ref{sec: Data}$). 
%SED redder. The slope. Excess flux in the red.
%where it makes the object look redder (extinct coefficient decreases with wavelength). 

%First just radii. size distribution later
\subsection{Particle Size Distributions}
\label{subsec: distribution}
In order to calculate effective extinction coefficients from the `raw' extinction coefficients directly computed from the Mie code, we need to choose a particle size distribution $n(r)$. We considered three different particle size distributions commonly used to model grains, clouds and hazes in various
settings:  power law, Gaussian, and Hansen distributions. Figure~$\ref{fig: dist}$, compares the shape of the three distributions. We describe the motivation to use these particle size distributions and results below. 

%Similarly, the mean mass of a forsterite grain is calculated by averaging over the same particle size distribution.
%\begin{equation}
%m = \frac{\int_{r_{\rm min}}^{r_{\rm max}}\! {\rho}\frac{4}{3}{\pi}r^2n(r)dr}{\int_{r_{\rm min}}^{r_{\rm max}}\! n(r)dr}
%\end{equation}
%where $\rho$ is the density of the dust grain. The derived mass is used to calculate the column mass of the small forsterite dust haze $mN$. (NOT DONE YET)
%\subsubsection{Power Law Distribution}
%\label{subsubsec: powerlaw}
%\subsubsection{Particle Size Distributions}
We considered a power law particle size distribution $n(r) \propto r^{p}$ with $p = -3$ and particle radius $r$ ranging between 0.01 -- 10 $\micron$ to model theoretical extinction due to the proposed small dust grains. 
Power law particle size distributions with $p \approx -3.5$ are typically used to characterize interstellar dust and grains in the circumstellar disks around young brown dwarfs \citep{Mathis77, Draine06, Luhman05, BF14}. Figure~$\ref{fig: dist}$ shows two different power law particle size distributions with $p =  -3$ and -3.5.
%\citet{Marocco14} used a power law grain size distribution $n(r) = r^{-2.5}$ for iron to de-redden ULAS J222711$-$004547 and found the maximum grain size to be 0.3 ${\mu}m$.

%\subsubsection{Gaussian Distribution}
We also considered a Gaussian particle size distribution $n(r) = \frac{1}{\sqrt{2\pi}\sigma}e^{-\frac{(x-\mu)^2}{2\sigma^2}}$ with the mean radius $\mu = 0.5 \micron$ and a width $\sqrt{2}\sigma = 0.1\times \mu$. This particle size distribution is adopted from \citet{Marocco14}, who de-reddened ULAS J222711$-$004547 using corundum and enstatite, and other red L dwarfs using corundum with Gaussian particle size distributions and found $\mu \thicksim 0.5 {\mu}m$.

%\citet{Marocco14} adopted a Gaussian particle size distribution to de-redden red L dwarfs using corundum and enstatite with a width $\sqrt{2}\sigma = 0.1\times \mu$ where $\mu$ is the characteristic grain radius. They de-reddened ULAS J222711$-$004547 using corundum and enstatite, and other red L dwarfs using corundum. They found $\mu \thicksim 0.5 {\mu}m$. 
%In Figure~$\ref{fig: Qext}$, we show forsterite extinction coefficients averaged over the two Gaussian particle size distributions. These extinction curves do not reproduce the shapes of the observed reddening as well as the extinction curves averaged over the Hansen particle size distributions.
%As shown in Figure~$\ref{fig: dist}$, the Gaussian particle size distribution is concentrated at $\mu$.

%\label{subsubsec: Hansen}
%\subsubsection{Hansen Distribution}
We also considered the Hansen particle size distribution, which is a variation of the gamma distribution and is expressed as follows.
\begin{equation}
n(r) = r^{\frac{1-3b}{b}}e^{-\frac{r}{ab}}
\label{eq: eq1}
\end{equation}
where $a$ is the mean effective radius and $b$ is the effective variance. Following \citet{Hansen}, $a$ and $b$ are defined as
\begin{equation}
a = \frac{\int_{0}^{\infty}\! r{\pi}r^2 n(r)dr}{\int_{0}^{\infty}\! {\pi}r^2 n(r)dr}
\end{equation}
\begin{equation}
b = \frac{\int_{0}^{\infty}\! (r - a)^2 {\pi}r^2 n(r)dr}{a^2 \int_{0}^{\infty}\! {\pi}r^2 n(r)dr}
\end{equation}
%\begin{equation}
%b = \frac{\int_{0}^{\infty}\! (r - \langle r \rangle _{eff})^2 {\pi}r^2 n(r)dr}{\langle r \rangle _{eff} ^2 \int_{0}^{\inf}\! {\pi}r^2 n(r)dr}
%\end{equation}
respectively.

%The Hansen particle size distribution is used to describe the particle size distribution of water droplets in Earth's clouds \citep{Hansen}. 
The Hansen particle size distribution successfully reproduces the observed particle size distributions of different types of water clouds in Earth's atmosphere (fair weather cumulus, altostratus, and stratus clouds) as shown in Figure 1 of \citet{Hansen}.
%The microphysics and structure of Earth's clouds may also apply to clouds in brown dwarf atmospheres, so we adopted the Hansen particle size distribution for our analysis.
%Based on its success in reproducing the measured cloud particle size distributions (see Figure 1 in \citet{Hansen}), the Hansen particle size distribution may also reflect cloud formation processes in brown dwarf atmospheres. 
Figure~$\ref{fig: dist}$ shows two Hansen particle size distributions with the mean effective radius $a = 0.2~\micron$ for effective variances $b$ = 0.1 and 0.5. The Hansen particle size distribution with a large variance (dashed green, $b = 0.5$) is similar to power law particle size distributions in the regime of small particle size ($\lesssim 0.1~\micron$).

\subsection{Computing Modeled Extinction Curves and Comparing to Observations}
\label{subsec: Qeff}

In order to account for a range of particle sizes and to smooth over the small-scale interference patterns in forsterite extinction coefficients, we computed effective extinction coefficients by averaging the raw extinction coefficients over a particle size distribution. 
%In order to model the reddening caused by the proposed dust haze, we must choose a grain size distribution. Mie extinction coefficients have small-scale interference patterns
%As a general trend, the greater number of large particles the dust haze contains, the flatter the forsterite extinction curves are (less reddening). 
%In order to compute "effective" extinction coefficients, we average forsterite extinction coefficients over the particle size distribution as described in \citet{Hansen}
Effective extinction coefficients are defined as
\begin{equation}
Q_{\rm ext}(\lambda) = \frac{\int_{r_{\rm min}}^{r_{\rm max}}\! {\pi}r^2Q_{\rm ext}( r, \lambda)n(r)dr}{\int_{r_{\rm min}}^{r_{\rm max}}\! {\pi}r^2n(r)dr}
\label{eq: eq4}
\end{equation}
where $n(r)dr$ is the number of particles per unit volume with radius between $r$ and $r+dr$. The integration limits we employed are 0.01$-10~\micron$. Particles smaller than this range would be too small (a few atoms) to scatter light and particles that exceed $10~\micron$ tend to be grey at all wavelengths (i.e. the extinction they cause will be independent of wavelength).

In order to compare the observed reddening to the modeled extinction, we assume that the observed flux, $I$, from a red L dwarf can be modeled as
%we introduce an equation for radiative transfer through an optically thick medium,
\begin{equation}
I(\lambda) = fI_{0}(\lambda)e^{-\tau(\lambda)}
\end{equation}
where $I_{0}(\lambda)$ is the flux of the field standard L dwarf, $f$ is a scaling factor, and $\tau(\lambda)$ is the optical depth of the dust haze in the red L dwarf atmosphere, assuming the red L dwarf can be described by the field L dwarf surrounded by the dust haze as in Figure 2.
The scaling factor $f$ is determined by the distances and sizes of the objects. 
\begin{equation}
f = \frac{{d_0}^2}{d^2}\frac{R^2}{{R_0}^2}
\end{equation}
where $d_0$ and $d$ are the distances to the field and the red L dwarfs, and $R$ and $R_0$ are the radii of the red and the field L dwarfs, respectively.

Solving Equation~$\ref{eq: eq1}$ for the optical depth we get
\begin{equation}
\tau(\lambda)  = \ln{f} + \ln{\frac{I_0(\lambda)}{I(\lambda)}}.
\end{equation} 
The optical depth $\tau(\lambda)$ is related to the Mie extinction coefficient $Q_{\rm ext}(\lambda)$ as
\begin{equation}
\tau(\lambda) = N{\pi}a^{2}Q_{\rm ext}(\lambda)
\end{equation}
where $N$ is the column density of the haze layer
% actually no this is not the case: 
%(including both the hypothesized dust haze and the main cloud deck) 
and $a$ is the effective radius of the scattering grains which we assume are composed of forsterite. $Q_{\rm ext}(\lambda)$ is the wavelength dependent forsterite grain extinction coefficient which we calculated using the Mie code \citep{Toon&Ackerman81}.  
%$Q_{\rm ext}(\lambda)$ is a function of wavelength, particle size $r$, and refractive index which also depends on wavelength. 
We averaged the forsterite coefficients over a particle size distribution $n(r)$ as described in Equation~\ref{eq: eq4}.
Combining Equations 7 and 8, we get
%\begin{equation}
%\ln{\frac{I_0(\lambda)}{I(\lambda)}} = N_{\rm small}{\pi}a_{\rm small}^{2}Q_{\rm small}(\lambda) + N_{\rm big}{\pi}a_{\rm big}^{2}Q_{\rm big} - \ln{f}
%\end{equation}
%The last two terms are constant with wavelength. So it can be written as
\begin{equation}
\ln{\frac{I_0(\lambda)}{I(\lambda)}}  = N{\pi}a^{2}Q_{\rm ext}(\lambda) + C
\label{eq: eq9}
\end{equation}
where $N$, $a$, and $Q_{\rm ext}(\lambda)$ are the column density of the dust haze, the mean effective radius found from the particle size distribution $n(r)$, and the effective forsterite extinction coefficients of small grains from Equation \ref{eq: eq4}. The constant term $C$ accounts for the scaling factor $f$ and any differences in grey atmospheric opacity between the L dwarfs. Since the reddening we observe is wavelength dependent, we relegate any grey component to the $C$ term. $C$ is independent of wavelength because $Q_{\rm ext}(\lambda)$ approaches a constant for $a >> \lambda$. 
%From here on, we will only discuss the effect of small particles and we will omit the subscript `small'.}
%We normalize both $N_{\rm small}{\pi}a_{\rm small}^{2}Q_{\rm small}(\lambda)$ and $\frac{I_0(\lambda)}{I(\lambda)}$ by subtracting the mean to get rid of the constant. (remove if unnormalizing)
%We fit $N_{\rm small}{\pi}a_{\rm small}^{2}Q_{\rm small}(\lambda)$ to the observed reddening to constrain the properties of the forsterite dust grains in the dust haze (Section~$\ref{sec: Method}$).
The RHS of Equation \ref{eq: eq9} is the modeled extinction and the LHS is the observed reddening.
As explained later in $\S\ref{sec: Method}$, we used MCMC methods to constrain the parameters in the modeled extinction ($N$, $a$) that best reproduce the observed reddening.

\subsubsection{Evaluation of Particle Size Distributions and Creating a Model Grid}
\begin{figure}
\plotone{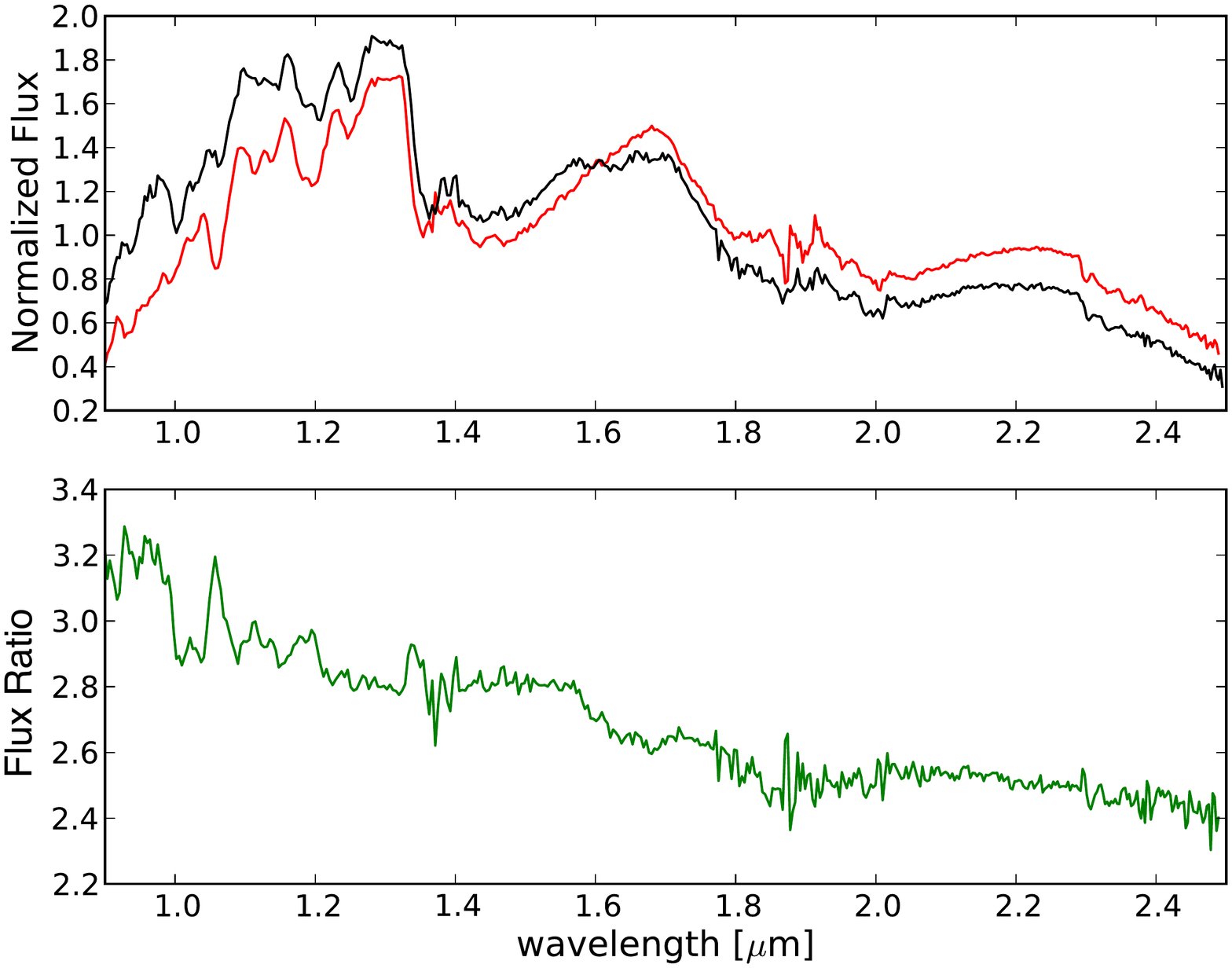}
\caption{The top panel shows SpeX Prism spectra of a red L0 dwarf 2M~0141-4633 (red) and the L0 field standard 2M~0345+2540 (black). Each spectrum is normalized by the mean flux of the entire spectrum. The red object has excess flux longward of 1.5 $\micron$. The bottom panel shows the observed reddening, which is the log of the flux ratio between 2M~0345+2540 and 2M~0141-4633.}
%dividing the spectrum of the standard object by the spectrum of the red object and taking the log of the flux ratio. The overall shape of the observed reddening resembles a power law curve.}
\label{fig: spec}
\end{figure}

\begin{figure}
\plotone{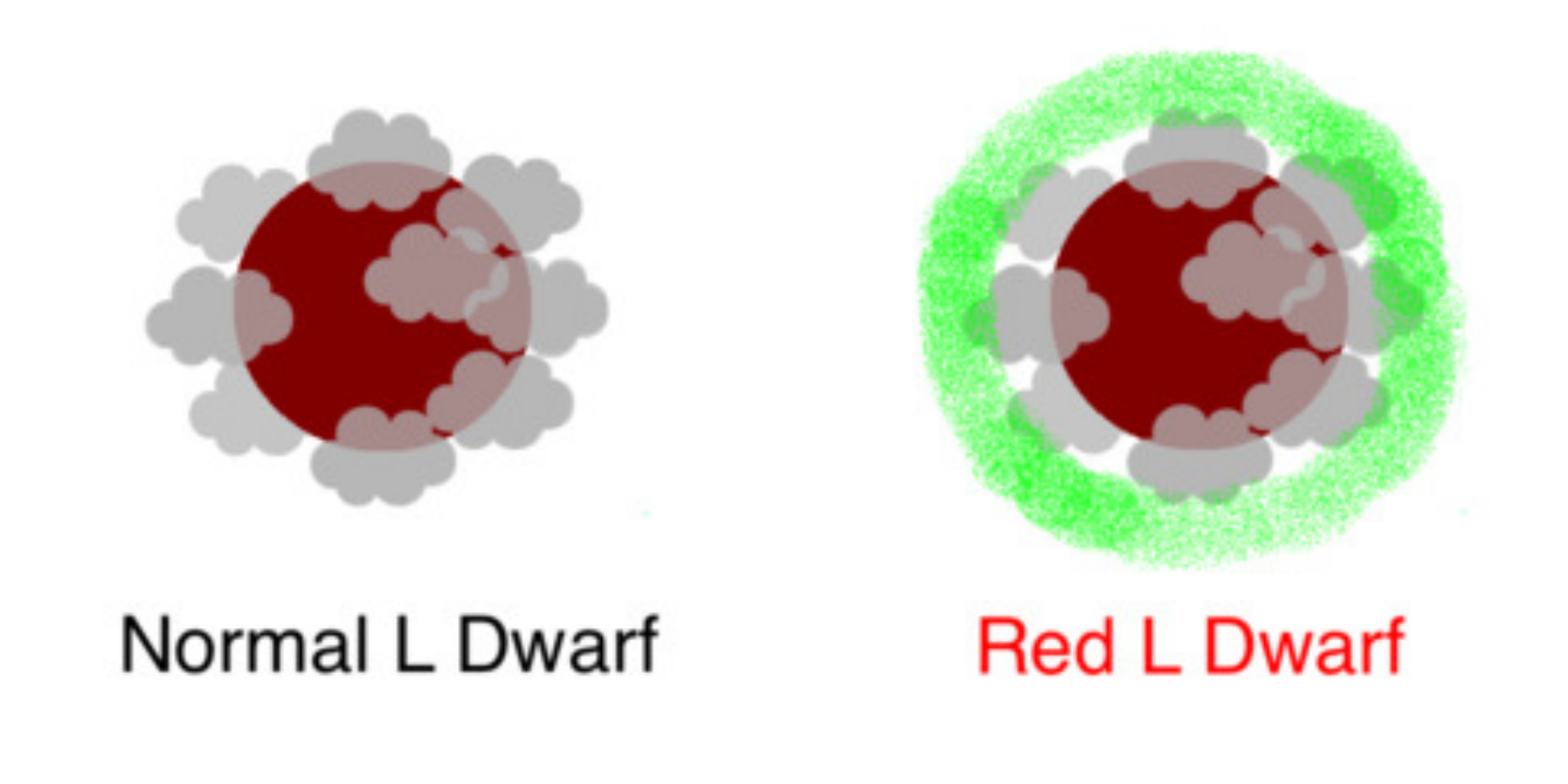}
\caption{A conceptual representation of our dust haze model. The regular clouds of large particles ($\sim 10~\micron$, grey cloud symbol) exist in both normal and red L dwarfs. An additional haze of small particles (green layer) is present in the red L dwarf atmosphere, which causes the observed reddening. We do not constrain the position or dimension of the haze any further than it lying above the main cloud deck.}
%While the specific altitude of the haze is currently unknown, it must be at an altitude with a sufficiently low temperature such that the dust grains do not emit in the NIR.}
%We do not know yet the specific location of the dust haze but it has to be at an altitude where the temperature is low enough so that the dust grains do not emit in the near Infrared.}
\label{fig: model}
\end{figure}

\begin{figure}
%\plotone{f4.pdf}
\plotone{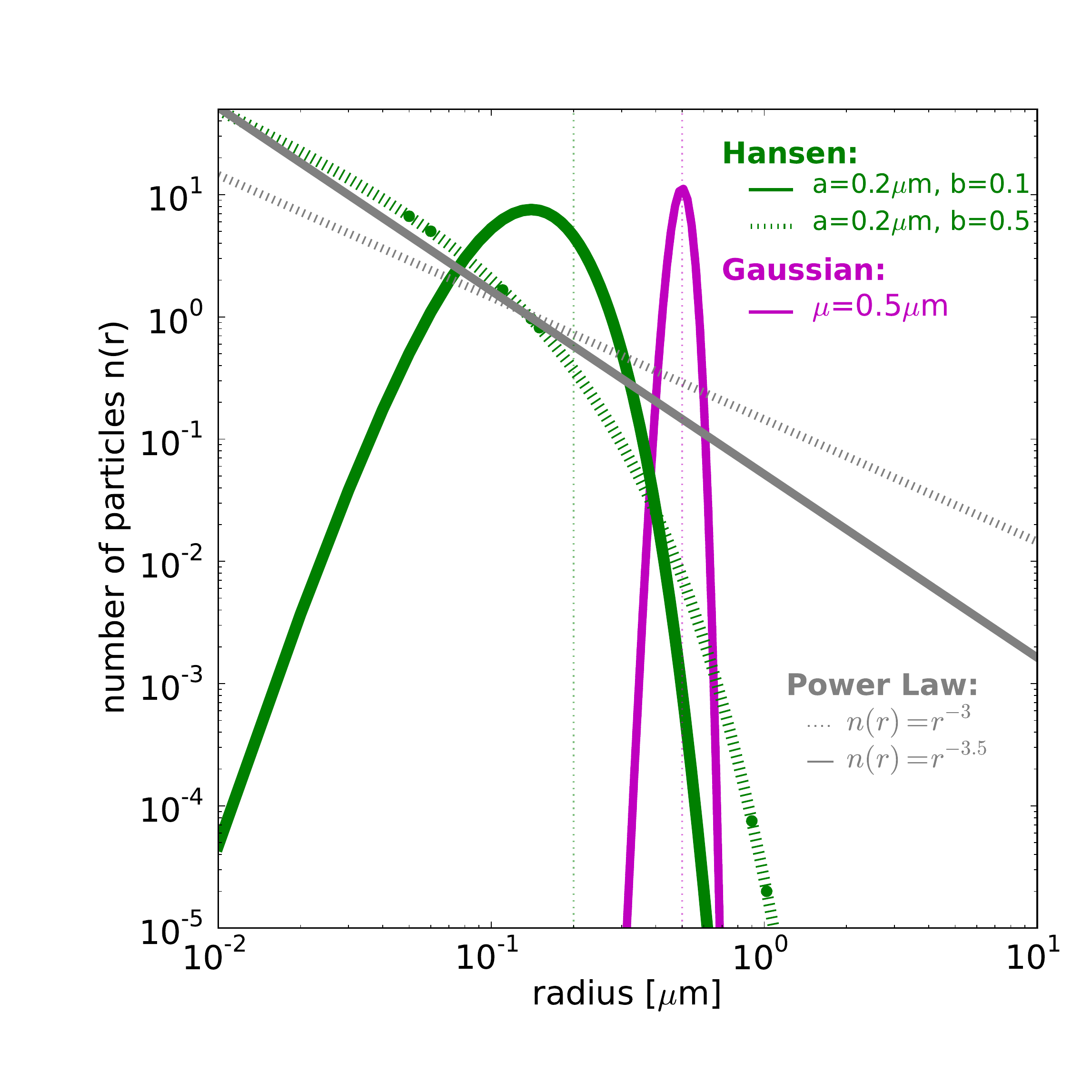}
\caption{Comparison of different particle size distributions. The gray lines are the power law distributions with power indices of -3 and -3.5. The magenta is a Gaussian distribution with characteristic grain size of 0.5 $\mu m$. The green lines are Hansen distributions for $a$ = 0.2 $\mu m$ and $b$ = 0.1, 0.5. }
%The Hansen distributions are a variation of the gamma distribution and they are wider than the Gaussian, but narrower than the power law distributions. The Hansen distributions can reproduce the shapes of the observed reddening better than the other distributions.}
\label{fig: dist} 
\end{figure}

\begin{figure}
\plotone{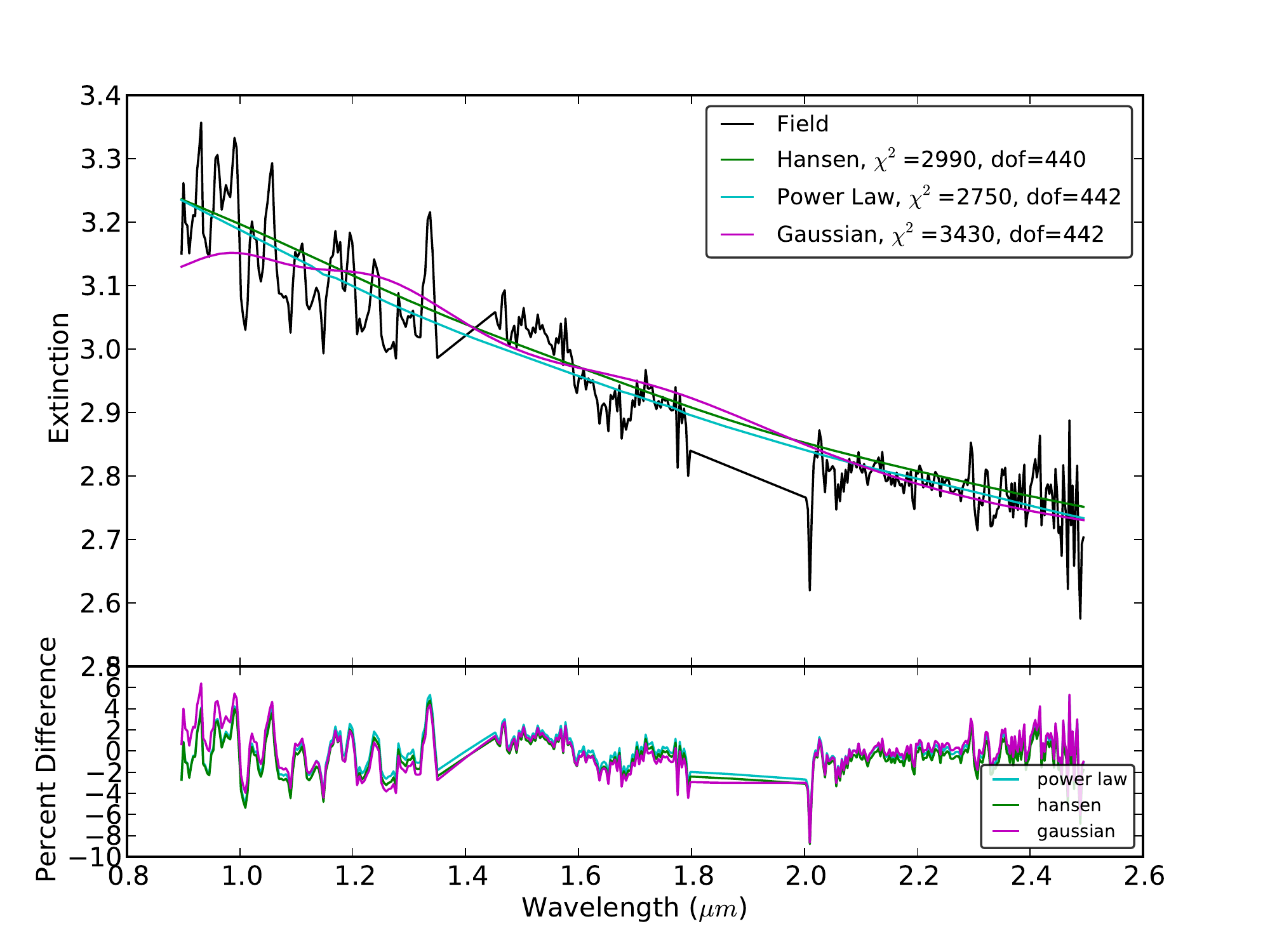}
\caption{Top panel shows best-fit models with three different particle size distributions. The black line is the observed reddening with the telluric bands removed, and the colored lines are the best-fit models. Bottom panel shows the percent difference between each model and the observed reddening. }
%All three particle size distributions are in good qualitative agreement.}
\label{fig: residual}
\end{figure}

Figure~$\ref{fig: residual}$ shows model fits to the observed extinction using the three different particle size distributions. The top panel shows the observed reddening (black, same as the bottom panel in Figure~$\ref{fig: spec}$), theoretical extinction curves using the Hansen, power law, and Gaussian particle size distributions. $\chi^2$ value and the degrees of freedom for each fit are reported in the legend. 
%The method for calculating the theoretical extinction is explained in $\S\ref{sec: Method}$. 
The bottom panel shows the residual between the observed reddening and the theoretical extinction curve as a percentage of the observed flux. 
%The telluric bands (1.35--1.45 $\micron$ and 1.8--2.0 $\micron$) are removed from the analysis.}
Figure~$\ref{fig: residual}$ demonstrates that all three modeled extinction curves fit the observed reddening reasonably. The Hansen and power law particle size distributions reproduce the smooth shape of the observed reddening, while the Gaussian particle size distribution results in a less smooth extinction curve. 

For the remainder of our analysis, we adopt the Hansen particle size distribution because of the favorable behavior shown in Figure~$\ref{fig: residual}$, the fact that it reflects the microphysics and structure of Earth's clouds, and because it has various helpful properties that are useful for algebraic manipulation \citep{Hansen}.

%We first computed effective forsterite extinction coefficients on a {\bf coarse} grid of mean effective radius $a$ and effective variance $b$. We then linearly interpolated the coefficients onto a finer grid. 

\begin{figure}
\plotone{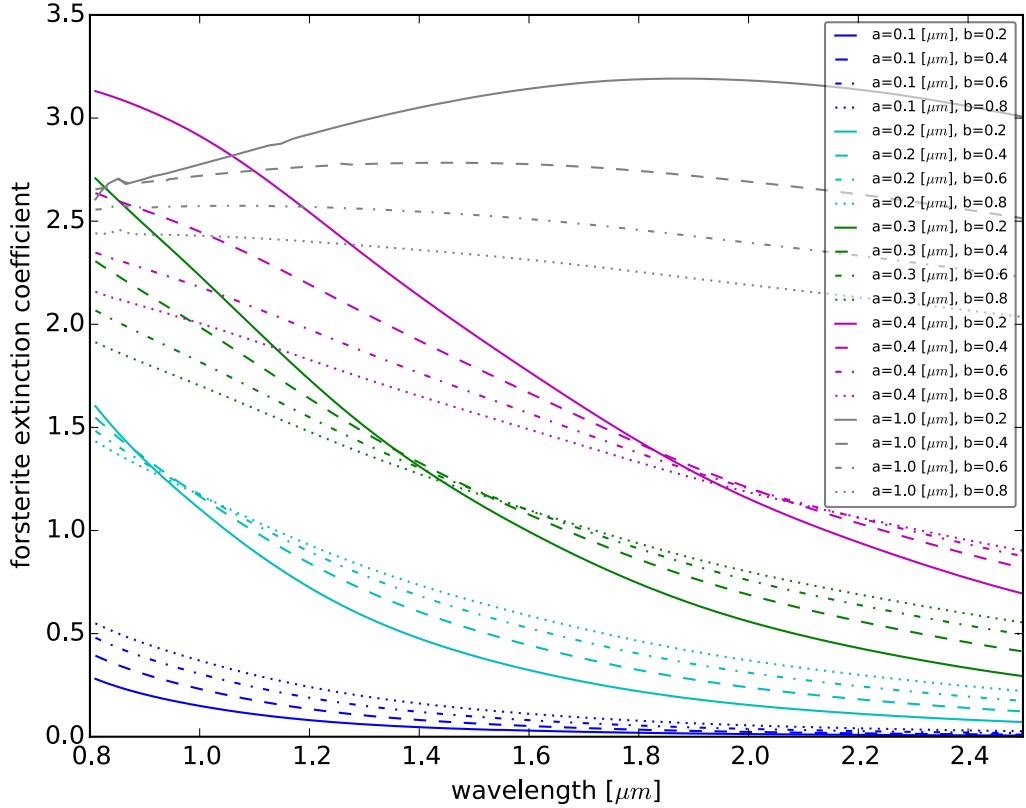}
\caption{Forsterite extinction coefficients according to Mie theory averaged over Hansen particle size distributions with various combinations of mean effective particle radius ($a$) and effective variance ($b$). Different colors correspond to different effective particle radii. Different line style correspond to different effective variance. The shapes of the extinction curves for smaller particles (0.1 -- 0.4 $\micron$) resemble the observed reddening while the 1.0 $\micron$ particles do not.}
\label{fig: Qext}
\end{figure}

In Figure~$\ref{fig: Qext}$, we show our model grid of forsterite extinction coefficients for various Hansen particle size distributions.
%Mie extinction coefficients for different particle sizes are plotted in Figure~$\ref{fig: Qext}$. 
%It shows that small grains have a smaller extinction (reddening) effect at shorter wavelengths and a larger effect at longer wavelengths. Large grains on the other hand, have a similar effect regardless of the wavelength and they would cause 'greying' instead of reddening. Therefore, small grains cause the reddening in $J-K$ color but large grains do not.
%We can tell from the figure that small particles resemble the shapes of the observed reddening (Figure~$\ref{fig: spec}) better than large particles. 
%small particles are steeper, more curved, concave up. Large particles are flatter.
%We tested if the Hansen particle size distribution can explain the observed reddening by visually comparing the shapes of forsterite extinction curves with the observed reddening (Section $\ref{sec: reddening}$). We compared the shapes of the observed reddening shown in Figure~$\ref{fig: spec}$ with various forsterite extinction curves shown in Figure~$\ref{fig: Qext}$. 
The effective forsterite extinction curves for small mean particle sizes ($a \leq 0.4\,\micron$) more closely resemble the observed reddening than those with a larger mean particle size.
%for large particle sizes ($a > 0.4\,\micron$). 
For these small mean particle sizes, extinction coefficients are wavelength dependent and resemble the curved shapes of the observed reddening (Figure~$\ref{fig: spec}$). For particles larger than 0.4 $\micron$, extinction coefficients are less wavelength dependent and the resulting extinction shapes are flat (`grey'), which do not fit the observed reddening. The curves for $a = 1.0 \micron$ are shown as representatives of the curves with $a > 0.4 \micron$.

The range of effective variance $b$ was determined based on the shape of the Hansen particle size distribution. Small effective variances ($b <  0.1$) result in a particle size distribution concentrated at the mean effective radius, and large effective variances  ($b > 1$) make the particle size distribution wide and resemble a power law particle size distribution. 
Thus, we decided to use Hansen particle distribution with a parameter grid of mean effective radius $a$ between 0.05 and 0.4 $\micron$, and effective variance $b$ between 0.1 and 1.0 as priors for the rest of our analysis. 

\section{Methods: Fitting the Models to the Observed Reddening}
\label{sec: Method}
%- Refer to Goodman $\&$ Weare?
% Steph's text
We use Markov-chain Monte Carlo (MCMC) fitting to estimate the best-fit parameters and their uncertainties. MCMC is a Bayesian inference method which provides a sampling approximation of the posterior probability distribution function (PDF). An MCMC run produces a chain of positions in parameter space, and a histogram of these positions provides the approximation of the posterior PDF.  MCMC allows for more in-depth probabilistic data analysis than $\chi^2$ minimization, for example, it efficiently approximates the full posterior PDF, which in turn provides uncertainties on and illustrates covariances between model parameters.  % need something about simple marginalization

%One of the most straight-forward MCMC algorithms is the Metropolis-Hastings (M-H) algorithm, which generates a single chain of parameter sets that approximate the posterior PDF. At each step in the fit, a trial position in the parameter space is randomly generated based on the last position in the chain.  A different step size is assigned for each parameter to determine how far away the trial position is from the last position.  The probabilities at the trial and last positions are compared, and if the trial position is more probable, then the corresponding parameters are added to the chain and the process repeats.  If the trial position is less probable, there is some probability that the corresponding parameters are added to the chain, and the rest of the time the last position is duplicated again.  This builds up the chain of parameter positions which approximates the posterior PDF. (see, e.g., \citet{ford2005} for a step-by-step description of the M-H algorithm.)  Although the M-H algorithm is powerful, the step sizes must be hand-tuned for efficiency; this may be impossible in cases with many covariant parameters or an otherwise complex probability space.  

The Goodman-Weare (G-W) algorithm improves upon the Metropolis-Hastings (M-H) algorithm by changing the method for choosing trial positions \citep{goodman2010}. The G-W algorithm deploys an ensemble of chains, known as ``walkers'', instead of a single chain. The trial position for each walker is chosen from the ensemble's location in parameter space, with some probability for choosing a position outside the occupied region. This method does not require hand-tuning the step size for each parameter, and the selection of trial positions can be parallelized. The G-W algorithm more efficient than the M-H algorithm in both human working hours and computation time. \citep{goodman2010, foreman-mackey2013} 

We use the open-source python implementation of the G-W algorithm, {\it emcee} \citep{foreman-mackey2013}, to fit the observed reddening with our model grid described in $\S\ref{subsec: Qeff}$. The modeled extinction curves are parameterized by the mean particle size $a$ and effective variance $b$ for the Hansen distribution and the column density $N$ of forsterite grains. We assume an unnormalized flat prior probability distribution for each parameter. The effective extinction coefficients are linearly interpolated and the modeled extinction at each wavelength point is calculated as Equation 9.

%\begin{equation}
%\frac{I_0(\lambda)}{I(\lambda)}  = N_{\rm small}{\pi}a_{\rm small}^{2}Q_{\rm small}(\lambda) + C
%\end{equation}

%where

%\begin{equation}
%Q_i = Q_{\rm ext}(\lambda) = \frac{\int_{r_{\rm min}}^{r_{\rm max}}\! {\pi}r^2Q_{\rm ext}( r, \lambda)n(r)dr}{\int_{r_{\rm min}}^{r_{\rm max}}\! {\pi}r^2n(r)dr}
%\end{equation}

%and
%\begin{equation}
%n(r) = r^{\frac{1-3b}{b}}e^{-\frac{r}{ab}}
%\end{equation}

%\begin{equation}
%Q_i = % The full equation for an extinction curve goes here, with the additional a terms, etc. 
%end{equation}

We also model the vertical offset between the observed reddening and the extinction curve with a constant $C$, % That might not be the right variable, just change as needed
and include a tolerance parameter $s$. The tolerance $s$ estimates the uncertainty in the model as a single value across the extinction curve; it accounts for the fact that the photon-noise uncertainties are smaller than the typical difference between each observed reddening point and the corresponding point on the extinction curve. If we denote observed reddening points as $r = \{r_i\}$ and the corresponding uncertainties as $\sigma_0 = \{\sigma_{0,i}\}$, we compute the natural logarithm of the likelihood function as
\begin{equation}
{\rm{ln}}\mathcal{L}( r|a,b,N,C,s,\sigma_0) = -\frac{1}{2} \sum_i \left[
\frac{(r_i - (N\pi a^2 Q_i^2 + C)}{(\sigma_{0,i}^2 + s^2)}
+ \ln(2\pi\ (\sigma_{0,i}^2 + s^2))
\right]
\end{equation}
% Check that all this actually matches what's in your code (I don't know exactly how you accounted for $c$, for example), so adjust as necessary.
The natural logarithm of the posterior PDF is given by
\begin{equation}
\ln({\rm {PDF}})(a,b,N,C,s|r,\sigma_0) = {\rm{ln}}\mathcal{L}(r|a,b,N,C,s,\sigma_0) + {\rm{ln}}\mathcal{P}(a,b,N,C,s,\sigma_0)
\end{equation}
We assume an unnormalized flat prior on each parameter, so $\mathcal{P}(a,b,N,C,s,\sigma_0)=1$ and ${\rm{ln}}\mathcal{P}(a,b,N,C,s,\sigma_0)=0$.  % This is technically wrong, but it mostly just means your posterior is improperly normalized

% FILL ALL OF THESE IN!!!
We pass a function for $\ln(\rm PDF)$ to {\it emcee}, which uses that function to determine acceptance of each step in parameter space. We typically use 100 walkers. After we iterate for 200 steps to generate a new set of initial positions for the walkers, we reset the walkers and restart from the new initial positions. We iterate for 2000 steps after a burn-in period of 200 steps.  

\section{Results and Discussion}
\label{sec: Results}
\subsection{Fitting Dust Haze Parameters}
\label{subsec: fitresult}

\begin{figure}
%\figurenum{5}
\plotone{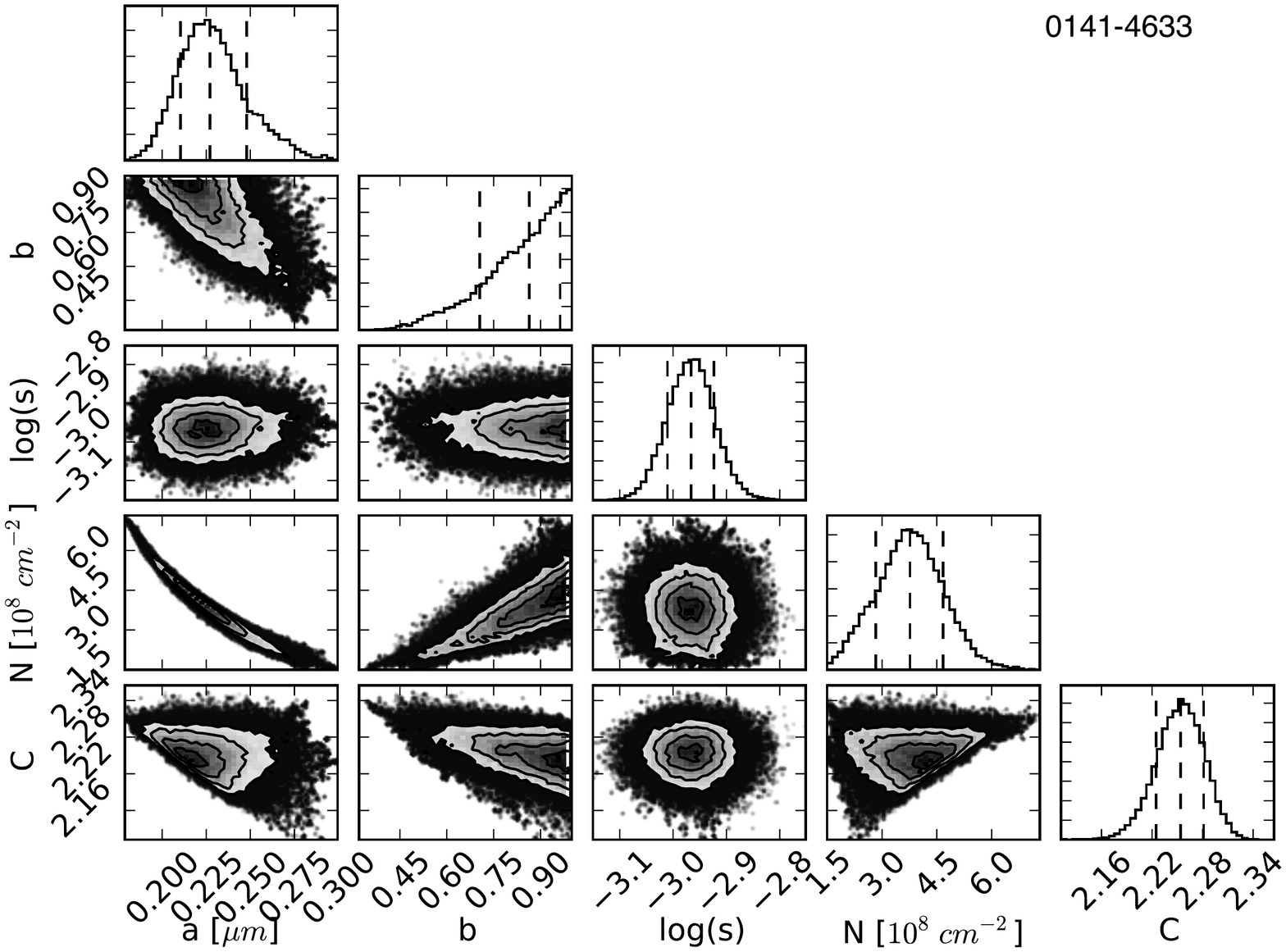}
\caption{Posterior distributions of mean effective radius, effective variance, tolerance, column density, and offset showing 1-D distributions for each parameter and 2-D distributions for each combination of parameters. Dashed lines in the 1-D distributions represent 16, 50, 84 percent quantiles, corresponding to the median and 1 $\sigma$ uncertainties.}
\label{fig: triangle}
\end{figure}

\begin{figure}
\plotone{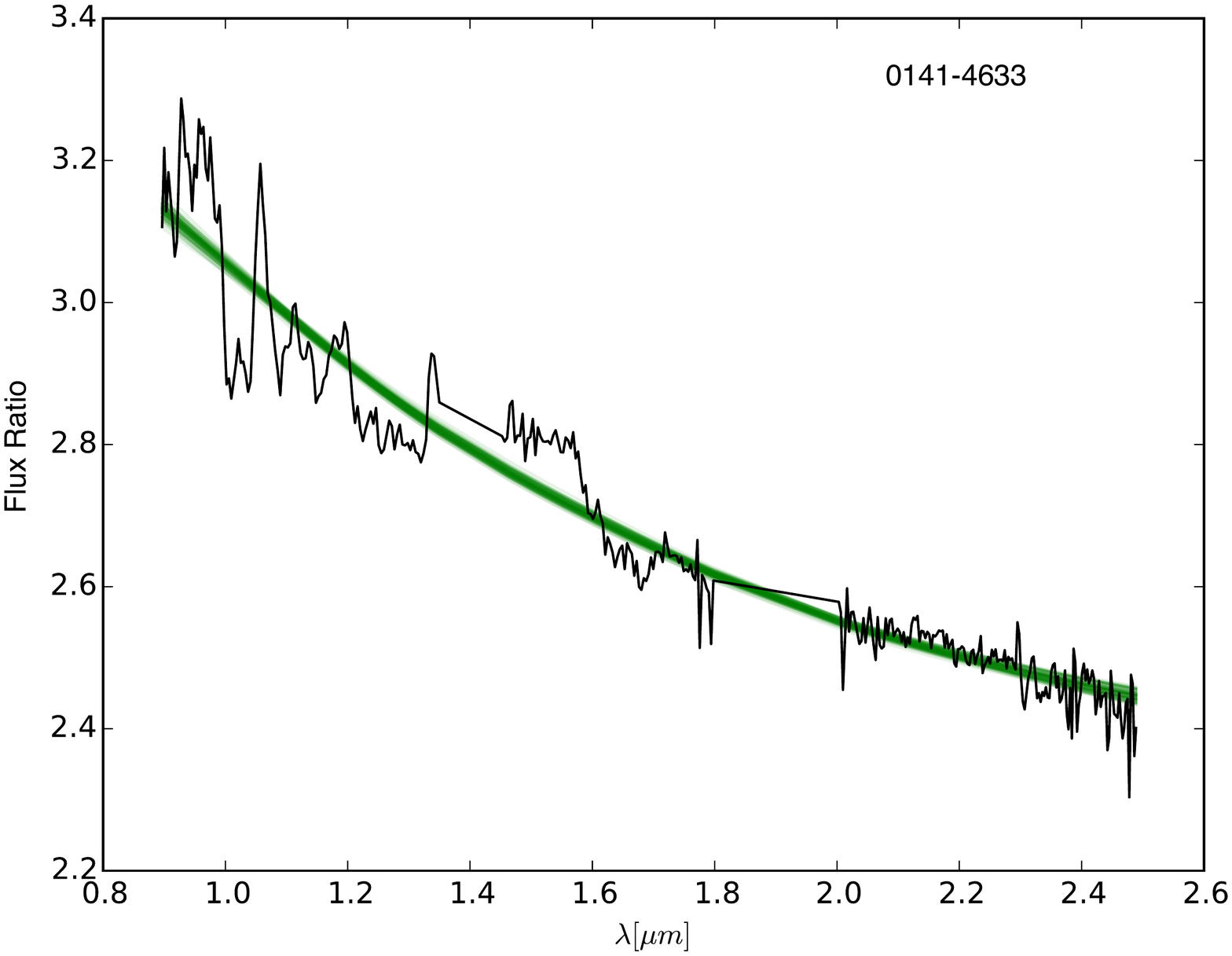}
\caption{Green lines are 100 models randomly drawn from the posterior distribution overplotted on the observed reddening in black. The telluric bands (1.35--1.45 $\micron$ and 1.8--2.0 $\micron$) are removed. The models reproduce the overall shape of the observed reddening well.}
\label{fig: fit}
\end{figure}

We used the MCMC method described in $\S\ref{sec: Method}$ to fit dust haze extinction models to the observed reddening curves for each object and constrain the physical properties of the proposed dust haze. We plot the 1D and 2D marginalized posterior PDFs for all parameters in Figure Set~$\ref{fig: triangle}$.  Models corresponding to 100 randomly-drawn parameter sets from the posterior PDF are shown with the data in Figure Set~$\ref{fig: fit}$. The constrained properties of the dust haze include mean effective radius, effective variance of the Hansen distribution ($\S\ref{subsec: distribution}$), and column density of the dust haze.
% Figure~$\ref{fig: triangle}$ and $\ref{fig: fit}$ show the posterior distribution for each parameter and model fits for each of the \textbf{56} objects in our sample.
% are figure sets of 61 plots each for all the objects in our sample described in Table 1. 

In Figure Set 6, we show the posterior distributions for each parameter. Each figure has 1-D distributions for the parameters and 2-D contours for each combination of parameters. Gaussian-like 1-D distributions and round 2-D contours indicate no covariances. Quantiles (16, 50, 84 $\%$) are shown with dashed lines and are used to report the uncertainties on the parameter fits. 
In many objects, the PDFs for the mean effective radius $a$, column density $N$, tolerance parameter $\log s$, and vertical offset constant $C$ have clear peaks and therefore are well constrained. The variance $b$, on the other hand, is not well constrained in most objects.

The column density values $N$ are comparable to the value of typical brown dwarf atmospheres ($\sim 10^8\,\rm cm^{-2}$), which indicates that our results are reasonably realistic.
However, there is a correlation between parameters $a$ and $N$ as seen in the 2-D contours. The relationship between $a$ and $N$ is shown in Equation \ref{eq: eq9}. In order to compute the optical depth, we multiply the column density $N$, scattering cross section $\pi a^2$, and effective extinction coefficient $Q_{\rm ext}(\lambda)$. The effective extinction curves are similar over a small range of grain radii, so we are constraining the product $Na^2$. Therefore, $a^2$ and $N$ are inversely proportional and this relationship appears in the posterior distributions.
%The term $N\pi a^2$ is the sum of the scattering cross sections of forsterite grains per unit area
%This is because $a^2$ and $N$ are inversely proportional (see Equation 10).

In Figure Set 7, we show the resulting model fits to the observed reddening. The black line is the observed reddening ($\S\ref{sec: reddening}$) and the green lines are 100 models randomly drawn from the posterior distributions. The models reproduce the overall shape of the observed reddening.

In Figure Set 8, we compare a de-reddened spectrum of a red L dwarf, the spectrum of the field standard L dwarf, and the original red L dwarf spectrum. The de-reddened spectrum is the spectrum of a red L dwarf corrected by the best-fit forsterite extinction curve determined by the MCMC analysis. 
$\chi^2$ values between the red and standard spectra, and between the de-reddened and standard spectra are reported. 
The $\chi^2$ values are used simply to quantitatively demonstrate that the de-reddened spectrum is a better fit to the standard spectrum than the original red L dwarf spectrum. 
%measure of the goodness of fit and not for statistical purposes. 
The tolerance parameter $s$ is not included in the calculation because it is not necessary for this purpose.
% and it is not clear what value of $s$ should be used for the calculation of $\chi^2$ between the original red L dwarf and the standard spectra.}
%The tolerance parameter $s$ is included in the calculation of $\chi^2$. 
In all cases the de-reddened spectrum fits the standard spectrum better than the original red L dwarf spectrum, as reflected by the significant improvement of $\chi^2$ value after the de-reddening.

In Figures~$\ref{fig: combo1}$ and~$\ref{fig: combo2}$, we show an example of the de-reddened spectra (Figure Set~8) and model fits (Figure Set~7) for each spectral type. In all spectral types, the de-reddened spectra looks much closer to the standard spectra than the original red L dwarf spectra. These results show that the submicron-sized dust haze prescription can successfully account for the red SED and $J - K_s$ colors of L dwarfs.

In Figure~$\ref{fig: dchisq}$, we show improvement in $\chi^2$ due to the proposed dust haze prescription. The ratio of $\chi^2$ before de-reddening to $\chi^2$ after de-reddening is plotted against $\Delta (J - K_s)$ color, which is the difference in $J - K_s$ color between the red L dwarf and the field standard L dwarf. We use $\Delta (J - K_s)$ because we compare the spectra of red L dwarfs to the standards to isolate the observed reddening. The value of $\chi^2$ is improved for all objects, which shows that the de-reddened spectra fit the field standards better, and in most cases substantially better ($> 10$x), than the original red L dwarf spectra.
%and the ratio is greater for objects with redder $J - K_s$ colors (?).

\subsection{Correlation with Gravity}
\label{subsec: correlation}

It has been widely noted that low-gravity L dwarfs have redder NIR spectral energy distributions compared to the field-gravity spectral standards. A discussion of how clouds behave differently at low and moderate gravity is given in \citet{Marley12}. In addition
to the cloud height differences discussed by \citet{Marley12}, 
one might expect to see a correlation between dust haze properties and low-gravity spectral features. We hypothesized that the proposed dust haze might dissipate over time due to grain growth by condensation. Large condensed particles are expected to fall out of the dust haze as a result of the sedimentation rate exceeding the remixing rate by eddy turbulence \citep{Marley99}. Reduced
convective velocities and perhaps less vigorous gravity wave excitation with age could also contribute to more efficient
dust settling. Thus, young, low gravity L dwarfs might have optically thicker dust hazes which may explain their red NIR colors. The proposed dust haze could also explain the reddening within field L dwarfs and the properties of the dust haze might be correlated with age.
%, but their dust hazes are expected to be different from those in young L dwarf atmospheres since the red field L dwarfs have higher surface gravities than young L dwarfs.
%Contrary to our hypothesis (theory? expectation?), 
%As shown in Figure~$\ref{fig: ascat}$ and $\ref{fig: nscat}$, both mean effective radius and column density behave similarly for young and red field objects and there is no apparent trend in young objects in comparison to red field objects. 

%In an attempt to determine if there is a correlation between the dust haze properties and gravity, we plotted the distributions of the computed dust haze parameters for both low-gravity and field objects.
%visualize and compare the distributions of the dust haze properties between low-gravity and field L dwarfs, we made scatter plots of each parameter 
%Even though a correlation between dust haze properties and gravity is expected, there is no statistical difference in $a$ and $N$ between low-gravity and red field L dwarfs. 
%(KS test to be done before submission)
In Figures~$\ref{fig: ascat}$ and $\ref{fig: nscat}$, we show scatter plots of mean effective radius $a$ and column density $N$ versus $\Delta(J-K)$ color, respectively. Green symbols denote low-gravity objects and magenta denote field-gravity. Circles denote objects with PDFs with strong constraints, while squares and diamonds denote objects with weak constraints for some of the parameters. There are visually noticeable differences between the distributions of the low-gravity and field-gravity objects in both Figure~$\ref{fig: ascat}$ and Figure~$\ref{fig: nscat}$. In Figure~$\ref{fig: ascat}$, we show that there is no correlation between mean effective radius $a$ and $\Delta(J-K)$ color for the low-gravity objects, while the distribution of the field-gravity objects show a correlation. In Figure~$\ref{fig: nscat}$, the distribution of the low-gravity objects span a wider range of column densities, while the field-gravity objects tend to have lower column densities. Even though there are visible differences between the distributions of the low-gravity and field-gravity objects, the two distributions partially overlap and the correlations are not strong.
%\textbf{There is no noticeable correlation between these dust properties and $\Delta(J - K_s)$ color. 

In order to quantitatively determine if the distributions of the low-gravity objects and the field-gravity objects are indeed different, we performed a two-dimensional K-S test on the two distributions in both Figure~$\ref{fig: ascat}$ and Figure~$\ref{fig: nscat}$. The probability of the two distributions in Figure~$\ref{fig: ascat}$ drawn from the same parent distribution is $2.32\cdot 10^{-4}$, and the probability of the two distributions in Figure~$\ref{fig: nscat}$ drawn from the same parent distribution is $5.11\cdot 10^{-6}$. These results show that the low-gravity and the field-gravity objects are most likely drawn from different distributions. 
%We also performed the same test on the objects with PDFs with clear peaks (circles). In both cases, 
%The results of the test indicate that there is no statistical difference in mean effective radius $a$ and column density $N$ between the low-gravity and field-gravity populations, even though a correlation between dust haze properties and gravity might be expected.}

%There are several possible explanations for this inconsistency
%[gap, conflict, inconsistency] 
 %between our hypothesis and results. 

The lack of strong correlations between the dust haze properties and gravity might be due to a model grid not spanning a wide enough range of effective variance, $b$. We consider 0.1 $< b <$ 1.0 because Hansen particle size distributions for large $b$ look like a power law size distribution (Figure~$\ref{fig: dist}$; Equation 1). 
%This appears to be the case for any range of $b$
%because the model grid was not wide enough. 
This makes the results for these objects somewhat unreliable. In most of those objects, $b$ tends to be close to the upper limit and therefore is not well constrained (Figure Set \ref{fig: triangle}). %(Cite object examples. It wants to go bigger. How many)
%SEEMS TO ME THAT THIS IMPLIES THAT A POWER LAW WOULD WORK BETTER? MAYBE SAY THIS?

Further, our initial assumptions about low-gravity and field L dwarfs may be unrealistic. We assumed that low-gravity and field L dwarfs with the same base spectral type (e.g. L2 and L2$\gamma$) have the same effective temperatures and the low-gravity L dwarf has the hypothetical dust haze of small grains in the upper atmosphere. However, recent evidence suggests that low-gravity L dwarfs do not necessarily share the same physical properties with field L dwarfs just because they share the same base spectral type \citep{Luhman12, Filippazzo15}. Young, low gravity L dwarfs might have cooler effective temperatures than field L dwarfs in the same spectral classification \citep{faherty13}. This suggests that an earlier type standard would be a better comparison. We could be comparing objects with different effective temperatures and thus, not setting accurate estimates of the dust haze properties. Since the overall shape of a spectrum is very sensitive to the effective temperature, comparing objects with the same effective temperatures might be more useful for future analysis.
%low-gravity L dwarfs should be compared to field L dwarfs with the same effective temperature regardless of the spectral type. 
%Our hypothesis and method would still be robust and reliable, except that it might be more appropriate to compare red objects to field objects with the same effective temperatures instead of field spectral standards. 

Finally, the overlapping distributions of $a$ and $N$ may indicate that the dust hazes of low-gravity and red field L dwarfs are not different from from one another. It might be the case that gravity does not play a major role in determining the properties of the dust haze and the same distribution of dust haze properties exist in both low-gravity and red field L dwarfs. Some red field L dwarfs might have a dust haze for reasons other than gravity, such as differing rotation rates, composition, or evolutionary history.

\subsection{Comparison to \citet{Marocco14}}
\label{subsec: Comparison}

%They did what and we did what. They got what and we got what.
Independently, \citet{Marocco14} (hereafter, M14) present results from a similar analysis. \citet{Bonn15} closely followed the approach of \citet{Marocco14} so we do not consider it further here.
%Our work is similar to \citet{Marocco14} even though the ideas are independent. 
Like us, they are motivated by the utility of the interstellar extinction law in de-reddening red L dwarf spectra, use Mie theory to characterize a high-altitude population of sub-micron dust grains, isolate the observed reddening by comparing red L dwarf spectra to spectral standards. 

%Although our ideas and motivations are similar to those of \citet{Marocco14}, our methods are different from theirs.
However, M14 consider corundum ($\rm Al_{2}O_3$), enstatite ($\rm MgSiO_3$), and iron while we use forsterite ($\rm Mg_{2}SiO_4$). We use forsterite as a test dust particle because extinction curves for forsterite, enstatite --both of which are silicates-- and corundum all behave similarly in the near infrared. We use silicate instead of iron because silicate grains form higher in the atmosphere than iron.
%We did not use iron because fewer refractive indices were available so it was harder to interpolate them to match the smooth shape of the observed reddening. 
Actual dust grains in the brown dwarfs atmospheres are most likely a mixture of these species.
% so we will consider enstatite and corundum in future work.

%As discussed in $\S \ref{subsec: distribution}$, M14 used Gaussian grain size distributions for enstatite and corundum to model the extinction due to the dust haze. For iron, they used a power law grain size distribution because the fit was poor with a Gaussian grain size distribution. As described in $\S \ref{subsubsec: powerlaw}$, we found that power law grain size distributions do not reproduce the observed reddening. Instead, we found that Hansen grain size distributions do provide good results. As shown in Figure $\ref{fig: dist}$, Hansen distributions include greater amounts of small particles compared to Gaussian distributions, but they resemble Gaussian for small variances.

%Both \citet{Marocco14} and we compared red L dwarf spectra to spectral standards to isolate the observed reddening. However, this method carelessly assumes that the red and standard objects have the same effective temperature and therefore may not be the best way to analyze the reddening as discussed earlier in this section.
%Both our work and \citet{Marocco14} considered similar ranges of particle radius (0.01/0.05 - 1. This is because 

Our model includes more parameters than M14. They have two parameters: characteristic grain radius $\it r$ and normalization of the extinction curve at 2.20 $\mu$m while our model includes mean effective grain radius $a$, effective grain size variance $b$, column density $N$, vertical offset $C$, and tolerance factor $s$ as described in $\S \ref{sec: Model}$. The $N$ and $C$ parameters account for normalization as shown in Equation 9.
M14 perform $\chi^2$ minimization to select the best fit parameters and therefore do not have uncertainties, while we use MCMC to determine best fit parameters and their uncertainties.

The range of the characteristic grain radius M14 found for corundum and enstatite are slightly larger than what we found. They obtained $r =$ 0.4 -- 0.6~$\micron$ for corundum and enstatite, while our results for the mean effective radius for forsterite are generally smaller ($0.15 - 0.35 \micron$). Their results for the maximum radius of iron are 0.15 -- 0.3 $\micron$, closer to the values we found for forsterite grains. 

M14 applied their method to five red L dwarfs, and we applied our method to 46 red L dwarfs including low-gravity and field-gravity objects. 
%(detail of objects. duplicates? standards?)
They used objects with later spectral types (L5 -- L7), so there is only one common object between their sample and ours. 2MASS~0355+1133 is used in both studies but compared to different spectral standards. M14 used SDSS~J0835+1953 as the L5 standard while we used 2MASS~1507-1627. They found the characteristic grain radius for 2MASS~0355+1133 to be 0.4 $\micron$ and we found the mean effective radius to be $0.3\substack{+0.03 \\ -0.02}~\micron$.

%$\num[statistical={+0.2,-0.3}, systematic=0.1] 1.24$
Regardless of the different methods, these two studies show similar results. The reproducibility of the results demonstrates the viability of sub-micron size dust grains, and warrants further study and inclusion of small dust grains in future atmosphere models.

\section{Conclusion and Summary}
\label{sec: Conclusion}
%Motivated by the success of the interstellar reddening law, we propose and test the validity of a prescription of a dust haze of small particles in the upper atmospheres of red L dwarfs that can possibly explain their red NIR colors.
%We studied and characterized the properties of the proposed dust haze in the atmospheres of red L dwarfs.
%There are L dwarfs whose spectra appear redder than normal, field-aged L dwarfs (ref). Red L dwarfs have excess flux in the near IR compared to normal, field-aged L dwarfs (Figure~$\ref{fig: spec}$). This observed reddening is not yet well understood or reproduced by current atmosphere models. We hypothesize that this observed reddening might be caused by small dust grains in the atmospheres of L dwarfs. 
The success of the interstellar reddening law in de-reddening L dwarf spectra led us to investigate the possibility of a population of small submicron-sized dust grains in red L dwarf atmospheres that can possibly explain their red NIR colors.

In order to isolate and characterize the reddening, we compare spectra of red L dwarfs to field spectral standards. The observed reddening is treated as smooth power-law shaped curves. 
We use Mie theory to model the dust haze of small particles with theoretical extinction curves. 
%We compared the theoretical extinction curves and the observed reddening curves and found that small grains ($< 0.5~\micron$) best reproduce the shapes of the red SEDs.
%, we assumed that there is a dust haze in the upper atmospheres of red L dwarfs above the regular clouds that consist of bigger particles that are also in the atmospheres of normal L dwarfs. In addition to the common regular clouds, we assumed that red L dwarfs are identical to normal L dwarfs except for the dust haze.
%To get the empirical reddening curve, we divided the flux of the standard L dwarf by the flux of the red L dwarf. 
%We found that Hansen grain size distributions reproduce the shape of the reddening curve better than power law grain size distributions.}
%, therefore we used Hansen grain size distributions along with Mie extinction coefficients to get theoretical extinction curves.
%reproduce the flux ratio of the red L dwarf to normal L dwarf, which quantifies the observed reddening.

We use a Markov-Chain Monte Carlo algorithm to fit the theoretical extinction curves to the observed reddening in order to find the best-fit values and uncertainties for the properties of the dust haze. 
%We found the best-fit values and uncertainties of our parameters including the mean effective grain radius $a$, effective grain size variance $b$, and column density $N$. 
We apply this method to 23 L dwarfs with low-gravity spectral features and 23 field L dwarfs with redder $J-K$ colors than the spectral standards.
We find that small forsterite grains ($< 0.5~\micron$) reproduce the observed reddening.
%We constrained the mean effective radius to be $a$ = 0.2 or 0.3 $\micron$(update values). 
%There is no apparent difference in grain properties between low-gravity and field L dwarfs.
There are differences in grain properties between low-gravity and field L dwarfs, which indicates that dust haze properties may be correlated with surface gravity.
%The best match value for $b$ increases with spectral type. This result shows that in order to reproduce the observed reddening, we need much smaller grains in the dust haze compared to grains in the regular clouds, which are around 10 $\micron$ (ref).
The best-fit column densities of forsterite grains are reasonable compared to typical brown dwarf atmospheres.
% ($\sim 10^8\,\rm cm^{-2}$).
%Mie extinction coefficients and Hansen distributions can reproduce the observed reddening
These results suggest that a dust haze of small particles with a Hansen particle size distribution can account for the observed red NIR spectral energy distributions of brown dwarfs. However, to rigorously explore this hypothesis other particle species and particle size distributions should
be explored as well.

This work is a proof of concept and it provides a strong motivation for including small dust grains in future atmosphere models of brown dwarfs. Future work on the role of small grains in brown dwarf atmospheres will include other grain species such as corundum and enstatite. 
%We will also compare the red L dwarfs to field L dwarfs with the same effective temperature instead of the spectral standard objects.
%Furthermore, we will add the proposed dust haze to atmosphere models of brown dwarfs.
The dust haze analysis can be applied to other studies including variability in brown dwarfs, exoplanet atmospheres, and the interstellar/intergalactic medium.

\vspace{12mm}

We thank our anonymous referee for thorough and helpful comments. This material is based upon work supported by the National Science Foundation under Grant No. AST-1313278. Support for this project was provided by a PSC-CUNY Award, jointly funded by The Professional Staff Congress and The City University of New York.

%\begin{figure}
%\plotone{f3.pdf}
%\caption{Top panel shows forsterite extinction coefficients according to Mie theory averaged over Hansen particle size distributions with various combinations of mean effective radius ($a$) and effective variance ($b$). Different colors correspond to different effective radii. Different line style correspond to different effective variance. The shapes of the extinction curves for smaller particles (0.1 -- 0.4 $\micron$) resemble the observed reddening but larger particles (1.0 $\micron$) do not.
%Bottom right panel shows forsterite extinction coefficients averaged over power law particle size distributions with indices of -3 and -3.5. The extinction curves are too flat and do not reproduce the observed reddening.
%Bottom right panel shows forsterite extinction coefficients averaged over Gaussian particle size distributions mean radius of 0.3 and 0.5 $\micron$.}
%\label{fig: Qext}
%\end{figure}

\figsetstart
\figsetnum{6}
\figsettitle{Posterior distributions for MCMC parameters}

\figsetgrpstart
\figsetgrpnum{6.1}
\figsetgrptitle{Figure of 0141-4633}
\figsetplot{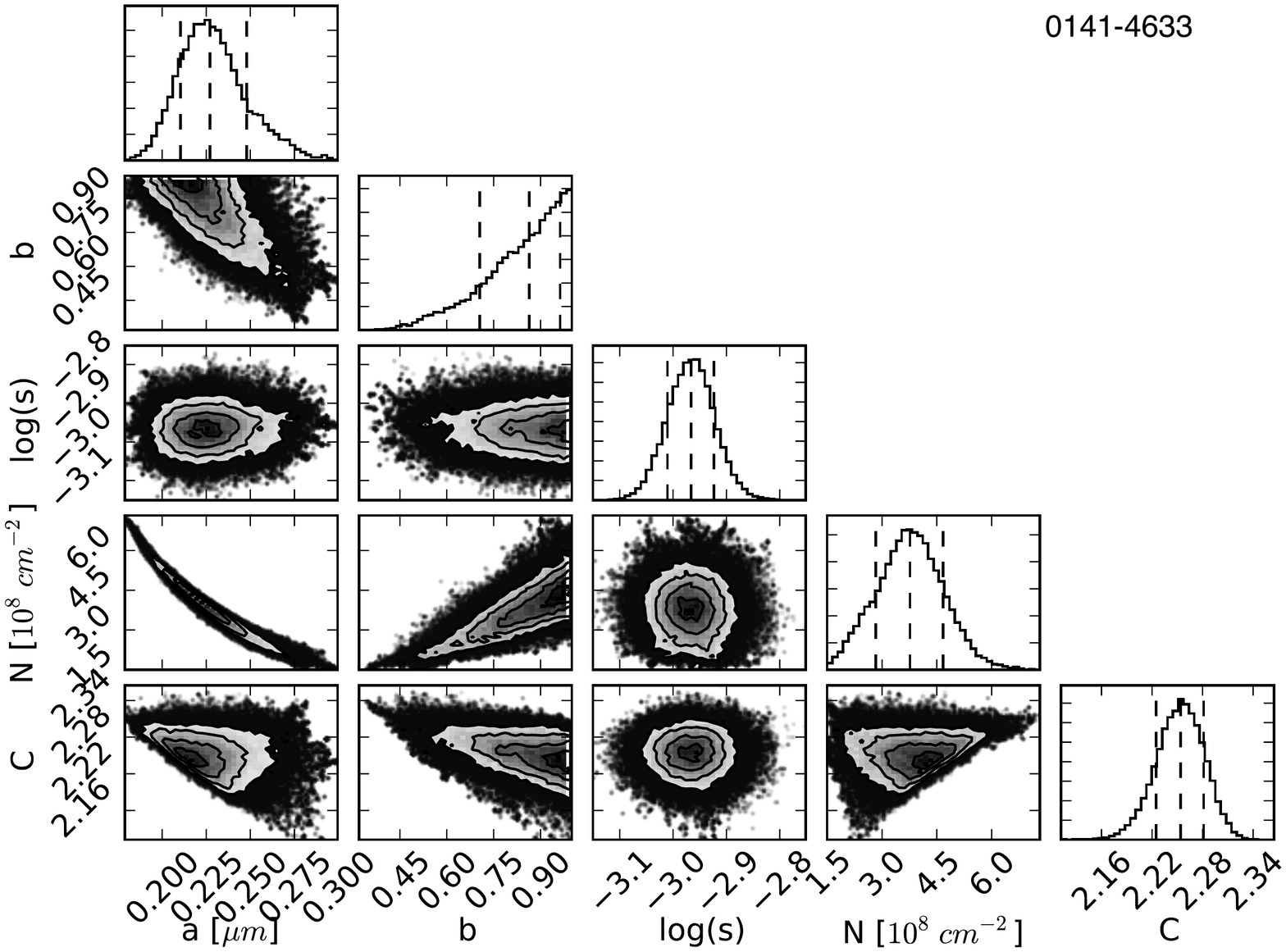}
\figsetgrpnote{Posterior distributions of mean effective radius, effective variance, tolerance, column density, and offset showing 1-D distributions for each parameter and 2-D distributions for each combination of parameters. Dashed lines in the 1-D distributions represent 16, 50, 84 percent quantiles, corresponding to the median and 1 $\sigma$ uncertainties.}
\figsetgrpend

\figsetgrpstart
\figsetgrpnum{6.2}
\figsetgrptitle{Figure of 0032-4405}
\figsetplot{f6_2.pdf}
\figsetgrpnote{Posterior distributions of mean effective radius, effective variance, tolerance, column density, and offset showing 1-D distributions for each parameter and 2-D distributions for each combination of parameters. Dashed lines in the 1-D distributions represent 16, 50, 84 percent quantiles, corresponding to the median and 1 $\sigma$ uncertainties.}
\figsetgrpend

\figsetgrpstart
\figsetgrpnum{6.3}
\figsetgrptitle{Figure of 0210-3015}
\figsetplot{f6_3.pdf}
\figsetgrpnote{Posterior distributions of mean effective radius, effective variance, tolerance, column density, and offset showing 1-D distributions for each parameter and 2-D distributions for each combination of parameters. Dashed lines in the 1-D distributions represent 16, 50, 84 percent quantiles, corresponding to the median and 1 $\sigma$ uncertainties.}
\figsetgrpend

\figsetgrpstart
\figsetgrpnum{6.4}
\figsetgrptitle{Figure of 0241-0326}
\figsetplot{f6_4.pdf}
\figsetgrpnote{Posterior distributions of mean effective radius, effective variance, tolerance, column density, and offset showing 1-D distributions for each parameter and 2-D distributions for each combination of parameters. Dashed lines in the 1-D distributions represent 16, 50, 84 percent quantiles, corresponding to the median and 1 $\sigma$ uncertainties.}
\figsetgrpend

\figsetgrpstart
\figsetgrpnum{6.5}
\figsetgrptitle{Figure of 0323-4631}
\figsetplot{f6_5.pdf}
\figsetgrpnote{Posterior distributions of mean effective radius, effective variance, tolerance, column density, and offset showing 1-D distributions for each parameter and 2-D distributions for each combination of parameters. Dashed lines in the 1-D distributions represent 16, 50, 84 percent quantiles, corresponding to the median and 1 $\sigma$ uncertainties.}
\figsetgrpend

\figsetgrpstart
\figsetgrpnum{6.6}
\figsetgrptitle{Figure of 2213-2136}
\figsetplot{f6_6.pdf}
\figsetgrpnote{Posterior distributions of mean effective radius, effective variance, tolerance, column density, and offset showing 1-D distributions for each parameter and 2-D distributions for each combination of parameters. Dashed lines in the 1-D distributions represent 16, 50, 84 percent quantiles, corresponding to the median and 1 $\sigma$ uncertainties.}
\figsetgrpend

\figsetgrpstart
\figsetgrpnum{6.7}
\figsetgrptitle{Figure of 2315+0617}
\figsetplot{f6_7.pdf}
\figsetgrpnote{Posterior distributions of mean effective radius, effective variance, tolerance, column density, and offset showing 1-D distributions for each parameter and 2-D distributions for each combination of parameters. Dashed lines in the 1-D distributions represent 16, 50, 84 percent quantiles, corresponding to the median and 1 $\sigma$ uncertainties.}
\figsetgrpend

\figsetgrpstart
\figsetgrpnum{6.8}
\figsetgrptitle{Figure of 1711+2326}
\figsetplot{f6_8.pdf}
\figsetgrpnote{Posterior distributions of mean effective radius, effective variance, tolerance, column density, and offset showing 1-D distributions for each parameter and 2-D distributions for each combination of parameters. Dashed lines in the 1-D distributions represent 16, 50, 84 percent quantiles, corresponding to the median and 1 $\sigma$ uncertainties.}
\figsetgrpend

\figsetgrpstart
\figsetgrpnum{6.9}
\figsetgrptitle{Figure of 1552+2948}
\figsetplot{f6_9.pdf}
\figsetgrpnote{Posterior distributions of mean effective radius, effective variance, tolerance, column density, and offset showing 1-D distributions for each parameter and 2-D distributions for each combination of parameters. Dashed lines in the 1-D distributions represent 16, 50, 84 percent quantiles, corresponding to the median and 1 $\sigma$ uncertainties.}
\figsetgrpend

\figsetgrpstart
\figsetgrpnum{6.10}
\figsetgrptitle{Figure of 0357-4417}
\figsetplot{f6_10.pdf}
\figsetgrpnote{Posterior distributions of mean effective radius, effective variance, tolerance, column density, and offset showing 1-D distributions for each parameter and 2-D distributions for each combination of parameters. Dashed lines in the 1-D distributions represent 16, 50, 84 percent quantiles, corresponding to the median and 1 $\sigma$ uncertainties.}
\figsetgrpend

\figsetgrpstart
\figsetgrpnum{6.11}
\figsetgrptitle{Figure of 0117-3403}
\figsetplot{f6_11.pdf}
\figsetgrpnote{Posterior distributions of mean effective radius, effective variance, tolerance, column density, and offset showing 1-D distributions for each parameter and 2-D distributions for each combination of parameters. Dashed lines in the 1-D distributions represent 16, 50, 84 percent quantiles, corresponding to the median and 1 $\sigma$ uncertainties.}
\figsetgrpend

\figsetgrpstart
\figsetgrpnum{6.12}
\figsetgrptitle{Figure of 0518-2756}
\figsetplot{f6_12.pdf}
\figsetgrpnote{Posterior distributions of mean effective radius, effective variance, tolerance, column density, and offset showing 1-D distributions for each parameter and 2-D distributions for each combination of parameters. Dashed lines in the 1-D distributions represent 16, 50, 84 percent quantiles, corresponding to the median and 1 $\sigma$ uncertainties.}
\figsetgrpend

\figsetgrpstart
\figsetgrpnum{6.13}
\figsetgrptitle{Figure of 0055+0134}
\figsetplot{f6_13.pdf}
\figsetgrpnote{Posterior distributions of mean effective radius, effective variance, tolerance, column density, and offset showing 1-D distributions for each parameter and 2-D distributions for each combination of parameters. Dashed lines in the 1-D distributions represent 16, 50, 84 percent quantiles, corresponding to the median and 1 $\sigma$ uncertainties.}
\figsetgrpend

\figsetgrpstart
\figsetgrpnum{6.14}
\figsetgrptitle{Figure of 0536-1920}
\figsetplot{f6_14.pdf}
\figsetgrpnote{Posterior distributions of mean effective radius, effective variance, tolerance, column density, and offset showing 1-D distributions for each parameter and 2-D distributions for each combination of parameters. Dashed lines in the 1-D distributions represent 16, 50, 84 percent quantiles, corresponding to the median and 1 $\sigma$ uncertainties.}
\figsetgrpend

\figsetgrpstart
\figsetgrpnum{6.15}
\figsetgrptitle{Figure of 1551+0941}
\figsetplot{f6_15.pdf}
\figsetgrpnote{Posterior distributions of mean effective radius, effective variance, tolerance, column density, and offset showing 1-D distributions for each parameter and 2-D distributions for each combination of parameters. Dashed lines in the 1-D distributions represent 16, 50, 84 percent quantiles, corresponding to the median and 1 $\sigma$ uncertainties.}
\figsetgrpend

\figsetgrpstart
\figsetgrpnum{6.16}
\figsetgrptitle{Figure of 1726+1538}
\figsetplot{f6_16.pdf}
\figsetgrpnote{Posterior distributions of mean effective radius, effective variance, tolerance, column density, and offset showing 1-D distributions for each parameter and 2-D distributions for each combination of parameters. Dashed lines in the 1-D distributions represent 16, 50, 84 percent quantiles, corresponding to the median and 1 $\sigma$ uncertainties.}
\figsetgrpend

\figsetgrpstart
\figsetgrpnum{6.17}
\figsetgrptitle{Figure of 0501-0010}
\figsetplot{f6_17.pdf}
\figsetgrpnote{Posterior distributions of mean effective radius, effective variance, tolerance, column density, and offset showing 1-D distributions for each parameter and 2-D distributions for each combination of parameters. Dashed lines in the 1-D distributions represent 16, 50, 84 percent quantiles, corresponding to the median and 1 $\sigma$ uncertainties.}
\figsetgrpend

\figsetgrpstart
\figsetgrpnum{6.18}
\figsetgrptitle{Figure of 2249+0044}
\figsetplot{f6_18.pdf}
\figsetgrpnote{Posterior distributions of mean effective radius, effective variance, tolerance, column density, and offset showing 1-D distributions for each parameter and 2-D distributions for each combination of parameters. Dashed lines in the 1-D distributions represent 16, 50, 84 percent quantiles, corresponding to the median and 1 $\sigma$ uncertainties.}
\figsetgrpend

\figsetgrpstart
\figsetgrpnum{6.19}
\figsetgrptitle{Figure of 0512-2949}
\figsetplot{f6_19.pdf}
\figsetgrpnote{Posterior distributions of mean effective radius, effective variance, tolerance, column density, and offset showing 1-D distributions for each parameter and 2-D distributions for each combination of parameters. Dashed lines in the 1-D distributions represent 16, 50, 84 percent quantiles, corresponding to the median and 1 $\sigma$ uncertainties.}
\figsetgrpend

\figsetgrpstart
\figsetgrpnum{6.20}
\figsetgrptitle{Figure of 0326-2102}
\figsetplot{f6_20.pdf}
\figsetgrpnote{Posterior distributions of mean effective radius, effective variance, tolerance, column density, and offset showing 1-D distributions for each parameter and 2-D distributions for each combination of parameters. Dashed lines in the 1-D distributions represent 16, 50, 84 percent quantiles, corresponding to the median and 1 $\sigma$ uncertainties.}
\figsetgrpend

\figsetgrpstart
\figsetgrpnum{6.21}
\figsetgrptitle{Figure of 2154-1055}
\figsetplot{f6_21.pdf}
\figsetgrpnote{Posterior distributions of mean effective radius, effective variance, tolerance, column density, and offset showing 1-D distributions for each parameter and 2-D distributions for each combination of parameters. Dashed lines in the 1-D distributions represent 16, 50, 84 percent quantiles, corresponding to the median and 1 $\sigma$ uncertainties.}
\figsetgrpend

\figsetgrpstart
\figsetgrpnum{6.22}
\figsetgrptitle{Figure of 0355+1133}
\figsetplot{f6_22.pdf}
\figsetgrpnote{Posterior distributions of mean effective radius, effective variance, tolerance, column density, and offset showing 1-D distributions for each parameter and 2-D distributions for each combination of parameters. Dashed lines in the 1-D distributions represent 16, 50, 84 percent quantiles, corresponding to the median and 1 $\sigma$ uncertainties.}
\figsetgrpend

\figsetgrpstart
\figsetgrpnum{6.23}
\figsetgrptitle{Figure of 1615+4953}
\figsetplot{f6_23.pdf}
\figsetgrpnote{Posterior distributions of mean effective radius, effective variance, tolerance, column density, and offset showing 1-D distributions for each parameter and 2-D distributions for each combination of parameters. Dashed lines in the 1-D distributions represent 16, 50, 84 percent quantiles, corresponding to the median and 1 $\sigma$ uncertainties.}
\figsetgrpend

\figsetgrpstart
\figsetgrpnum{6.24}
\figsetgrptitle{Figure of 0235-2331}
\figsetplot{f6_24.pdf}
\figsetgrpnote{Posterior distributions of mean effective radius, effective variance, tolerance, column density, and offset showing 1-D distributions for each parameter and 2-D distributions for each combination of parameters. Dashed lines in the 1-D distributions represent 16, 50, 84 percent quantiles, corresponding to the median and 1 $\sigma$ uncertainties.}
\figsetgrpend

\figsetgrpstart
\figsetgrpnum{6.25}
\figsetgrptitle{Figure of 0543+6422}
\figsetplot{f6_25.pdf}
\figsetgrpnote{Posterior distributions of mean effective radius, effective variance, tolerance, column density, and offset showing 1-D distributions for each parameter and 2-D distributions for each combination of parameters. Dashed lines in the 1-D distributions represent 16, 50, 84 percent quantiles, corresponding to the median and 1 $\sigma$ uncertainties.}
\figsetgrpend

\figsetgrpstart
\figsetgrpnum{6.26}
\figsetgrptitle{Figure of 0602+6336}
\figsetplot{f6_26.pdf}
\figsetgrpnote{Posterior distributions of mean effective radius, effective variance, tolerance, column density, and offset showing 1-D distributions for each parameter and 2-D distributions for each combination of parameters. Dashed lines in the 1-D distributions represent 16, 50, 84 percent quantiles, corresponding to the median and 1 $\sigma$ uncertainties.}
\figsetgrpend

\figsetgrpstart
\figsetgrpnum{6.27}
\figsetgrptitle{Figure of 0016-4056}
\figsetplot{f6_27.pdf}
\figsetgrpnote{Posterior distributions of mean effective radius, effective variance, tolerance, column density, and offset showing 1-D distributions for each parameter and 2-D distributions for each combination of parameters. Dashed lines in the 1-D distributions represent 16, 50, 84 percent quantiles, corresponding to the median and 1 $\sigma$ uncertainties.}
\figsetgrpend

\figsetgrpstart
\figsetgrpnum{6.28}
\figsetgrptitle{Figure of 2339+3507}
\figsetplot{f6_28.pdf}
\figsetgrpnote{Posterior distributions of mean effective radius, effective variance, tolerance, column density, and offset showing 1-D distributions for each parameter and 2-D distributions for each combination of parameters. Dashed lines in the 1-D distributions represent 16, 50, 84 percent quantiles, corresponding to the median and 1 $\sigma$ uncertainties.}
\figsetgrpend

\figsetgrpstart
\figsetgrpnum{6.29}
\figsetgrptitle{Figure of 1100+4957}
\figsetplot{f6_29.pdf}
\figsetgrpnote{Posterior distributions of mean effective radius, effective variance, tolerance, column density, and offset showing 1-D distributions for each parameter and 2-D distributions for each combination of parameters. Dashed lines in the 1-D distributions represent 16, 50, 84 percent quantiles, corresponding to the median and 1 $\sigma$ uncertainties.}
\figsetgrpend

\figsetgrpstart
\figsetgrpnum{6.30}
\figsetgrptitle{Figure of 0051-1544}
\figsetplot{f6_30.pdf}
\figsetgrpnote{Posterior distributions of mean effective radius, effective variance, tolerance, column density, and offset showing 1-D distributions for each parameter and 2-D distributions for each combination of parameters. Dashed lines in the 1-D distributions represent 16, 50, 84 percent quantiles, corresponding to the median and 1 $\sigma$ uncertainties.}
\figsetgrpend

\figsetgrpstart
\figsetgrpnum{6.31}
\figsetgrptitle{Figure of 2317-4838}
\figsetplot{f6_31.pdf}
\figsetgrpnote{Posterior distributions of mean effective radius, effective variance, tolerance, column density, and offset showing 1-D distributions for each parameter and 2-D distributions for each combination of parameters. Dashed lines in the 1-D distributions represent 16, 50, 84 percent quantiles, corresponding to the median and 1 $\sigma$ uncertainties.}
\figsetgrpend

\figsetgrpstart
\figsetgrpnum{6.32}
\figsetgrptitle{Figure of 0337-1758}
\figsetplot{f6_32.pdf}
\figsetgrpnote{Posterior distributions of mean effective radius, effective variance, tolerance, column density, and offset showing 1-D distributions for each parameter and 2-D distributions for each combination of parameters. Dashed lines in the 1-D distributions represent 16, 50, 84 percent quantiles, corresponding to the median and 1 $\sigma$ uncertainties.}
\figsetgrpend

\figsetgrpstart
\figsetgrpnum{6.33}
\figsetgrptitle{Figure of 0208+2737}
\figsetplot{f6_33.pdf}
\figsetgrpnote{Posterior distributions of mean effective radius, effective variance, tolerance, column density, and offset showing 1-D distributions for each parameter and 2-D distributions for each combination of parameters. Dashed lines in the 1-D distributions represent 16, 50, 84 percent quantiles, corresponding to the median and 1 $\sigma$ uncertainties.}
\figsetgrpend

\figsetgrpstart
\figsetgrpnum{6.34}
\figsetgrptitle{Figure of 0835-0819}
\figsetplot{f6_34.pdf}
\figsetgrpnote{Posterior distributions of mean effective radius, effective variance, tolerance, column density, and offset showing 1-D distributions for each parameter and 2-D distributions for each combination of parameters. Dashed lines in the 1-D distributions represent 16, 50, 84 percent quantiles, corresponding to the median and 1 $\sigma$ uncertainties.}
\figsetgrpend

\figsetgrpstart
\figsetgrpnum{6.35}
\figsetgrptitle{Figure of 0358-4116}
\figsetplot{f6_35.pdf}
\figsetgrpnote{Posterior distributions of mean effective radius, effective variance, tolerance, column density, and offset showing 1-D distributions for each parameter and 2-D distributions for each combination of parameters. Dashed lines in the 1-D distributions represent 16, 50, 84 percent quantiles, corresponding to the median and 1 $\sigma$ uncertainties.}
\figsetgrpend

\figsetgrpstart
\figsetgrpnum{6.36}
\figsetgrptitle{Figure of 0905+5623}
\figsetplot{f6_36.pdf}
\figsetgrpnote{Posterior distributions of mean effective radius, effective variance, tolerance, column density, and offset showing 1-D distributions for each parameter and 2-D distributions for each combination of parameters. Dashed lines in the 1-D distributions represent 16, 50, 84 percent quantiles, corresponding to the median and 1 $\sigma$ uncertainties.}
\figsetgrpend

\figsetgrpstart
\figsetgrpnum{6.37}
\figsetgrptitle{Figure of 1228-1547}
\figsetplot{f6_37.pdf}
\figsetgrpnote{Posterior distributions of mean effective radius, effective variance, tolerance, column density, and offset showing 1-D distributions for each parameter and 2-D distributions for each combination of parameters. Dashed lines in the 1-D distributions represent 16, 50, 84 percent quantiles, corresponding to the median and 1 $\sigma$ uncertainties.}
\figsetgrpend

\figsetgrpstart
\figsetgrpnum{6.38}
\figsetgrptitle{Figure of 1239+5515}
\figsetplot{f6_38.pdf}
\figsetgrpnote{Posterior distributions of mean effective radius, effective variance, tolerance, column density, and offset showing 1-D distributions for each parameter and 2-D distributions for each combination of parameters. Dashed lines in the 1-D distributions represent 16, 50, 84 percent quantiles, corresponding to the median and 1 $\sigma$ uncertainties.}
\figsetgrpend

\figsetgrpstart
\figsetgrpnum{6.39}
\figsetgrptitle{Figure of 0310-2756}
\figsetplot{f6_39.pdf}
\figsetgrpnote{Posterior distributions of mean effective radius, effective variance, tolerance, column density, and offset showing 1-D distributions for each parameter and 2-D distributions for each combination of parameters. Dashed lines in the 1-D distributions represent 16, 50, 84 percent quantiles, corresponding to the median and 1 $\sigma$ uncertainties.}
\figsetgrpend

\figsetgrpstart
\figsetgrpnum{6.40}
\figsetgrptitle{Figure of 0624-4521}
\figsetplot{f6_40.pdf}
\figsetgrpnote{Posterior distributions of mean effective radius, effective variance, tolerance, column density, and offset showing 1-D distributions for each parameter and 2-D distributions for each combination of parameters. Dashed lines in the 1-D distributions represent 16, 50, 84 percent quantiles, corresponding to the median and 1 $\sigma$ uncertainties.}
\figsetgrpend

\figsetgrpstart
\figsetgrpnum{6.41}
\figsetgrptitle{Figure of 0652+4710}
\figsetplot{f6_41.pdf}
\figsetgrpnote{Posterior distributions of mean effective radius, effective variance, tolerance, column density, and offset showing 1-D distributions for each parameter and 2-D distributions for each combination of parameters. Dashed lines in the 1-D distributions represent 16, 50, 84 percent quantiles, corresponding to the median and 1 $\sigma$ uncertainties.}
\figsetgrpend

\figsetgrpstart
\figsetgrpnum{6.42}
\figsetgrptitle{Figure of 1438+5722}
\figsetplot{f6_42.pdf}
\figsetgrpnote{Posterior distributions of mean effective radius, effective variance, tolerance, column density, and offset showing 1-D distributions for each parameter and 2-D distributions for each combination of parameters. Dashed lines in the 1-D distributions represent 16, 50, 84 percent quantiles, corresponding to the median and 1 $\sigma$ uncertainties.}
\figsetgrpend

\figsetgrpstart
\figsetgrpnum{6.43}
\figsetgrptitle{Figure of 1326-0038}
\figsetplot{f6_43.pdf}
\figsetgrpnote{Posterior distributions of mean effective radius, effective variance, tolerance, column density, and offset showing 1-D distributions for each parameter and 2-D distributions for each combination of parameters. Dashed lines in the 1-D distributions represent 16, 50, 84 percent quantiles, corresponding to the median and 1 $\sigma$ uncertainties.}
\figsetgrpend

\figsetgrpstart
\figsetgrpnum{6.44}
\figsetgrptitle{Figure of 2212+3430}
\figsetplot{f6_44.pdf}
\figsetgrpnote{Posterior distributions of mean effective radius, effective variance, tolerance, column density, and offset showing 1-D distributions for each parameter and 2-D distributions for each combination of parameters. Dashed lines in the 1-D distributions represent 16, 50, 84 percent quantiles, corresponding to the median and 1 $\sigma$ uncertainties.}
\figsetgrpend

\figsetgrpstart
\figsetgrpnum{6.45}
\figsetgrptitle{Figure of 2148+4003}
\figsetplot{f6_45.pdf}
\figsetgrpnote{Posterior distributions of mean effective radius, effective variance, tolerance, column density, and offset showing 1-D distributions for each parameter and 2-D distributions for each combination of parameters. Dashed lines in the 1-D distributions represent 16, 50, 84 percent quantiles, corresponding to the median and 1 $\sigma$ uncertainties.}
\figsetgrpend

\figsetgrpstart
\figsetgrpnum{6.46}
\figsetgrptitle{Figure of 2244+2043}
\figsetplot{f6_46.pdf}
\figsetgrpnote{Posterior distributions of mean effective radius, effective variance, tolerance, column density, and offset showing 1-D distributions for each parameter and 2-D distributions for each combination of parameters. Dashed lines in the 1-D distributions represent 16, 50, 84 percent quantiles, corresponding to the median and 1 $\sigma$ uncertainties.}
\figsetgrpend

\figsetend

\figsetstart
\figsetnum{7}
\figsettitle{Model fits}

\figsetgrpstart
\figsetgrpnum{7.1}
\figsetgrptitle{Figure of 0141-4633}
\figsetplot{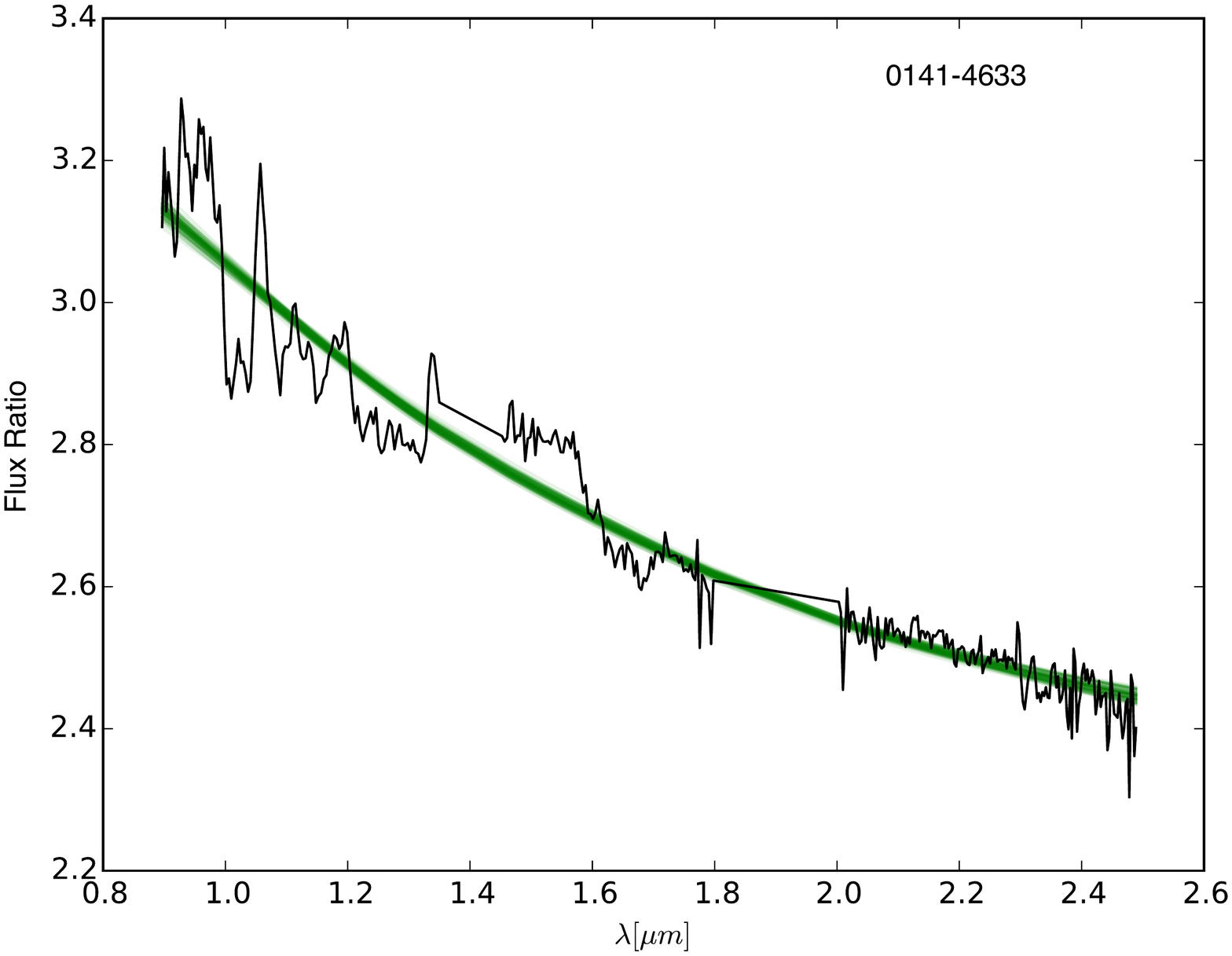}
\figsetgrpnote{Green lines are 100 models randomly drawn from the posterior distribution overplotted on the observed reddening in black. The telluric bands (1.35--1.45 $\micron$ and 1.8--2.0 $\micron$) are removed. The models reproduce the overall shape of the observed reddening well.}
\figsetgrpend

\figsetgrpstart
\figsetgrpnum{7.2}
\figsetgrptitle{Figure of 0032-4405}
\figsetplot{f7_2.pdf}
\figsetgrpnote{Green lines are 100 models randomly drawn from the posterior distribution overplotted on the observed reddening in black. The telluric bands (1.35--1.45 $\micron$ and 1.8--2.0 $\micron$) are removed. The models reproduce the overall shape of the observed reddening well.}
\figsetgrpend

\figsetgrpstart
\figsetgrpnum{7.3}
\figsetgrptitle{Figure of 0210-3015}
\figsetplot{f7_3.pdf}
\figsetgrpnote{Green lines are 100 models randomly drawn from the posterior distribution overplotted on the observed reddening in black. The telluric bands (1.35--1.45 $\micron$ and 1.8--2.0 $\micron$) are removed. The models reproduce the overall shape of the observed reddening well.}
\figsetgrpend

\figsetgrpstart
\figsetgrpnum{7.4}
\figsetgrptitle{Figure of 0241-0326}
\figsetplot{f7_4.pdf}
\figsetgrpnote{Green lines are 100 models randomly drawn from the posterior distribution overplotted on the observed reddening in black. The telluric bands (1.35--1.45 $\micron$ and 1.8--2.0 $\micron$) are removed. The models reproduce the overall shape of the observed reddening well.}
\figsetgrpend

\figsetgrpstart
\figsetgrpnum{7.5}
\figsetgrptitle{Figure of 0323-4631}
\figsetplot{f7_5.pdf}
\figsetgrpnote{Green lines are 100 models randomly drawn from the posterior distribution overplotted on the observed reddening in black. The telluric bands (1.35--1.45 $\micron$ and 1.8--2.0 $\micron$) are removed. The models reproduce the overall shape of the observed reddening well.}
\figsetgrpend

\figsetgrpstart
\figsetgrpnum{7.6}
\figsetgrptitle{Figure of 2213-2136}
\figsetplot{f7_6.pdf}
\figsetgrpnote{Green lines are 100 models randomly drawn from the posterior distribution overplotted on the observed reddening in black. The telluric bands (1.35--1.45 $\micron$ and 1.8--2.0 $\micron$) are removed. The models reproduce the overall shape of the observed reddening well.}
\figsetgrpend

\figsetgrpstart
\figsetgrpnum{7.7}
\figsetgrptitle{Figure of 2315+0617}
\figsetplot{f7_7.pdf}
\figsetgrpnote{Green lines are 100 models randomly drawn from the posterior distribution overplotted on the observed reddening in black. The telluric bands (1.35--1.45 $\micron$ and 1.8--2.0 $\micron$) are removed. The models reproduce the overall shape of the observed reddening well.}
\figsetgrpend

\figsetgrpstart
\figsetgrpnum{7.8}
\figsetgrptitle{Figure of 1711+2326}
\figsetplot{f7_8.pdf}
\figsetgrpnote{Green lines are 100 models randomly drawn from the posterior distribution overplotted on the observed reddening in black. The telluric bands (1.35--1.45 $\micron$ and 1.8--2.0 $\micron$) are removed. The models reproduce the overall shape of the observed reddening well.}
\figsetgrpend

\figsetgrpstart
\figsetgrpnum{7.9}
\figsetgrptitle{Figure of 1552+2948}
\figsetplot{f7_9.pdf}
\figsetgrpnote{Green lines are 100 models randomly drawn from the posterior distribution overplotted on the observed reddening in black. The telluric bands (1.35--1.45 $\micron$ and 1.8--2.0 $\micron$) are removed. The models reproduce the overall shape of the observed reddening well.}
\figsetgrpend

\figsetgrpstart
\figsetgrpnum{7.10}
\figsetgrptitle{Figure of 0357-4417}
\figsetplot{f7_10.pdf}
\figsetgrpnote{Green lines are 100 models randomly drawn from the posterior distribution overplotted on the observed reddening in black. The telluric bands (1.35--1.45 $\micron$ and 1.8--2.0 $\micron$) are removed. The models reproduce the overall shape of the observed reddening well.}
\figsetgrpend

\figsetgrpstart
\figsetgrpnum{7.11}
\figsetgrptitle{Figure of 0117-3403}
\figsetplot{f7_11.pdf}
\figsetgrpnote{Green lines are 100 models randomly drawn from the posterior distribution overplotted on the observed reddening in black. The telluric bands (1.35--1.45 $\micron$ and 1.8--2.0 $\micron$) are removed. The models reproduce the overall shape of the observed reddening well.}
\figsetgrpend

\figsetgrpstart
\figsetgrpnum{7.12}
\figsetgrptitle{Figure of 0518-2756}
\figsetplot{f7_12.pdf}
\figsetgrpnote{Green lines are 100 models randomly drawn from the posterior distribution overplotted on the observed reddening in black. The telluric bands (1.35--1.45 $\micron$ and 1.8--2.0 $\micron$) are removed. The models reproduce the overall shape of the observed reddening well.}
\figsetgrpend

\figsetgrpstart
\figsetgrpnum{7.13}
\figsetgrptitle{Figure of 0055+0134}
\figsetplot{f7_13.pdf}
\figsetgrpnote{Green lines are 100 models randomly drawn from the posterior distribution overplotted on the observed reddening in black. The telluric bands (1.35--1.45 $\micron$ and 1.8--2.0 $\micron$) are removed. The models reproduce the overall shape of the observed reddening well.}
\figsetgrpend

\figsetgrpstart
\figsetgrpnum{7.14}
\figsetgrptitle{Figure of 0536-1920}
\figsetplot{f7_14.pdf}
\figsetgrpnote{Green lines are 100 models randomly drawn from the posterior distribution overplotted on the observed reddening in black. The telluric bands (1.35--1.45 $\micron$ and 1.8--2.0 $\micron$) are removed. The models reproduce the overall shape of the observed reddening well.}
\figsetgrpend

\figsetgrpstart
\figsetgrpnum{7.15}
\figsetgrptitle{Figure of 1551+0941}
\figsetplot{f7_15.pdf}
\figsetgrpnote{Green lines are 100 models randomly drawn from the posterior distribution overplotted on the observed reddening in black. The telluric bands (1.35--1.45 $\micron$ and 1.8--2.0 $\micron$) are removed. The models reproduce the overall shape of the observed reddening well.}
\figsetgrpend

\figsetgrpstart
\figsetgrpnum{7.16}
\figsetgrptitle{Figure of 1726+1538}
\figsetplot{f7_16.pdf}
\figsetgrpnote{Green lines are 100 models randomly drawn from the posterior distribution overplotted on the observed reddening in black. The telluric bands (1.35--1.45 $\micron$ and 1.8--2.0 $\micron$) are removed. The models reproduce the overall shape of the observed reddening well.}
\figsetgrpend

\figsetgrpstart
\figsetgrpnum{7.17}
\figsetgrptitle{Figure of 0501-0010}
\figsetplot{f7_17.pdf}
\figsetgrpnote{Green lines are 100 models randomly drawn from the posterior distribution overplotted on the observed reddening in black. The telluric bands (1.35--1.45 $\micron$ and 1.8--2.0 $\micron$) are removed. The models reproduce the overall shape of the observed reddening well.}
\figsetgrpend

\figsetgrpstart
\figsetgrpnum{7.18}
\figsetgrptitle{Figure of 2249+0044}
\figsetplot{f7_18.pdf}
\figsetgrpnote{Green lines are 100 models randomly drawn from the posterior distribution overplotted on the observed reddening in black. The telluric bands (1.35--1.45 $\micron$ and 1.8--2.0 $\micron$) are removed. The models reproduce the overall shape of the observed reddening well.}
\figsetgrpend

\figsetgrpstart
\figsetgrpnum{7.19}
\figsetgrptitle{Figure of 0512-2949}
\figsetplot{f7_19.pdf}
\figsetgrpnote{Green lines are 100 models randomly drawn from the posterior distribution overplotted on the observed reddening in black. The telluric bands (1.35--1.45 $\micron$ and 1.8--2.0 $\micron$) are removed. The models reproduce the overall shape of the observed reddening well.}
\figsetgrpend

\figsetgrpstart
\figsetgrpnum{7.20}
\figsetgrptitle{Figure of 0326-2102}
\figsetplot{f7_20.pdf}
\figsetgrpnote{Green lines are 100 models randomly drawn from the posterior distribution overplotted on the observed reddening in black. The telluric bands (1.35--1.45 $\micron$ and 1.8--2.0 $\micron$) are removed. The models reproduce the overall shape of the observed reddening well.}
\figsetgrpend

\figsetgrpstart
\figsetgrpnum{7.21}
\figsetgrptitle{Figure of 2154-1055}
\figsetplot{f7_21.pdf}
\figsetgrpnote{Green lines are 100 models randomly drawn from the posterior distribution overplotted on the observed reddening in black. The telluric bands (1.35--1.45 $\micron$ and 1.8--2.0 $\micron$) are removed. The models reproduce the overall shape of the observed reddening well.}
\figsetgrpend

\figsetgrpstart
\figsetgrpnum{7.22}
\figsetgrptitle{Figure of 0355+1133}
\figsetplot{f7_22.pdf}
\figsetgrpnote{Green lines are 100 models randomly drawn from the posterior distribution overplotted on the observed reddening in black. The telluric bands (1.35--1.45 $\micron$ and 1.8--2.0 $\micron$) are removed. The models reproduce the overall shape of the observed reddening well.}
\figsetgrpend

\figsetgrpstart
\figsetgrpnum{7.23}
\figsetgrptitle{Figure of 1615+4953}
\figsetplot{f7_23.pdf}
\figsetgrpnote{Posterior distributions of mean effective radius, effective variance, tolerance, column density, and offset showing 1-D distributions for each parameter and 2-D distributions for each combination of parameters. Dashed lines in the 1-D distributions represent 16, 7.0, 84 percent quantiles, corresponding to the median and 1 $\sigma$ uncertainties.}
\figsetgrpend

\figsetgrpstart
\figsetgrpnum{7.24}
\figsetgrptitle{Figure of 0235-2331}
\figsetplot{f7_24.pdf}
\figsetgrpnote{Green lines are 100 models randomly drawn from the posterior distribution overplotted on the observed reddening in black. The telluric bands (1.35--1.45 $\micron$ and 1.8--2.0 $\micron$) are removed. The models reproduce the overall shape of the observed reddening well.}
\figsetgrpend

\figsetgrpstart
\figsetgrpnum{7.25}
\figsetgrptitle{Figure of 0543+6422}
\figsetplot{f7_25.pdf}
\figsetgrpnote{Green lines are 100 models randomly drawn from the posterior distribution overplotted on the observed reddening in black. The telluric bands (1.35--1.45 $\micron$ and 1.8--2.0 $\micron$) are removed. The models reproduce the overall shape of the observed reddening well.}
\figsetgrpend

\figsetgrpstart
\figsetgrpnum{7.26}
\figsetgrptitle{Figure of 0602+6336}
\figsetplot{f7_26.pdf}
\figsetgrpnote{Green lines are 100 models randomly drawn from the posterior distribution overplotted on the observed reddening in black. The telluric bands (1.35--1.45 $\micron$ and 1.8--2.0 $\micron$) are removed. The models reproduce the overall shape of the observed reddening well.}
\figsetgrpend

\figsetgrpstart
\figsetgrpnum{7.27}
\figsetgrptitle{Figure of 0016-4056}
\figsetplot{f7_27.pdf}
\figsetgrpnote{Green lines are 100 models randomly drawn from the posterior distribution overplotted on the observed reddening in black. The telluric bands (1.35--1.45 $\micron$ and 1.8--2.0 $\micron$) are removed. The models reproduce the overall shape of the observed reddening well.}
\figsetgrpend

\figsetgrpstart
\figsetgrpnum{7.28}
\figsetgrptitle{Figure of 2339+3507}
\figsetplot{f7_28.pdf}
\figsetgrpnote{Green lines are 100 models randomly drawn from the posterior distribution overplotted on the observed reddening in black. The telluric bands (1.35--1.45 $\micron$ and 1.8--2.0 $\micron$) are removed. The models reproduce the overall shape of the observed reddening well.}
\figsetgrpend

\figsetgrpstart
\figsetgrpnum{7.29}
\figsetgrptitle{Figure of 1100+4957}
\figsetplot{f7_29.pdf}
\figsetgrpnote{Green lines are 100 models randomly drawn from the posterior distribution overplotted on the observed reddening in black. The telluric bands (1.35--1.45 $\micron$ and 1.8--2.0 $\micron$) are removed. The models reproduce the overall shape of the observed reddening well.}
\figsetgrpend

\figsetgrpstart
\figsetgrpnum{7.30}
\figsetgrptitle{Figure of 0051-1544}
\figsetplot{f7_30.pdf}
\figsetgrpnote{Green lines are 100 models randomly drawn from the posterior distribution overplotted on the observed reddening in black. The telluric bands (1.35--1.45 $\micron$ and 1.8--2.0 $\micron$) are removed. The models reproduce the overall shape of the observed reddening well.}
\figsetgrpend

\figsetgrpstart
\figsetgrpnum{7.31}
\figsetgrptitle{Figure of 2317-4838}
\figsetplot{f7_31.pdf}
\figsetgrpnote{Green lines are 100 models randomly drawn from the posterior distribution overplotted on the observed reddening in black. The telluric bands (1.35--1.45 $\micron$ and 1.8--2.0 $\micron$) are removed. The models reproduce the overall shape of the observed reddening well.}
\figsetgrpend

\figsetgrpstart
\figsetgrpnum{7.32}
\figsetgrptitle{Figure of 0337-1758}
\figsetplot{f7_32.pdf}
\figsetgrpnote{Green lines are 100 models randomly drawn from the posterior distribution overplotted on the observed reddening in black. The telluric bands (1.35--1.45 $\micron$ and 1.8--2.0 $\micron$) are removed. The models reproduce the overall shape of the observed reddening well.}
\figsetgrpend

\figsetgrpstart
\figsetgrpnum{7.33}
\figsetgrptitle{Figure of 0208+2737}
\figsetplot{f7_33.pdf}
\figsetgrpnote{Green lines are 100 models randomly drawn from the posterior distribution overplotted on the observed reddening in black. The telluric bands (1.35--1.45 $\micron$ and 1.8--2.0 $\micron$) are removed. The models reproduce the overall shape of the observed reddening well.}
\figsetgrpend

\figsetgrpstart
\figsetgrpnum{7.34}
\figsetgrptitle{Figure of 0835-0819}
\figsetplot{f7_34.pdf}
\figsetgrpnote{Green lines are 100 models randomly drawn from the posterior distribution overplotted on the observed reddening in black. The telluric bands (1.35--1.45 $\micron$ and 1.8--2.0 $\micron$) are removed. The models reproduce the overall shape of the observed reddening well.}
\figsetgrpend

\figsetgrpstart
\figsetgrpnum{7.35}
\figsetgrptitle{Figure of 0358-4116}
\figsetplot{f7_357_.pdf}
\figsetgrpnote{Green lines are 100 models randomly drawn from the posterior distribution overplotted on the observed reddening in black. The telluric bands (1.35--1.45 $\micron$ and 1.8--2.0 $\micron$) are removed. The models reproduce the overall shape of the observed reddening well.}
\figsetgrpend

\figsetgrpstart
\figsetgrpnum{7.36}
\figsetgrptitle{Figure of 0905+5623}
\figsetplot{f7_36.pdf}
\figsetgrpnote{Green lines are 100 models randomly drawn from the posterior distribution overplotted on the observed reddening in black. The telluric bands (1.35--1.45 $\micron$ and 1.8--2.0 $\micron$) are removed. The models reproduce the overall shape of the observed reddening well.}
\figsetgrpend

\figsetgrpstart
\figsetgrpnum{7.37}
\figsetgrptitle{Figure of 1228-1547}
\figsetplot{f7_37.pdf}
\figsetgrpnote{Green lines are 100 models randomly drawn from the posterior distribution overplotted on the observed reddening in black. The telluric bands (1.35--1.45 $\micron$ and 1.8--2.0 $\micron$) are removed. The models reproduce the overall shape of the observed reddening well.}
\figsetgrpend

\figsetgrpstart
\figsetgrpnum{7.38}
\figsetgrptitle{Figure of 1239+5515}
\figsetplot{f7_38.pdf}
\figsetgrpnote{Green lines are 100 models randomly drawn from the posterior distribution overplotted on the observed reddening in black. The telluric bands (1.35--1.45 $\micron$ and 1.8--2.0 $\micron$) are removed. The models reproduce the overall shape of the observed reddening well.}
\figsetgrpend

\figsetgrpstart
\figsetgrpnum{7.39}
\figsetgrptitle{Figure of 0310-2756}
\figsetplot{f7_39.pdf}
\figsetgrpnote{Green lines are 100 models randomly drawn from the posterior distribution overplotted on the observed reddening in black. The telluric bands (1.35--1.45 $\micron$ and 1.8--2.0 $\micron$) are removed. The models reproduce the overall shape of the observed reddening well.}
\figsetgrpend

\figsetgrpstart
\figsetgrpnum{7.40}
\figsetgrptitle{Figure of 0624-4521}
\figsetplot{f7_40.pdf}
\figsetgrpnote{Green lines are 100 models randomly drawn from the posterior distribution overplotted on the observed reddening in black. The telluric bands (1.35--1.45 $\micron$ and 1.8--2.0 $\micron$) are removed. The models reproduce the overall shape of the observed reddening well.}
\figsetgrpend

\figsetgrpstart
\figsetgrpnum{7.41}
\figsetgrptitle{Figure of 0652+4710}
\figsetplot{f7_41.pdf}
\figsetgrpnote{Green lines are 100 models randomly drawn from the posterior distribution overplotted on the observed reddening in black. The telluric bands (1.35--1.45 $\micron$ and 1.8--2.0 $\micron$) are removed. The models reproduce the overall shape of the observed reddening well.}
\figsetgrpend

\figsetgrpstart
\figsetgrpnum{7.42}
\figsetgrptitle{Figure of 1438+5722}
\figsetplot{f7_42.pdf}
\figsetgrpnote{Green lines are 100 models randomly drawn from the posterior distribution overplotted on the observed reddening in black. The telluric bands (1.35--1.45 $\micron$ and 1.8--2.0 $\micron$) are removed. The models reproduce the overall shape of the observed reddening well.}
\figsetgrpend

\figsetgrpstart
\figsetgrpnum{7.43}
\figsetgrptitle{Figure of 1326-0038}
\figsetplot{f7_43.pdf}
\figsetgrpnote{Green lines are 100 models randomly drawn from the posterior distribution overplotted on the observed reddening in black. The telluric bands (1.35--1.45 $\micron$ and 1.8--2.0 $\micron$) are removed. The models reproduce the overall shape of the observed reddening well.}
\figsetgrpend

\figsetgrpstart
\figsetgrpnum{7.44}
\figsetgrptitle{Figure of 2212+3430}
\figsetplot{f7_44.pdf}
\figsetgrpnote{Green lines are 100 models randomly drawn from the posterior distribution overplotted on the observed reddening in black. The telluric bands (1.35--1.45 $\micron$ and 1.8--2.0 $\micron$) are removed. The models reproduce the overall shape of the observed reddening well.}
\figsetgrpend

\figsetgrpstart
\figsetgrpnum{7.45}
\figsetgrptitle{Figure of 2148+4003}
\figsetplot{f7_45.pdf}
\figsetgrpnote{Green lines are 100 models randomly drawn from the posterior distribution overplotted on the observed reddening in black. The telluric bands (1.35--1.45 $\micron$ and 1.8--2.0 $\micron$) are removed. The models reproduce the overall shape of the observed reddening well.}
\figsetgrpend

\figsetgrpstart
\figsetgrpnum{7.46}
\figsetgrptitle{Figure of 2244+2043}
\figsetplot{f7_46.pdf}
\figsetgrpnote{Green lines are 100 models randomly drawn from the posterior distribution overplotted on the observed reddening in black. The telluric bands (1.35--1.45 $\micron$ and 1.8--2.0 $\micron$) are removed. The models reproduce the overall shape of the observed reddening well.}
\figsetgrpend

\figsetend

\figsetstart
\figsetnum{8}
\figsettitle{De-reddened spectra}

\figsetgrpstart
\figsetgrpnum{8.1}
\figsetgrptitle{Figure of 0141-4633}
\figsetplot{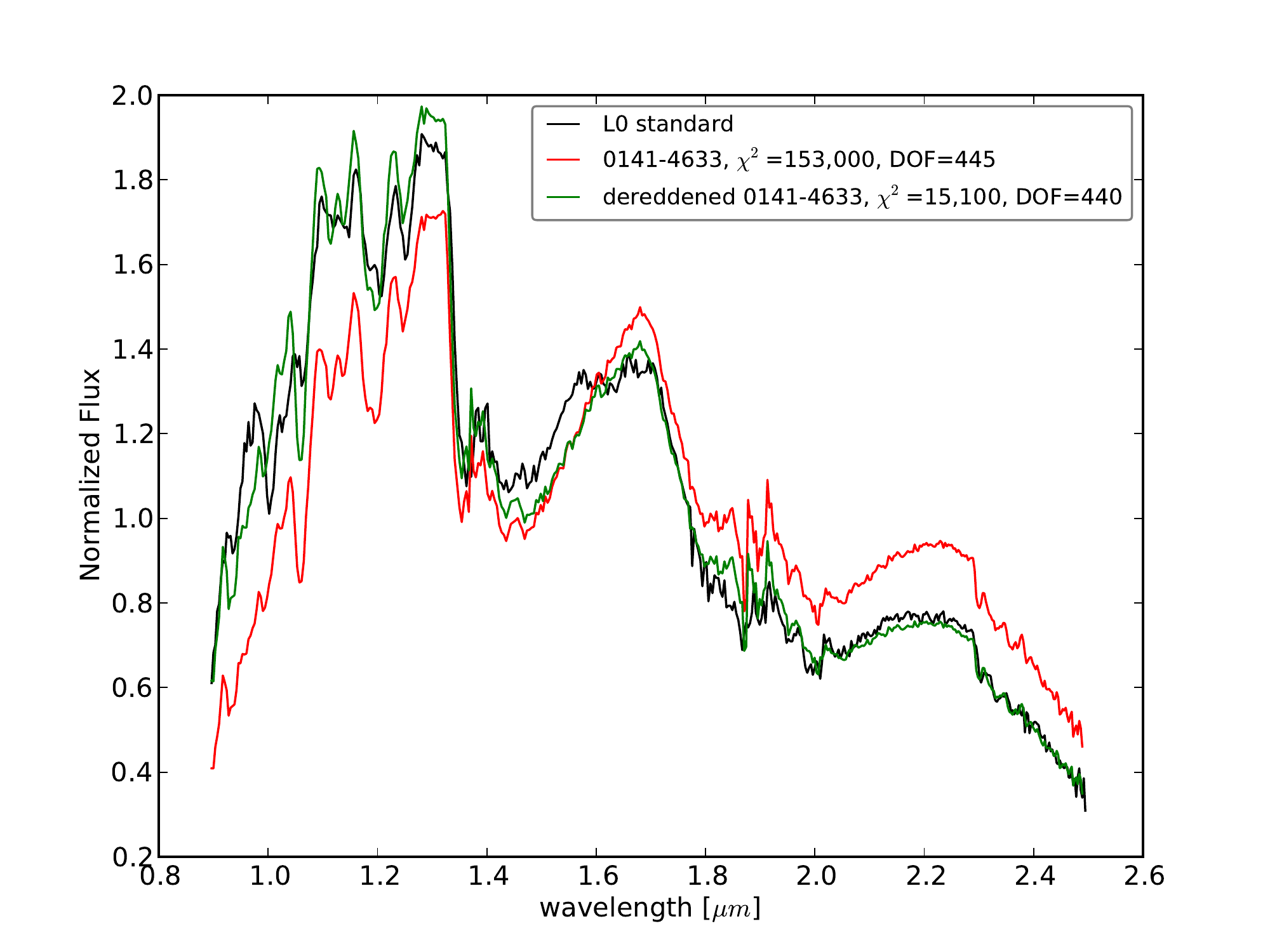}
\figsetgrpnote{The spectrum of a red L dwarf de-reddened by the dust haze prescription is compared to the spectra of the field standard object. Black is the standard, red is the red object, and green is the spectrum of the red object de-reddened by the forsterite extinction curve for the best fit parameters from MCMC fits. $\chi^2$ values before and after the de-reddening are also shown. The de-reddened spectrum fits the standard spectrum much better and has a smaller $\chi^2$ value than the original red spectrum. This shows that the proposed dust haze prescription successfully corrects the red NIR slopes of red L dwarfs to look like the standard object.}
\figsetgrpend

\figsetgrpstart
\figsetgrpnum{8.2}
\figsetgrptitle{Figure of 0032-4405}
\figsetplot{f8_2.pdf}
\figsetgrpnote{The spectrum of a red L dwarf de-reddened by the dust haze prescription is compared to the spectra of the field standard object. Black is the standard, red is the red object, and green is the spectrum of the red object de-reddened by the forsterite extinction curve for the best fit parameters from MCMC fits. $\chi^2$ values before and after the de-reddening are also shown. The de-reddened spectrum fits the standard spectrum much better and has a smaller $\chi^2$ value than the original red spectrum. This shows that the proposed dust haze prescription successfully corrects the red NIR slopes of red L dwarfs to look like the standard object.}
\figsetgrpend

\figsetgrpstart
\figsetgrpnum{8.3}
\figsetgrptitle{Figure of 0210-3015}
\figsetplot{f8_3.pdf}
\figsetgrpnote{The spectrum of a red L dwarf de-reddened by the dust haze prescription is compared to the spectra of the field standard object. Black is the standard, red is the red object, and green is the spectrum of the red object de-reddened by the forsterite extinction curve for the best fit parameters from MCMC fits. $\chi^2$ values before and after the de-reddening are also shown. The de-reddened spectrum fits the standard spectrum much better and has a smaller $\chi^2$ value than the original red spectrum. This shows that the proposed dust haze prescription successfully corrects the red NIR slopes of red L dwarfs to look like the standard object.}
\figsetgrpend

\figsetgrpstart
\figsetgrpnum{8.4}
\figsetgrptitle{Figure of 0241-0326}
\figsetplot{f8_4.pdf}
\figsetgrpnote{The spectrum of a red L dwarf de-reddened by the dust haze prescription is compared to the spectra of the field standard object. Black is the standard, red is the red object, and green is the spectrum of the red object de-reddened by the forsterite extinction curve for the best fit parameters from MCMC fits. $\chi^2$ values before and after the de-reddening are also shown. The de-reddened spectrum fits the standard spectrum much better and has a smaller $\chi^2$ value than the original red spectrum. This shows that the proposed dust haze prescription successfully corrects the red NIR slopes of red L dwarfs to look like the standard object.}
\figsetgrpend

\figsetgrpstart
\figsetgrpnum{8.5}
\figsetgrptitle{Figure of 0323-4631}
\figsetplot{f8_5.pdf}
\figsetgrpnote{The spectrum of a red L dwarf de-reddened by the dust haze prescription is compared to the spectra of the field standard object. Black is the standard, red is the red object, and green is the spectrum of the red object de-reddened by the forsterite extinction curve for the best fit parameters from MCMC fits. $\chi^2$ values before and after the de-reddening are also shown. The de-reddened spectrum fits the standard spectrum much better and has a smaller $\chi^2$ value than the original red spectrum. This shows that the proposed dust haze prescription successfully corrects the red NIR slopes of red L dwarfs to look like the standard object.}
\figsetgrpend

\figsetgrpstart
\figsetgrpnum{8.6}
\figsetgrptitle{Figure of 2213-2136}
\figsetplot{f8_6.pdf}
\figsetgrpnote{The spectrum of a red L dwarf de-reddened by the dust haze prescription is compared to the spectra of the field standard object. Black is the standard, red is the red object, and green is the spectrum of the red object de-reddened by the forsterite extinction curve for the best fit parameters from MCMC fits. $\chi^2$ values before and after the de-reddening are also shown. The de-reddened spectrum fits the standard spectrum much better and has a smaller $\chi^2$ value than the original red spectrum. This shows that the proposed dust haze prescription successfully corrects the red NIR slopes of red L dwarfs to look like the standard object.}
\figsetgrpend

\figsetgrpstart
\figsetgrpnum{8.7}
\figsetgrptitle{Figure of 2315+0617}
\figsetplot{f8_7.pdf}
\figsetgrpnote{The spectrum of a red L dwarf de-reddened by the dust haze prescription is compared to the spectra of the field standard object. Black is the standard, red is the red object, and green is the spectrum of the red object de-reddened by the forsterite extinction curve for the best fit parameters from MCMC fits. $\chi^2$ values before and after the de-reddening are also shown. The de-reddened spectrum fits the standard spectrum much better and has a smaller $\chi^2$ value than the original red spectrum. This shows that the proposed dust haze prescription successfully corrects the red NIR slopes of red L dwarfs to look like the standard object.}
\figsetgrpend

\figsetgrpstart
\figsetgrpnum{8.8}
\figsetgrptitle{Figure of 1711+2326}
\figsetplot{f8_8.pdf}
\figsetgrpnote{The spectrum of a red L dwarf de-reddened by the dust haze prescription is compared to the spectra of the field standard object. Black is the standard, red is the red object, and green is the spectrum of the red object de-reddened by the forsterite extinction curve for the best fit parameters from MCMC fits. $\chi^2$ values before and after the de-reddening are also shown. The de-reddened spectrum fits the standard spectrum much better and has a smaller $\chi^2$ value than the original red spectrum. This shows that the proposed dust haze prescription successfully corrects the red NIR slopes of red L dwarfs to look like the standard object.}
\figsetgrpend

\figsetgrpstart
\figsetgrpnum{8.9}
\figsetgrptitle{Figure of 1552+2948}
\figsetplot{f8_9.pdf}
\figsetgrpnote{The spectrum of a red L dwarf de-reddened by the dust haze prescription is compared to the spectra of the field standard object. Black is the standard, red is the red object, and green is the spectrum of the red object de-reddened by the forsterite extinction curve for the best fit parameters from MCMC fits. $\chi^2$ values before and after the de-reddening are also shown. The de-reddened spectrum fits the standard spectrum much better and has a smaller $\chi^2$ value than the original red spectrum. This shows that the proposed dust haze prescription successfully corrects the red NIR slopes of red L dwarfs to look like the standard object.}
\figsetgrpend

\figsetgrpstart
\figsetgrpnum{8.10}
\figsetgrptitle{Figure of 0357-4417}
\figsetplot{f8_10.pdf}
\figsetgrpnote{The spectrum of a red L dwarf de-reddened by the dust haze prescription is compared to the spectra of the field standard object. Black is the standard, red is the red object, and green is the spectrum of the red object de-reddened by the forsterite extinction curve for the best fit parameters from MCMC fits. $\chi^2$ values before and after the de-reddening are also shown. The de-reddened spectrum fits the standard spectrum much better and has a smaller $\chi^2$ value than the original red spectrum. This shows that the proposed dust haze prescription successfully corrects the red NIR slopes of red L dwarfs to look like the standard object.}
\figsetgrpend

\figsetgrpstart
\figsetgrpnum{8.11}
\figsetgrptitle{Figure of 0117-3403}
\figsetplot{f8_11.pdf}
\figsetgrpnote{The spectrum of a red L dwarf de-reddened by the dust haze prescription is compared to the spectra of the field standard object. Black is the standard, red is the red object, and green is the spectrum of the red object de-reddened by the forsterite extinction curve for the best fit parameters from MCMC fits. $\chi^2$ values before and after the de-reddening are also shown. The de-reddened spectrum fits the standard spectrum much better and has a smaller $\chi^2$ value than the original red spectrum. This shows that the proposed dust haze prescription successfully corrects the red NIR slopes of red L dwarfs to look like the standard object.}
\figsetgrpend

\figsetgrpstart
\figsetgrpnum{8.12}
\figsetgrptitle{Figure of 0518-2756}
\figsetplot{f8_12.pdf}
\figsetgrpnote{The spectrum of a red L dwarf de-reddened by the dust haze prescription is compared to the spectra of the field standard object. Black is the standard, red is the red object, and green is the spectrum of the red object de-reddened by the forsterite extinction curve for the best fit parameters from MCMC fits. $\chi^2$ values before and after the de-reddening are also shown. The de-reddened spectrum fits the standard spectrum much better and has a smaller $\chi^2$ value than the original red spectrum. This shows that the proposed dust haze prescription successfully corrects the red NIR slopes of red L dwarfs to look like the standard object.}
\figsetgrpend

\figsetgrpstart
\figsetgrpnum{8.13}
\figsetgrptitle{Figure of 0055+0134}
\figsetplot{f8_13.pdf}
\figsetgrpnote{The spectrum of a red L dwarf de-reddened by the dust haze prescription is compared to the spectra of the field standard object. Black is the standard, red is the red object, and green is the spectrum of the red object de-reddened by the forsterite extinction curve for the best fit parameters from MCMC fits. $\chi^2$ values before and after the de-reddening are also shown. The de-reddened spectrum fits the standard spectrum much better and has a smaller $\chi^2$ value than the original red spectrum. This shows that the proposed dust haze prescription successfully corrects the red NIR slopes of red L dwarfs to look like the standard object.}
\figsetgrpend

\figsetgrpstart
\figsetgrpnum{8.14}
\figsetgrptitle{Figure of 0536-1920}
\figsetplot{f8_14.pdf}
\figsetgrpnote{The spectrum of a red L dwarf de-reddened by the dust haze prescription is compared to the spectra of the field standard object. Black is the standard, red is the red object, and green is the spectrum of the red object de-reddened by the forsterite extinction curve for the best fit parameters from MCMC fits. $\chi^2$ values before and after the de-reddening are also shown. The de-reddened spectrum fits the standard spectrum much better and has a smaller $\chi^2$ value than the original red spectrum. This shows that the proposed dust haze prescription successfully corrects the red NIR slopes of red L dwarfs to look like the standard object.}
\figsetgrpend

\figsetgrpstart
\figsetgrpnum{8.15}
\figsetgrptitle{Figure of 1551+0941}
\figsetplot{f8_15.pdf}
\figsetgrpnote{The spectrum of a red L dwarf de-reddened by the dust haze prescription is compared to the spectra of the field standard object. Black is the standard, red is the red object, and green is the spectrum of the red object de-reddened by the forsterite extinction curve for the best fit parameters from MCMC fits. $\chi^2$ values before and after the de-reddening are also shown. The de-reddened spectrum fits the standard spectrum much better and has a smaller $\chi^2$ value than the original red spectrum. This shows that the proposed dust haze prescription successfully corrects the red NIR slopes of red L dwarfs to look like the standard object.}
\figsetgrpend

\figsetgrpstart
\figsetgrpnum{8.16}
\figsetgrptitle{Figure of 1726+1538}
\figsetplot{f8_16.pdf}
\figsetgrpnote{The spectrum of a red L dwarf de-reddened by the dust haze prescription is compared to the spectra of the field standard object. Black is the standard, red is the red object, and green is the spectrum of the red object de-reddened by the forsterite extinction curve for the best fit parameters from MCMC fits. $\chi^2$ values before and after the de-reddening are also shown. The de-reddened spectrum fits the standard spectrum much better and has a smaller $\chi^2$ value than the original red spectrum. This shows that the proposed dust haze prescription successfully corrects the red NIR slopes of red L dwarfs to look like the standard object.}
\figsetgrpend

\figsetgrpstart
\figsetgrpnum{8.17}
\figsetgrptitle{Figure of 0501-0010}
\figsetplot{f8_17.pdf}
\figsetgrpnote{The spectrum of a red L dwarf de-reddened by the dust haze prescription is compared to the spectra of the field standard object. Black is the standard, red is the red object, and green is the spectrum of the red object de-reddened by the forsterite extinction curve for the best fit parameters from MCMC fits. $\chi^2$ values before and after the de-reddening are also shown. The de-reddened spectrum fits the standard spectrum much better and has a smaller $\chi^2$ value than the original red spectrum. This shows that the proposed dust haze prescription successfully corrects the red NIR slopes of red L dwarfs to look like the standard object.}
\figsetgrpend

\figsetgrpstart
\figsetgrpnum{8.18}
\figsetgrptitle{Figure of 2249+0044}
\figsetplot{f8_18.pdf}
\figsetgrpnote{The spectrum of a red L dwarf de-reddened by the dust haze prescription is compared to the spectra of the field standard object. Black is the standard, red is the red object, and green is the spectrum of the red object de-reddened by the forsterite extinction curve for the best fit parameters from MCMC fits. $\chi^2$ values before and after the de-reddening are also shown. The de-reddened spectrum fits the standard spectrum much better and has a smaller $\chi^2$ value than the original red spectrum. This shows that the proposed dust haze prescription successfully corrects the red NIR slopes of red L dwarfs to look like the standard object.}
\figsetgrpend

\figsetgrpstart
\figsetgrpnum{8.19}
\figsetgrptitle{Figure of 0512-2949}
\figsetplot{f8_19.pdf}
\figsetgrpnote{The spectrum of a red L dwarf de-reddened by the dust haze prescription is compared to the spectra of the field standard object. Black is the standard, red is the red object, and green is the spectrum of the red object de-reddened by the forsterite extinction curve for the best fit parameters from MCMC fits. $\chi^2$ values before and after the de-reddening are also shown. The de-reddened spectrum fits the standard spectrum much better and has a smaller $\chi^2$ value than the original red spectrum. This shows that the proposed dust haze prescription successfully corrects the red NIR slopes of red L dwarfs to look like the standard object.}
\figsetgrpend

\figsetgrpstart
\figsetgrpnum{8.20}
\figsetgrptitle{Figure of 0326-2102}
\figsetplot{f8_20.pdf}
\figsetgrpnote{The spectrum of a red L dwarf de-reddened by the dust haze prescription is compared to the spectra of the field standard object. Black is the standard, red is the red object, and green is the spectrum of the red object de-reddened by the forsterite extinction curve for the best fit parameters from MCMC fits. $\chi^2$ values before and after the de-reddening are also shown. The de-reddened spectrum fits the standard spectrum much better and has a smaller $\chi^2$ value than the original red spectrum. This shows that the proposed dust haze prescription successfully corrects the red NIR slopes of red L dwarfs to look like the standard object.}
\figsetgrpend

\figsetgrpstart
\figsetgrpnum{8.21}
\figsetgrptitle{Figure of 2154-1055}
\figsetplot{f8_21.pdf}
\figsetgrpnote{The spectrum of a red L dwarf de-reddened by the dust haze prescription is compared to the spectra of the field standard object. Black is the standard, red is the red object, and green is the spectrum of the red object de-reddened by the forsterite extinction curve for the best fit parameters from MCMC fits. $\chi^2$ values before and after the de-reddening are also shown. The de-reddened spectrum fits the standard spectrum much better and has a smaller $\chi^2$ value than the original red spectrum. This shows that the proposed dust haze prescription successfully corrects the red NIR slopes of red L dwarfs to look like the standard object.}
\figsetgrpend

\figsetgrpstart
\figsetgrpnum{8.22}
\figsetgrptitle{Figure of 0355+1133}
\figsetplot{f8_22.pdf}
\figsetgrpnote{The spectrum of a red L dwarf de-reddened by the dust haze prescription is compared to the spectra of the field standard object. Black is the standard, red is the red object, and green is the spectrum of the red object de-reddened by the forsterite extinction curve for the best fit parameters from MCMC fits. $\chi^2$ values before and after the de-reddening are also shown. The de-reddened spectrum fits the standard spectrum much better and has a smaller $\chi^2$ value than the original red spectrum. This shows that the proposed dust haze prescription successfully corrects the red NIR slopes of red L dwarfs to look like the standard object.}
\figsetgrpend

\figsetgrpstart
\figsetgrpnum{8.23}
\figsetgrptitle{Figure of 1615+4953}
\figsetplot{f8_23.pdf}
\figsetgrpnote{The spectrum of a red L dwarf de-reddened by the dust haze prescription is compared to the spectra of the field standard object. Black is the standard, red is the red object, and green is the spectrum of the red object de-reddened by the forsterite extinction curve for the best fit parameters from MCMC fits. $\chi^2$ values before and after the de-reddening are also shown. The de-reddened spectrum fits the standard spectrum much better and has a smaller $\chi^2$ value than the original red spectrum. This shows that the proposed dust haze prescription successfully corrects the red NIR slopes of red L dwarfs to look like the standard object.}
\figsetgrpend

\figsetgrpstart
\figsetgrpnum{8.24}
\figsetgrptitle{Figure of 0235-2331}
\figsetplot{f8_24.pdf}
\figsetgrpnote{The spectrum of a red L dwarf de-reddened by the dust haze prescription is compared to the spectra of the field standard object. Black is the standard, red is the red object, and green is the spectrum of the red object de-reddened by the forsterite extinction curve for the best fit parameters from MCMC fits. $\chi^2$ values before and after the de-reddening are also shown. The de-reddened spectrum fits the standard spectrum much better and has a smaller $\chi^2$ value than the original red spectrum. This shows that the proposed dust haze prescription successfully corrects the red NIR slopes of red L dwarfs to look like the standard object.}
\figsetgrpend

\figsetgrpstart
\figsetgrpnum{8.25}
\figsetgrptitle{Figure of 0543+6422}
\figsetplot{f8_25.pdf}
\figsetgrpnote{The spectrum of a red L dwarf de-reddened by the dust haze prescription is compared to the spectra of the field standard object. Black is the standard, red is the red object, and green is the spectrum of the red object de-reddened by the forsterite extinction curve for the best fit parameters from MCMC fits. $\chi^2$ values before and after the de-reddening are also shown. The de-reddened spectrum fits the standard spectrum much better and has a smaller $\chi^2$ value than the original red spectrum. This shows that the proposed dust haze prescription successfully corrects the red NIR slopes of red L dwarfs to look like the standard object.}
\figsetgrpend

\figsetgrpstart
\figsetgrpnum{8.26}
\figsetgrptitle{Figure of 0602+6336}
\figsetplot{f8_26.pdf}
\figsetgrpnote{The spectrum of a red L dwarf de-reddened by the dust haze prescription is compared to the spectra of the field standard object. Black is the standard, red is the red object, and green is the spectrum of the red object de-reddened by the forsterite extinction curve for the best fit parameters from MCMC fits. $\chi^2$ values before and after the de-reddening are also shown. The de-reddened spectrum fits the standard spectrum much better and has a smaller $\chi^2$ value than the original red spectrum. This shows that the proposed dust haze prescription successfully corrects the red NIR slopes of red L dwarfs to look like the standard object.}
\figsetgrpend

\figsetgrpstart
\figsetgrpnum{8.27}
\figsetgrptitle{Figure of 0016-4056}
\figsetplot{f8_27.pdf}
\figsetgrpnote{The spectrum of a red L dwarf de-reddened by the dust haze prescription is compared to the spectra of the field standard object. Black is the standard, red is the red object, and green is the spectrum of the red object de-reddened by the forsterite extinction curve for the best fit parameters from MCMC fits. $\chi^2$ values before and after the de-reddening are also shown. The de-reddened spectrum fits the standard spectrum much better and has a smaller $\chi^2$ value than the original red spectrum. This shows that the proposed dust haze prescription successfully corrects the red NIR slopes of red L dwarfs to look like the standard object.}
\figsetgrpend

\figsetgrpstart
\figsetgrpnum{8.28}
\figsetgrptitle{Figure of 2339+3507}
\figsetplot{f8_28.pdf}
\figsetgrpnote{The spectrum of a red L dwarf de-reddened by the dust haze prescription is compared to the spectra of the field standard object. Black is the standard, red is the red object, and green is the spectrum of the red object de-reddened by the forsterite extinction curve for the best fit parameters from MCMC fits. $\chi^2$ values before and after the de-reddening are also shown. The de-reddened spectrum fits the standard spectrum much better and has a smaller $\chi^2$ value than the original red spectrum. This shows that the proposed dust haze prescription successfully corrects the red NIR slopes of red L dwarfs to look like the standard object.}
\figsetgrpend

\figsetgrpstart
\figsetgrpnum{8.29}
\figsetgrptitle{Figure of 1100+4957}
\figsetplot{f8_29.pdf}
\figsetgrpnote{The spectrum of a red L dwarf de-reddened by the dust haze prescription is compared to the spectra of the field standard object. Black is the standard, red is the red object, and green is the spectrum of the red object de-reddened by the forsterite extinction curve for the best fit parameters from MCMC fits. $\chi^2$ values before and after the de-reddening are also shown. The de-reddened spectrum fits the standard spectrum much better and has a smaller $\chi^2$ value than the original red spectrum. This shows that the proposed dust haze prescription successfully corrects the red NIR slopes of red L dwarfs to look like the standard object.}
\figsetgrpend

\figsetgrpstart
\figsetgrpnum{8.30}
\figsetgrptitle{Figure of 0051-1544}
\figsetplot{f8_30.pdf}
\figsetgrpnote{The spectrum of a red L dwarf de-reddened by the dust haze prescription is compared to the spectra of the field standard object. Black is the standard, red is the red object, and green is the spectrum of the red object de-reddened by the forsterite extinction curve for the best fit parameters from MCMC fits. $\chi^2$ values before and after the de-reddening are also shown. The de-reddened spectrum fits the standard spectrum much better and has a smaller $\chi^2$ value than the original red spectrum. This shows that the proposed dust haze prescription successfully corrects the red NIR slopes of red L dwarfs to look like the standard object.}
\figsetgrpend

\figsetgrpstart
\figsetgrpnum{8.31}
\figsetgrptitle{Figure of 2317-4838}
\figsetplot{f8_31.pdf}
\figsetgrpnote{The spectrum of a red L dwarf de-reddened by the dust haze prescription is compared to the spectra of the field standard object. Black is the standard, red is the red object, and green is the spectrum of the red object de-reddened by the forsterite extinction curve for the best fit parameters from MCMC fits. $\chi^2$ values before and after the de-reddening are also shown. The de-reddened spectrum fits the standard spectrum much better and has a smaller $\chi^2$ value than the original red spectrum. This shows that the proposed dust haze prescription successfully corrects the red NIR slopes of red L dwarfs to look like the standard object.}
\figsetgrpend

\figsetgrpstart
\figsetgrpnum{8.32}
\figsetgrptitle{Figure of 0337-1758}
\figsetplot{f8_32.pdf}
\figsetgrpnote{The spectrum of a red L dwarf de-reddened by the dust haze prescription is compared to the spectra of the field standard object. Black is the standard, red is the red object, and green is the spectrum of the red object de-reddened by the forsterite extinction curve for the best fit parameters from MCMC fits. $\chi^2$ values before and after the de-reddening are also shown. The de-reddened spectrum fits the standard spectrum much better and has a smaller $\chi^2$ value than the original red spectrum. This shows that the proposed dust haze prescription successfully corrects the red NIR slopes of red L dwarfs to look like the standard object.}
\figsetgrpend

\figsetgrpstart
\figsetgrpnum{8.33}
\figsetgrptitle{Figure of 0208+2737}
\figsetplot{f8_33.pdf}
\figsetgrpnote{The spectrum of a red L dwarf de-reddened by the dust haze prescription is compared to the spectra of the field standard object. Black is the standard, red is the red object, and green is the spectrum of the red object de-reddened by the forsterite extinction curve for the best fit parameters from MCMC fits. $\chi^2$ values before and after the de-reddening are also shown. The de-reddened spectrum fits the standard spectrum much better and has a smaller $\chi^2$ value than the original red spectrum. This shows that the proposed dust haze prescription successfully corrects the red NIR slopes of red L dwarfs to look like the standard object.}
\figsetgrpend

\figsetgrpstart
\figsetgrpnum{8.34}
\figsetgrptitle{Figure of 0835-0819}
\figsetplot{f8_34.pdf}
\figsetgrpnote{The spectrum of a red L dwarf de-reddened by the dust haze prescription is compared to the spectra of the field standard object. Black is the standard, red is the red object, and green is the spectrum of the red object de-reddened by the forsterite extinction curve for the best fit parameters from MCMC fits. $\chi^2$ values before and after the de-reddening are also shown. The de-reddened spectrum fits the standard spectrum much better and has a smaller $\chi^2$ value than the original red spectrum. This shows that the proposed dust haze prescription successfully corrects the red NIR slopes of red L dwarfs to look like the standard object.}
\figsetgrpend

\figsetgrpstart
\figsetgrpnum{8.35}
\figsetgrptitle{Figure of 0358-4116}
\figsetplot{f8_35.pdf}
\figsetgrpnote{The spectrum of a red L dwarf de-reddened by the dust haze prescription is compared to the spectra of the field standard object. Black is the standard, red is the red object, and green is the spectrum of the red object de-reddened by the forsterite extinction curve for the best fit parameters from MCMC fits. $\chi^2$ values before and after the de-reddening are also shown. The de-reddened spectrum fits the standard spectrum much better and has a smaller $\chi^2$ value than the original red spectrum. This shows that the proposed dust haze prescription successfully corrects the red NIR slopes of red L dwarfs to look like the standard object.}
\figsetgrpend

\figsetgrpstart
\figsetgrpnum{8.36}
\figsetgrptitle{Figure of 0905+5623  }
\figsetplot{f8_36.pdf}
\figsetgrpnote{The spectrum of a red L dwarf de-reddened by the dust haze prescription is compared to the spectra of the field standard object. Black is the standard, red is the red object, and green is the spectrum of the red object de-reddened by the forsterite extinction curve for the best fit parameters from MCMC fits. $\chi^2$ values before and after the de-reddening are also shown. The de-reddened spectrum fits the standard spectrum much better and has a smaller $\chi^2$ value than the original red spectrum. This shows that the proposed dust haze prescription successfully corrects the red NIR slopes of red L dwarfs to look like the standard object.}
\figsetgrpend

\figsetgrpstart
\figsetgrpnum{8.37}
\figsetgrptitle{Figure of 1228-1547}
\figsetplot{f8_37.pdf}
\figsetgrpnote{The spectrum of a red L dwarf de-reddened by the dust haze prescription is compared to the spectra of the field standard object. Black is the standard, red is the red object, and green is the spectrum of the red object de-reddened by the forsterite extinction curve for the best fit parameters from MCMC fits. $\chi^2$ values before and after the de-reddening are also shown. The de-reddened spectrum fits the standard spectrum much better and has a smaller $\chi^2$ value than the original red spectrum. This shows that the proposed dust haze prescription successfully corrects the red NIR slopes of red L dwarfs to look like the standard object.}
\figsetgrpend

\figsetgrpstart
\figsetgrpnum{8.38}
\figsetgrptitle{Figure of 1239+5515}
\figsetplot{f8_38.pdf}
\figsetgrpnote{The spectrum of a red L dwarf de-reddened by the dust haze prescription is compared to the spectra of the field standard object. Black is the standard, red is the red object, and green is the spectrum of the red object de-reddened by the forsterite extinction curve for the best fit parameters from MCMC fits. $\chi^2$ values before and after the de-reddening are also shown. The de-reddened spectrum fits the standard spectrum much better and has a smaller $\chi^2$ value than the original red spectrum. This shows that the proposed dust haze prescription successfully corrects the red NIR slopes of red L dwarfs to look like the standard object.}
\figsetgrpend

\figsetgrpstart
\figsetgrpnum{8.39}
\figsetgrptitle{Figure of 0310-2756}
\figsetplot{f8_39.pdf}
\figsetgrpnote{The spectrum of a red L dwarf de-reddened by the dust haze prescription is compared to the spectra of the field standard object. Black is the standard, red is the red object, and green is the spectrum of the red object de-reddened by the forsterite extinction curve for the best fit parameters from MCMC fits. $\chi^2$ values before and after the de-reddening are also shown. The de-reddened spectrum fits the standard spectrum much better and has a smaller $\chi^2$ value than the original red spectrum. This shows that the proposed dust haze prescription successfully corrects the red NIR slopes of red L dwarfs to look like the standard object.}
\figsetgrpend

\figsetgrpstart
\figsetgrpnum{8.40}
\figsetgrptitle{Figure of 0624-4521}
\figsetplot{f8_40.pdf}
\figsetgrpnote{The spectrum of a red L dwarf de-reddened by the dust haze prescription is compared to the spectra of the field standard object. Black is the standard, red is the red object, and green is the spectrum of the red object de-reddened by the forsterite extinction curve for the best fit parameters from MCMC fits. $\chi^2$ values before and after the de-reddening are also shown. The de-reddened spectrum fits the standard spectrum much better and has a smaller $\chi^2$ value than the original red spectrum. This shows that the proposed dust haze prescription successfully corrects the red NIR slopes of red L dwarfs to look like the standard object.}
\figsetgrpend

\figsetgrpstart
\figsetgrpnum{8.41}
\figsetgrptitle{Figure of 0652+4710}
\figsetplot{f8_41.pdf}
\figsetgrpnote{The spectrum of a red L dwarf de-reddened by the dust haze prescription is compared to the spectra of the field standard object. Black is the standard, red is the red object, and green is the spectrum of the red object de-reddened by the forsterite extinction curve for the best fit parameters from MCMC fits. $\chi^2$ values before and after the de-reddening are also shown. The de-reddened spectrum fits the standard spectrum much better and has a smaller $\chi^2$ value than the original red spectrum. This shows that the proposed dust haze prescription successfully corrects the red NIR slopes of red L dwarfs to look like the standard object.}
\figsetgrpend

\figsetgrpstart
\figsetgrpnum{8.42}
\figsetgrptitle{Figure of 1438+5722}
\figsetplot{f8_42.pdf}
\figsetgrpnote{The spectrum of a red L dwarf de-reddened by the dust haze prescription is compared to the spectra of the field standard object. Black is the standard, red is the red object, and green is the spectrum of the red object de-reddened by the forsterite extinction curve for the best fit parameters from MCMC fits. $\chi^2$ values before and after the de-reddening are also shown. The de-reddened spectrum fits the standard spectrum much better and has a smaller $\chi^2$ value than the original red spectrum. This shows that the proposed dust haze prescription successfully corrects the red NIR slopes of red L dwarfs to look like the standard object.}
\figsetgrpend

\figsetgrpstart
\figsetgrpnum{8.43}
\figsetgrptitle{Figure of 1326-0038}
\figsetplot{f8_43.pdf}
\figsetgrpnote{The spectrum of a red L dwarf de-reddened by the dust haze prescription is compared to the spectra of the field standard object. Black is the standard, red is the red object, and green is the spectrum of the red object de-reddened by the forsterite extinction curve for the best fit parameters from MCMC fits. $\chi^2$ values before and after the de-reddening are also shown. The de-reddened spectrum fits the standard spectrum much better and has a smaller $\chi^2$ value than the original red spectrum. This shows that the proposed dust haze prescription successfully corrects the red NIR slopes of red L dwarfs to look like the standard object.}
\figsetgrpend

\figsetgrpstart
\figsetgrpnum{8.44}
\figsetgrptitle{Figure of 2212+3430}
\figsetplot{f8_44.pdf}
\figsetgrpnote{The spectrum of a red L dwarf de-reddened by the dust haze prescription is compared to the spectra of the field standard object. Black is the standard, red is the red object, and green is the spectrum of the red object de-reddened by the forsterite extinction curve for the best fit parameters from MCMC fits. $\chi^2$ values before and after the de-reddening are also shown. The de-reddened spectrum fits the standard spectrum much better and has a smaller $\chi^2$ value than the original red spectrum. This shows that the proposed dust haze prescription successfully corrects the red NIR slopes of red L dwarfs to look like the standard object.}
\figsetgrpend

\figsetgrpstart
\figsetgrpnum{8.45}
\figsetgrptitle{Figure of 2148+4003}
\figsetplot{f8_45.pdf}
\figsetgrpnote{The spectrum of a red L dwarf de-reddened by the dust haze prescription is compared to the spectra of the field standard object. Black is the standard, red is the red object, and green is the spectrum of the red object de-reddened by the forsterite extinction curve for the best fit parameters from MCMC fits. $\chi^2$ values before and after the de-reddening are also shown. The de-reddened spectrum fits the standard spectrum much better and has a smaller $\chi^2$ value than the original red spectrum. This shows that the proposed dust haze prescription successfully corrects the red NIR slopes of red L dwarfs to look like the standard object.}
\figsetgrpend

\figsetgrpstart
\figsetgrpnum{8.46}
\figsetgrptitle{Figure of 2244+2043}
\figsetplot{f8_46.pdf}
\figsetgrpnote{The spectrum of a red L dwarf de-reddened by the dust haze prescription is compared to the spectra of the field standard object. Black is the standard, red is the red object, and green is the spectrum of the red object de-reddened by the forsterite extinction curve for the best fit parameters from MCMC fits. $\chi^2$ values before and after the de-reddening are also shown. The de-reddened spectrum fits the standard spectrum much better and has a smaller $\chi^2$ value than the original red spectrum. This shows that the proposed dust haze prescription successfully corrects the red NIR slopes of red L dwarfs to look like the standard object.}
\figsetgrpend

\figsetend

\begin{figure}
%\figurenum{8}
\plotone{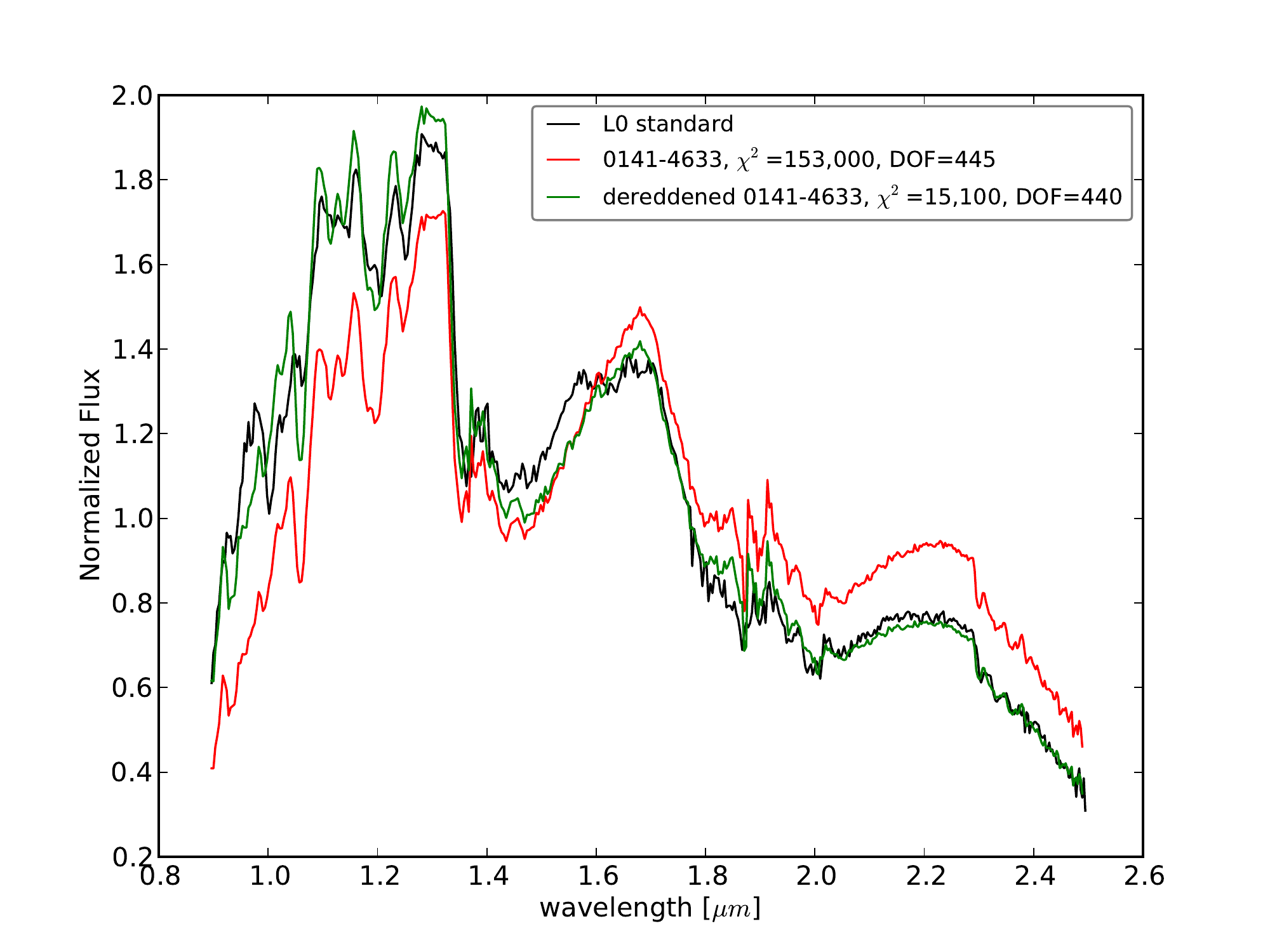}
\caption{The spectrum of a red L dwarf de-reddened by the dust haze prescription is compared to the spectra of the field standard object. Black is the standard, red is the red object, and green is the spectrum of the red object de-reddened by the forsterite extinction curve for the best fit parameters from MCMC fits. $\chi^2$ values before and after the de-reddening are also shown. The de-reddened spectrum fits the standard spectrum much better and has a smaller $\chi^2$ value than the original red spectrum. This shows that the proposed dust haze prescription successfully corrects the red NIR slopes of red L dwarfs to look like the standard object.}
\end{figure}

\begin{figure}
%\plotone{combo.pdf}
%\includegraphics[max size={\textwidth}{\textheight}]{combo.pdf}
%\figurenum{9.1}
%\plotone{../Plots/combo_s1.pdf}
\includegraphics[width=\textwidth,height=\textheight,keepaspectratio]{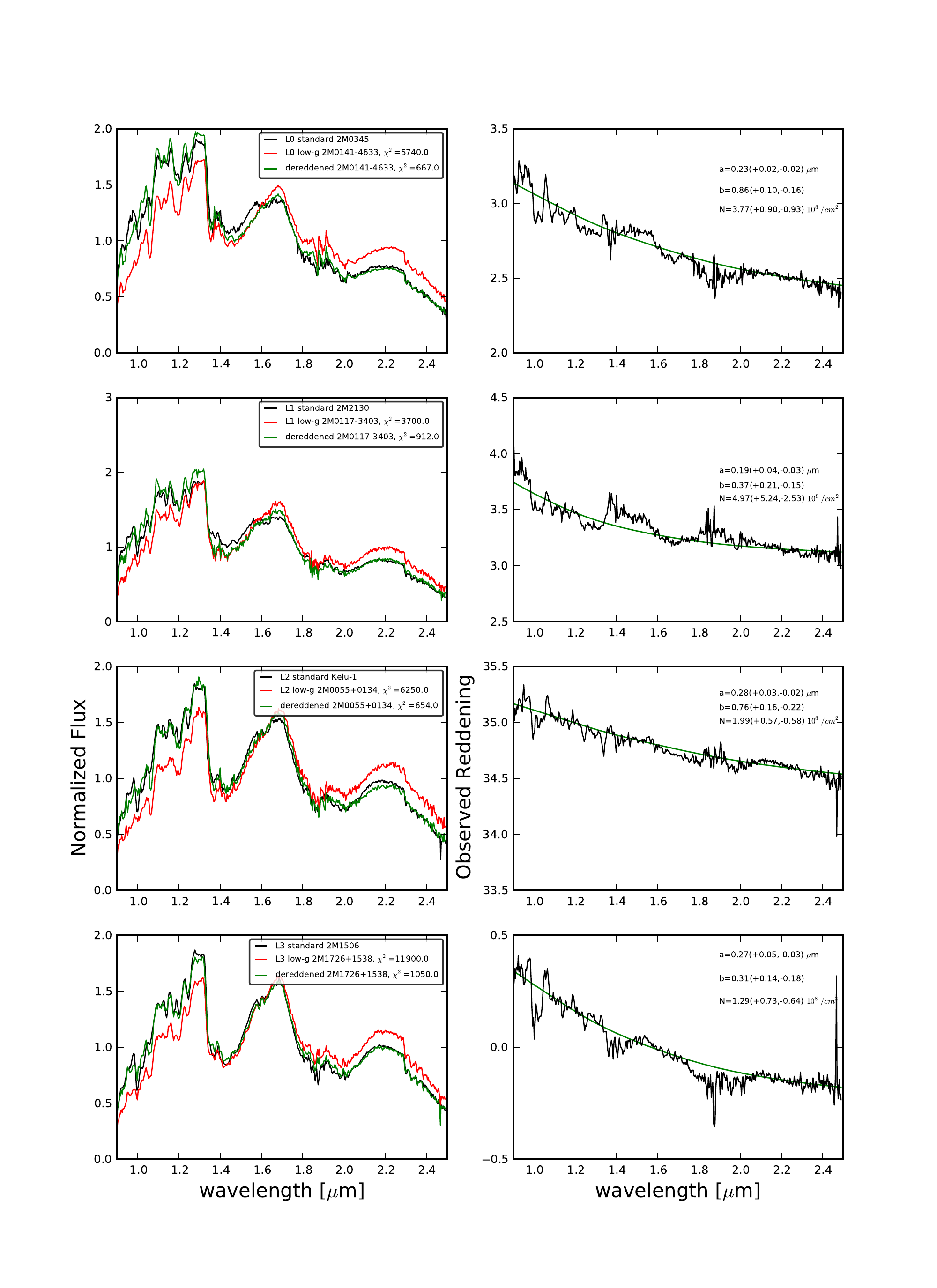}
\caption{An example of de-reddened spectra (equivalent to Figure Set~8) and model fits (equivalent to Figure Set~7) for spectral types L0--L3. The left column shows the spectral standard (black), the red L dwarf (red), and the de-reddened (green) spectra. The right column shows the best-fit model (green) and the observed reddening (black).}
\label{fig: combo1}
\end{figure}

\begin{figure}
%\plotone{combo.pdf}
%\includegraphics[max size={\textwidth}{\textheight}]{combo.pdf}
\figurenum{9.2}
%\plotone{../Plots/combo_s1.pdf}
\includegraphics[width=\textwidth,height=\textheight,keepaspectratio]{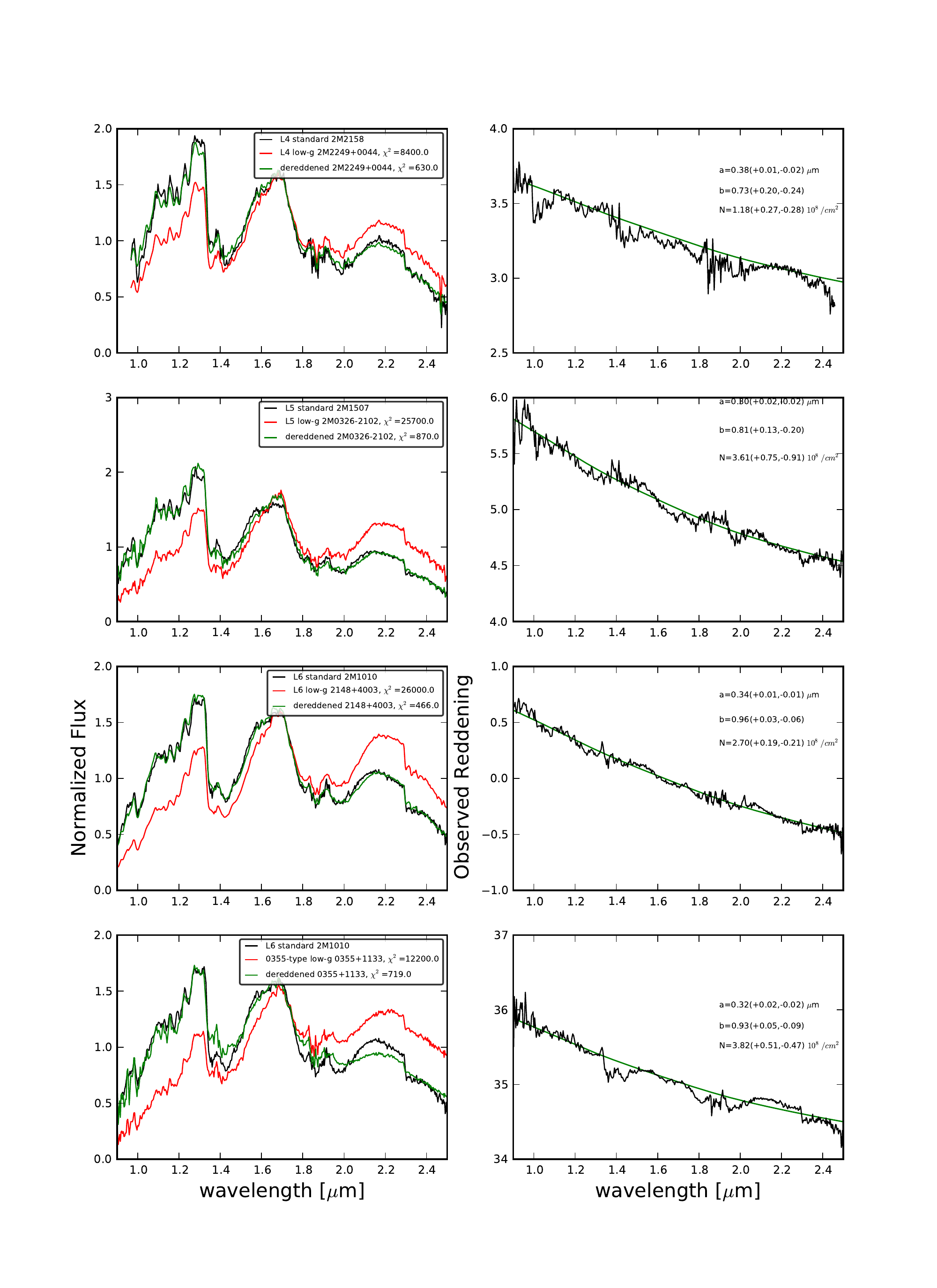}
\caption{Same as Figure~\ref{fig: combo1} for spectral types L4--L6 including 0355-type. The left column shows the spectral standard (black), the red L dwarf (red), and the de-reddened (green) spectra. The right column shows the best-fit model (green) and the observed reddening (black).}
\label{fig: combo2}
\end{figure}

\begin{figure}
\plotone{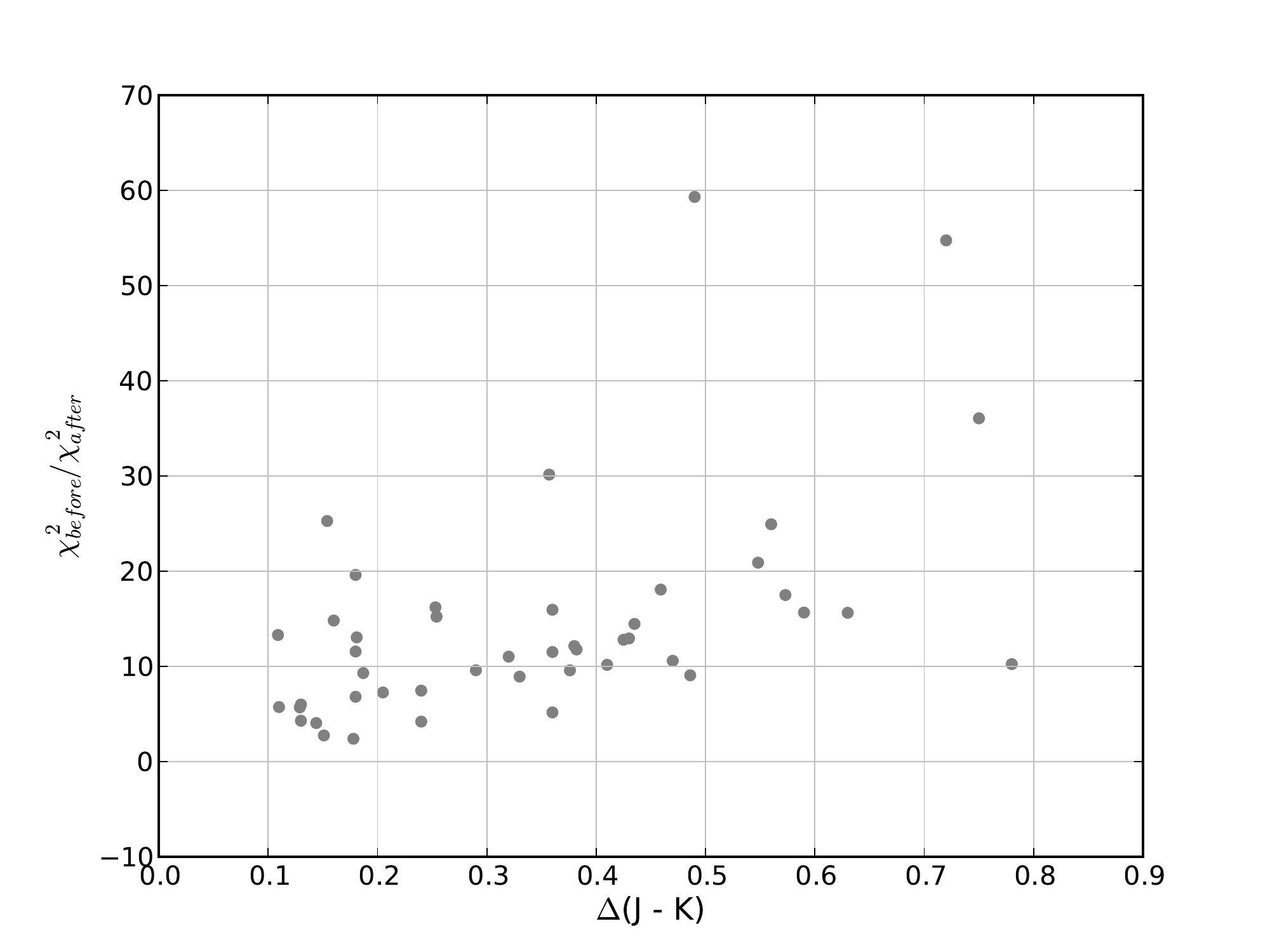}
\caption{Improvement on $\chi^2$ due to the dust haze prescription. The ratio of $\chi^2_{\rm before}$ to $\chi^2_{\rm after}$ is plotted against $\Delta(J-K)$ color. $\chi^2_{\rm before}$ is $\chi^2$ between standard and red L dwarf spectra, and $\chi^2_{\rm after}$ is $\chi^2$ between standard and corrected red L dwarf spectra. $\chi^2$ value is reduced after the correction for all the objects. }
\label{fig: dchisq}
\end{figure}

\begin{figure}
%\plotone{f9.pdf}
\plotone{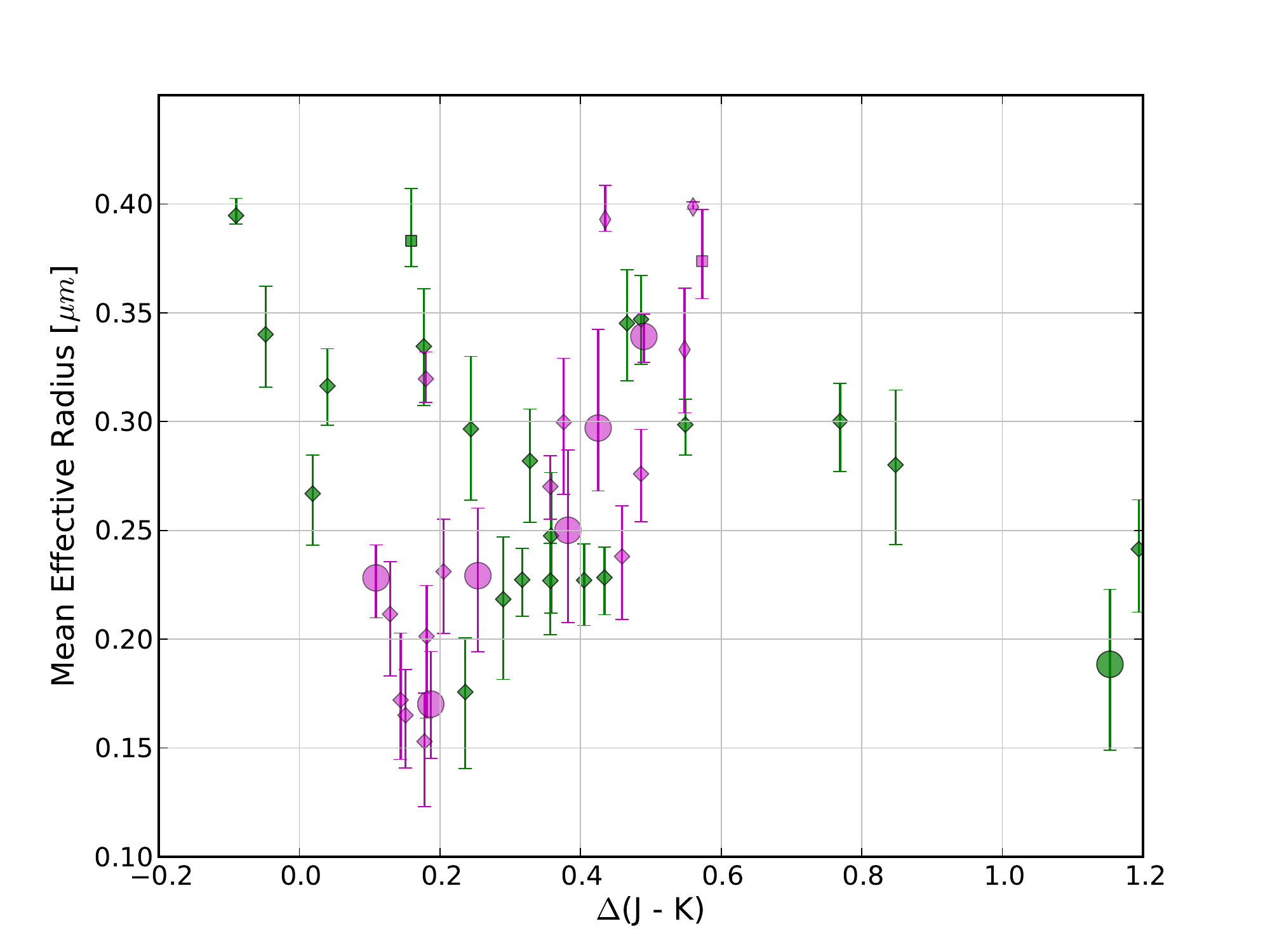}
\caption{A scatter plot of mean effective radius $a~ [\rm\micron]$ against $\Delta(J-K)$ color. Green markers denote low-gravity L dwarfs and magenta markers denote field-aged red L dwarfs both ranging between L0 -- L5. Circles denote objects with PDFs with clear peaks. Diamonds denote objects with PDFs for $b$ close to the limit. Thin diamonds denote objects with PDFs for $a$ close to the limit. Squares denote objects with PDFs for both $a$ and $b$ close to the limits. Our sample includes object with $\Delta(J-K)$ $>$ 0.1. There appears to be a linear correlation between radius and $\Delta(J-K)$ for the field-gravity objects, while there is no noticeable trend for the low-gravity objects. The distributions of low-gravity and field-gravity objects are distinct.}
%There is no noticeable trend in radius in relation to color. There is also no visible difference between young and red field L dwarfs. 
%Combine with next figure?

\label{fig: ascat}
\end{figure}

\begin{figure}
%\plotone{f10.pdf}
\plotone{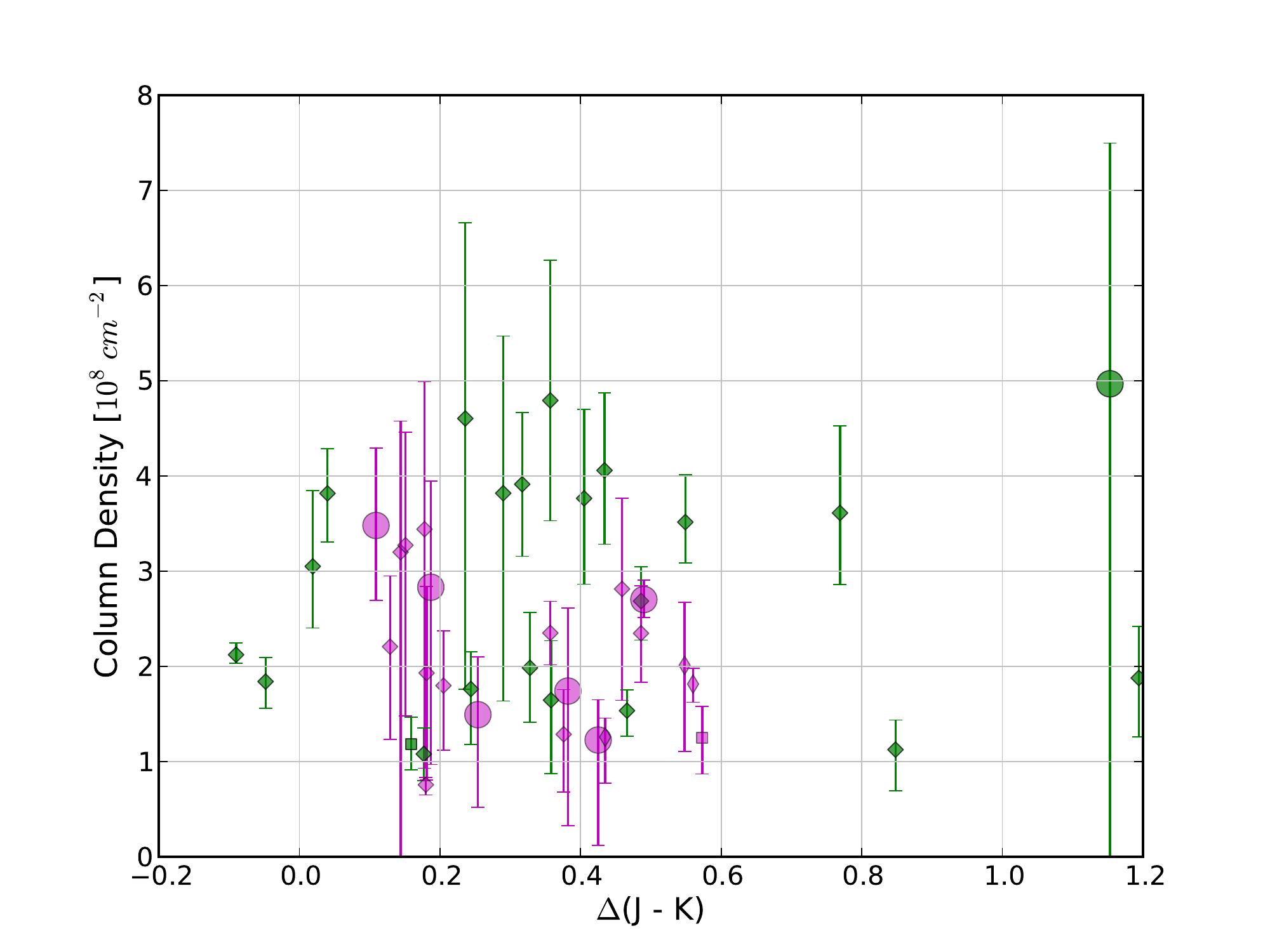}
\caption{A scatter plot of column density $N~[10^8\rm cm^{-2}]$ against $\Delta(J-K)$ color. Green markers denote low-gravity L dwarfs and magenta markers denote field-aged red L dwarfs both ranging between L0 -- L5. Circles denote objects with PDFs with clear peaks. Diamonds denote objects with PDFs for $b$ close to the limit. Thin diamonds denote objects with PDFs for $a$ close to the limit. Squares denote objects with PDFs for both $a$ and $b$ close to the limits. Our sample includes object with $\Delta(J-K)$ $>$ 0.1. The field-gravity objects appear to have lower column densities than the low-gravity objects. There is no visible trend in column density in relation to $\Delta(J-K)$ color for both low-gravity and field-gravity objects.}
%There is no noticeable trend in radius in relation to color. There is also no visible difference between young and red field L dwarfs. }
\label{fig: nscat}
\end{figure}

% Rotate table
\begin{deluxetable}{llllllll}
\rotate
\tablecaption{L dwarfs used in this paper}
\tablecolumns{8}
\tablewidth{630pt}
\tabletypesize{\scriptsize}
\tablehead{\colhead{2MASS} & \colhead{Sp. Type} & \colhead{2MASS} & \colhead{2MASS} & \colhead{Discovery} & \colhead{Spectral Type} & \colhead{SpeX Prism} & \colhead{SpeX Prism}\\
\colhead{Designation} & \colhead{Type} & \colhead{$J-K_s$} & \colhead{$\Delta$$(J-K_s)$\tablenotemark{a}} & \colhead{Reference} & \colhead{Reference} & \colhead{Reference} & \colhead{Observation Date}}
\startdata
\cutinhead{Spectral Standard L Dwarfs}
0345432+254023 & L0 & 1.33 & 0 & \citet{Kirkpatrick97} &\citet{Kirkpatrick99} & \citet{BM06}\\
% DB says Cruz et al. in prep  \textbf{ASK KELLE}\\
2130446-084520 & L1 & 1.33  & 0  &  \citet{Kirkpatrick08}
%(not in db) 
&  \citet{Kirkpatrick08} & \citet{Kirkpatrick10}\\%\textbf{NOT IN DB}\\
13054019-2541059& L2 & 1.67  & 0 &  \citet{Ruiz97} %\textbf{NOT IN DB}
& \citet{Kirkpatrick99}& \citet{Burg07}\\ %2009ApJS..185..289R (Rayn09),
1506544+132106 & L3 & 1.63  & 0 & \citet{Gizis00}&\citet{Gizis00} & \citet{Burg07}\\ % DB says 2005ApJ...623.1115C (Cush05)\\
21580457-1550098 & L4 & 1.86  & 0 & \citet{Kirkpatrick08}&\citet{Kirkpatrick08} & \citet{Kirkpatrick10}\\
1507476-162738 & L5 & 1.46  & 0 & \citet{Reid00} & \citet{Kirkpatrick00}& \citet{Burg07}\\%\textbf{NOT IN DB?}\\
10101480-0406499 & L6 & 1.89 & 0 & \citet{Cruz03} & \citet{Cruz03} & \citet{Reid06}\\
\cutinhead{Low-gravity L Dwarfs}
01415823-4633574 & L0$\gamma$ & 1.74 & 0.410 & \citet{Kirkpatrick06} & \citet{cruz09} & \citet{Kirkpatrick06}\\
00325584-4405058 & L0$\gamma$ & 1.51 & 0.178 & \citet{Gold99} & \citet{cruz09} & \citet{BG14} \\
%& 2007 Nov 13\\
02103857-3015313 & L0$\gamma$ & 1.57 & 0.236 & Cruz et al. in prep & Cruz et al. in prep & This Paper & 2003 Sep 04\\
0241115-032658 & L0$\gamma$ & 1.76 & 0.434 &\citet{Cruz07} & \citet{cruz09} & This Paper & 2006 Aug 21\\
03231002-4631237& L0$\gamma$ & 1.69 & 0.357 &\citet{Reid08} & \citet{Cruz07} & This Paper & 2007 Nov 13\\
2213449-213607 & L0$\gamma$ & 1.62 & 0.290 & \citet{Cruz07} & \citet{cruz09} & \citet{BG14} \\
%& 2008 Aug 29\\
23153135+0617146 & L0$\gamma$ & 1.80 & 0.466 &  Cruz et al. in prep  & Cruz et al. in prep & This Paper & 2007 Nov 14\\
1711135+232633 & L0$\gamma$  & 1.44 & 0.113 & %420
\citet{Cruz07} & Cruz et al. in prep & This Paper &  2008 Jul 13\\
15525906+2948485 & L0$\beta$  & 1.46 & 0.126 & %80
\citet{Wilson03} & \citet{cruz09} & \citet{BG14}\\
%& 2008 Jul 30\\
%0241-0326 & L0  & 1.764 & 0.434 & %98
%\citet{Cruz07} &  Cruz et al. in prep \textbf{ASK KELLE}\\
03572695-4417305 & L0$\beta$  & 1.46 & 0.127 & %225
\citet{Bouy03} & \citet{cruz09} & \citet{BG14}\\
% & 2007 Sep 16\\
0117474-340325 & L1$\gamma$ & 1.69 & 0.358 & \citet{Cruz03} & \citet{Gagne15} & This Paper & 2006 Aug 21\\
0518461-275645& L1$\gamma$ & 1.65 & 0.317 & \citet{Cruz07} & \citet{Gagne15} & \citet{BG14}\\
% & 2011 Dec 08\\
%U20037$\_$2M0045+1634$\_$adam & L2 & \\
00550564+0134365 & L2$\gamma$ & 2.00 & 0.328 & Cruz et al. in prep & Cruz et al. in prep & This Paper & 2003 Sep 04\\
0536199-192039 & L2$\gamma$ & 1.91 & 0.244 & \citet{Cruz07} & \citet{Gagne15} & This Paper & 2003 Sep 03\\
%  & 2008 Jul 1\\
%0001+1535 & L3 & 1.812 & 0.182 & \citet{Knap04} &? & This Paper &  2003 Sep 03\\
%Spex Prism$\_$2208+2921$\_$U40004 & L3 & \\
15515237+0941148 & L3:$\gamma$ & 2.01 & 0.149 & \citet{Reid08} & \citet{Gagne15} & This Paper & 2008 Jul 13\\
1726000+153819 & L3.5$\gamma$ & 1.81 & 0.380 & \citet{Kirkpatrick00} & \citet{Allers13} & \citet{BG14}\\
%2massj0126+1428spex & L4 & \\
05012406-0010452 & L4$\gamma$ & 2.02 & 0.159 & \citet{Reid08} & \citet{cruz09}& \citet{Filippazzo15}\\
%U20622$\_$1538-1953 & L4 & \\
%Spex Prism$\_$2206-4217$\_$080714 & L4 & \\
22495345+0044046 & L4$\gamma$ & 2.22 & 0.369 & \citet{Geba02} & \citet{Gagne15} & \citet{Allers10}\\
05120636-2949540 & L5$\gamma$ & 2.18 & 0.718 & \citet{Cruz03} & \citet{Gagne15}& \citet{BG14}\\
% & 2011 Dec 07\\
03264225-21020572 & L5$\beta/\gamma$ & 2.21 & 0.752 & %872
\citet{Gizis03} & \citet{Gagne15} & This Paper & 2007 Nov 13\\
21543454-1055308 & L5$\beta/\gamma$ & 2.24 & 0.381 & %313
\citet{Gagne14} & \citet{Gagne15} & This Paper & 2003 Aug 11\\
%20025073-0521524 & L5--L7$\gamma$ & 1.90 & 0.439 & \citet{Cruz07} &\citet{Gagne15}& \citet{Burg08d}\\
%\enddata
03552337+1133437 & 0355-type\tablenotemark{b} & 2.52 & 1.06 & \citet{Reid08} &\citet{Gagne15} & \citet{faherty13}\\
1615425+495321& 0355-type\tablenotemark{b} & 2.48 & 0.623 & \citet{Cruz07} &\citet{Gagne15}& This Paper & 2005 Mar 23 \\
%01033203+1935361 & L6$\beta$ & 2.14 & 0.25 & \citet{Kirkpatrick00} & \citet{Kirkpatrick00} & \citet{BG14}\\
\cutinhead{Red field L dwarfs}
%\tablehead{\colhead{Object} & \colhead{Spectral Type} & \colhead{$J-K_s$} & \colhead{$\Delta$$J-K_s$} & \colhead{Source ID}}
%\startdata
%J115442.09-340038.8 & L0  & 1.345 & \\
%J144137.01-094559.1 & L0
% (L0.5)
% & 1.359\\
%J232029.43+412342.2 & L1 & 1.393 & \\
%12123389+0206280 & L1 & 1.936 & 0.606 & %316
%2012A$\&$A...548A..53L (Lodi12),2003yCat.2246....0C(Cutri03) & 2009AJ....137....1F(Fahe09) & This Paper? & 2005 Mar 23\\
%J192230.79+661020.6 & L1 & 1.409\\
%J153921.29+650237.1 & L1 & 1.404\\
%02192196+0506306 & L1 & 1.491 & 0.161 & %472
%2012A$\&$A...548A..53L (Lodi12),cutri03? &  & This Paper &  2003 Sep 05\\
%1531134+164128 & L1 & 1.779 & 0.449 & %494
%2012A$\&$A...548A..53L (Lodi12) & 2009AJ....137....1F(Fahe09) & This Paper & 2003 Sep 04 \\
%J101707.58+130838.6 & L1 & 1.386\\
02355993-2331205 & L1 & 1.48 & 0.154 & %567
\citet{Burg08b} & \citet{Gizis01} & \citet{Burg08b}\\
%1512+3403 & L1 & 1.627 & 0.297 & %588
% & \textbf{NOT IN DB}\\
%J072314.69+572705.6 & L1 & 1.357\\
%None & L1 & 1.339\\
%J205754.09-025231.1 & L1 & 1.397\\
 % & 2011 Dec 08\\
05431887+6422528 & L1 & 1.52 & 0.187 & %768
\citet{Reid08} & \citet{Reid08} &\citet{BG14}\\
06022216+6336391 & L1: & 1.58 & 0.263 & %766
 \citet{Reid08}& \citet{Reid08} &\citet{BG14}\\
% & 2011 Dec 08\\
%J18071593+5015316 & L1
% (L1.5)
% & 1.332\\
%g1963b & L2 & 2.053\\
%J092839.57-160312.3  & L2 & 1.707\\
%J231531.39+061714.2  & L2 & 1.796 & 0.126\\
%J213952.22+214839.4 & L3  & 1.819 & 0.189\\
%05161597-3332046 & L3:  & 1.892 & 0.262 & %448
%2012A$\&$A...548A..53L (Lodi12) & 2014AJ....147...34S(Schn14) & \citet{Burg10a} & 2004 Nov 07\\
%0327409-314815 & L3  & 1.942 & 0.312 & %589
%2012A$\&$A...548A..53L (Lodi12) &? & This Paper & 2003 Sep 04\\
%J071716.23+570543.6 & L3  & 1.691\\
00165953-4056541 & L3 
%(L3.5) 
 & 1.88 & 0.254 &%143
\citet{Kirkpatrick08} & \citet{Kirkpatrick08} &\citet{Burg10c}\\
23392527+3507165 & L3.5 
%(L3.5) 
& 1.77 & 0.144 & %145
\citet{Reid08}  & \citet{Reid08} & This Paper & 2003 Sep 04\\
11000965+4957470  & L3.5 
%(L3.5) 
& 1.81 & 0.178 & %445
\citet{Reid08} & \citet{Reid08} & This Paper& 2004 Nov 08\\
00511078-1544169 & L3.5 
%(L3.5)
 & 1.81 & 0.181 & %583
\citet{Kirkpatrick00}  & \citet{Kirkpatrick00} & \citet{Burg10a}\\
%J2224438-015852 & L3/L4 
%(L3.5/L4.5)
% & 2.051\\
%J144825.80+103157.9 & L4 & 1.876\\
%040707.56+154645.1 & L4 & 1.919\\
23174712-4838501 & L4pec & 1.97 & 0.109 & %432
\citet{Reid08} & \citet{Reid08} & \citet{Kirkpatrick10} \\ %\textbf{NOT IN DB}\\
%J002503.65+475919.1 & L4 & 1.938\\
%J003324.08-152130.5 & L4 & 1.876\\
%J184108.66+311728.5 & L4 & 1.938\\
%2M01614+4953 & L4 & 2.479\\
0337036-175807 & L4.5 
%(L4.5)
 & 2.04 & 0.180 & %252 
\citet{Kirkpatrick00}  & \citet{Kirkpatrick00} & \citet{BG14}\\
% & 2011 Dec 08\\ 
%0447430-193604 & L5 & 1.957 & 0.497 & %136
%2012A$\&$A...548A..53L (Lodi12) & ? & This Paper & 2003 Sep 04\\
02082363+2737400 & L5 & 1.84 & 0.382 & %174
 \citet{Kirkpatrick00}&  \citet{Kirkpatrick00} & \citet{Burg10a}\\
0835425-081923 & L5 & 2.03 & 0.573 & %216
\citet{Cruz03} & \citet{Cruz03} & This Paper & 2004 Nov 08\\
03582255-4116060  & L5 & 2.01 & 0.548 & %223
\citet{Reid08} & \citet{Reid08} & This Paper & 2007 Nov 14\\
09054654+5623117 & L5 & 1.67 & 0.205 & %239
\citet{Reid08} & \citet{Reid08} & \citet{Burg10a}\\
12281523-1547342 & L5 & 1.61 & 0.151 & %269
\citet{Delf97} & \citet{Burg10c}& \citet{Burg10a}\\
12392727+5515371 & L5 & 1.92 & 0.459 & %281
\citet{Kirkpatrick00} & \citet{Burg10a} & \citet{Burg10a}\\
03101401-2756452 & L5 & 1.84 & 0.376 & %326
\citet{Cruz07} & \citet{Cruz07} & \citet{BG14}\\
%0835+1953 & L5 & 1.775 & 0.315 & %356
%\citet{Chiu06} & \\
06244595-4521548 & L5 & 1.89 & 0.425 & %376
\citet{Reid08} & \citet{Reid08} & \citet{BG14}\\
% & 2005 Mar 24\\
0652307+471034 & L5 & 1.82 & 0.357 & %395
\citet{Cruz03} & \citet{Cruz03} & This Paper & 2004 Nov 07\\
14383259+5722168 & L5 & 1.59 & 0.129 & %655
\citet{Zhang09} & \citet{Zhang09} & \citet{BG14}\\
1326298-003831 & L5 
%(L5.5)
 & 1.90 & 0.435 & %401 
\citet{Fan00} & \citet{Cruz03} & \citet{Geba02}\\
%0107+0041 & L5
%(L5.5)
%  & 2.115 & 0.665 & %686 
%\citet{Geba02} & & \citet{Burg10c}\\
22120703+3430351& L5: & 1.95 & 0.486 & %724
\citet{Reid08} & \citet{Reid08} & This Paper & 2004 Sep 05\\
%08503593+1057156 & L6 & 1.99 & 0.10 & \citet{Kirkpatrick99} &  \citet{Kirkpatrick99} & \citet{Burg11} \\
%21321145+1341584 & L6 & 1.96 & 0.07 & \citet{Cruz07} & \citet{Cruz07} & \citet{Siegler07}\\
21481633+4003594 & L6 & 2.38 & 0.49 & \citet{Looper08} &  \citet{Looper08} &  \citet{Looper08}\\
22443167+2043433 & L6.5 &  2.45 & 0.56 & \citet{Dahn02} & \citet{Kirkpatrick08} & \citet{Looper08}\\

\enddata
\vspace{-0.3cm}
%If an object has a non-integer spectral type (shown in parentheses), it was rounded down (I need to rephrase this). 
%Spectral standards and red field L dwarfs are from \citet{Kirkpatrick10}. Low-gravity L dwarfs are from \citep{cruz09}.
\tablenotetext{a}{$\Delta$$J-K_s$ is the difference in $J-K_s$ colors between the red L dwarf and the corresponding spectral standard object.}
\tablenotetext{b}{0355-type objects are defined by \citet{Gagne15}. A conservative estimate of L3--L6$\gamma$ is adopted for the spectral type range. For our analysis, 0355-type objects are compared to the L6 spectral standard object.}
%, investigate the object 'None' which has neither name nor designation. Maybe add $\Delta$$J-K$}
\label{table: data}
\end{deluxetable}

\end{document}